\documentclass[12pt]{article}

\usepackage[left=2.8cm,right=2.8cm,top=2.8cm,bottom=2.8cm]{geometry}

\usepackage[colorlinks=true,linkcolor=black,citecolor=black,urlcolor=blue,filecolor=black]{hyperref}

\usepackage{amsmath}
\usepackage{amsfonts}
\usepackage[nosort]{cite}
\usepackage[normalem]{ulem}

 \csname
@addtoreset\endcsname{equation}{section}

\def\a{\alpha}
\def\b{\beta}
\def\g{\gamma}

\def\d{\delta}

\def\e{\epsilon}

\def\z{\zeta}
\def\h{\eta}
\def\th{\theta}

\def\k{\kappa}
\def\l{\lambda}
\def\L{\Lambda}
\def\m{\mu}
\def\n{\nu}
\def\x{\xi}

\def\p{\pi}

\def\r{\rho}

\def\s{\sigma}

\def\vf{\varphi}

\def\o{\omega}

\def\cF{{\cal F}}

\def\cJ{{\cal J}}

\def\cL{{\cal L}}
\def\cM{{\cal M}}

\def\cO{{\cal O}}

\def\cU{{\cal U}}

\def\cW{{\cal W}}

\def\cZ{{\cal Z}}

\def\be{\begin{equation}}
\def\ee{\end{equation}}
\def\bea{\begin{eqnarray}}
\def\eea{\end{eqnarray}}
\def\ba{\begin{array}}
\def\ea{\end{array}}

\def\nn{\nonumber}

\def\tr{\text{tr}}

\def\ww{\wedge}

\def\comma{\,,\,}

\def\pe{\prime}
\def\12{\frac{1}{2}}
\def\pr{\partial}
\def\prd{\partial \cdot}

\begin{document}


\vspace{30pt}

\begin{center}


{\Large\sc Asymptotic symmetries of three-dimensional\\[6pt] higher-spin gravity: the metric approach} \\


\vspace{25pt}
{\sc Andrea Campoleoni${}^{\; a,}$\footnote{Postdoctoral Researcher of the Fund for Scientific Research-FNRS Belgium.} and Marc Henneaux${}^{\; a, b}$}

\vspace{10pt}
{${}^a$\sl\small
Universit{\'e} Libre de Bruxelles\\
and International Solvay Institutes\\
ULB-Campus Plaine CP231\\
1050 Brussels,\ Belgium
\vspace{10pt}

${}^b$\sl \small Centro de Estudios Cient{\'\i}ficos (CECs)\\ Casilla 1469, Valdivia, Chile\\
\vspace{10pt}

{\it andrea.campoleoni@ulb.ac.be, henneaux@ulb.ac.be} 
}

\vspace{50pt} {\sc\large Abstract} \end{center}

\noindent
The asymptotic structure of three-dimensional higher-spin anti-de Sitter gravity is analyzed in the metric approach, in which the fields are described by completely symmetric tensors and the dynamics is determined by the standard Einstein-Fronsdal action improved by higher order terms that secure gauge invariance.  Precise boundary conditions are given on the fields.  The asymptotic symmetries are computed and shown to form a non-linear $W$-algebra, in complete agreement with what was found in the Chern-Simons formulation. The $W$-symmetry generators are two-dimensional traceless and divergenceless rank-$s$ symmetric tensor densities of weight $s$ ($s = 2, 3, \cdots$), while asymptotic symmetries emerge at infinity through the conformal Killing vector and conformal Killing tensor equations on the two-dimensional boundary, the solution space of which is infinite-dimensional.  For definiteness, only the spin 3 and spin 4 cases are considered, but these illustrate the features of the general case: emergence of the $W$-extended conformal structure,  importance of the improvement terms in the action that maintain gauge invariance, necessity of the higher spin gauge transformations of the metric, role of field redefinitions.

\newpage


\tableofcontents

\newpage

\section{Introduction}\label{sec:intro}

The asymptotic symmetries of three-dimensional higher-spin gravity \cite{Blencowe1,Blencowe2,Vasiliev:1995dn,Prokushkin:1998bq} have been shown recently to be remarkably rich and to be described by the direct sum of two copies of a nonlinear $W$-algebra, one for each chiral sector \cite{HR,spin3,Wlambda}.  This generalizes the earlier result of \cite{BH} for pure three-dimensional anti-de Sitter gravity, for which one gets two copies of the Virasoro algebra.  The emergence at infinity of the $W$-symmetry paved the way to new insight into the AdS/CFT correspondence \cite{GG,GH,minimal-rev}.

The  derivation of the asymptotic symmetries was performed in \cite{HR,spin3} using the Chern-Simons formulation of the higher spin theory.  While extremely powerful, this approach is clearly tailored to three spacetime dimensions since the Chern-Simons reformulation is not available in four or higher spacetime dimensions.  For this reason it is useful to investigate the asymptotic properties of three-dimensional higher-spin gravity in terms of the metric and the higher spin fields, described by the Einstein and Fronsdal-like actions \cite{Fronsdal,metric-like}, which are also relevant to higher spacetime dimensions.  This paper fulfills this goal.\footnote{It is true that there exist connection-based first-order formulations also for $D > 3$, but these do not possess a ``standard'' action principle.}  The use of the metric approach might also shed light on matter couplings \cite{Prokushkin:1998bq}, or on the introduction of a topological mass \cite{Damour:1987vm,Chen:2011vp,Bagchi:2011vr,Campoleoni:2011tn}. 

Given an action, there is no systematic procedure for deriving a unique set of consistent boundary conditions. The obtention of the boundary conditions is somewhat of an art.  Indeed, there can be different consistent sets of boundary conditions for a given action, corresponding to different physical situations. In the search for consistent boundary conditions, one is guided by a few principles:
\begin{itemize}
\item The boundary conditions should contain the physical solutions that one wants to investigate.
\item They should be invariant under a group of transformations that contains (and may be bigger than) the group of expected symmetries, e.g., the Poincar\'e group for asymptotically flat spaces, or the anti-de Sitter group for asymptotically anti-de Sitter space.
\item The boundary conditions should ensure that the charges generating the infinitesimal asymptotic symmetries mentioned in the previous point are finite.
\end{itemize}
We provide here boundary conditions on the metric $g_{\l \m}$ and higher spin fields $\phi_{\l_1 \l_2 \cdots \l_s}$ which obey these principles: (i) they contain the solutions described in \cite{spin3,metric-like};  (ii) they are invariant under the $W$-symmetry; (iii) the $W$-charges are finite.

To achieve this task, we proceed as follows.  First we motivate a set of boundary conditions within the metric formulation through various considerations, namely the form of the boundary conditions in the pure spin-2 case and the behavior at infinity of the known solutions, and how they transform under the exact symmetries of the background described by the $AdS_3$ Killing tensors, which have a definite fall-off.  Following the philosophy that ``the proof of the pudding is in the eating", we then explicitly verify that these boundary conditions fulfill the three requirements listed above.  The check of the first two requirements is rather straightforward.  The proof of the third requirement -- that the charges are finite -- requires first the identification of the charges.  This could be done using Noether theorem, but we use here a shortcut: we identify them by direct comparison with the Chern-Simons formulation. Once the charges have been determined, one can not only verify that they are finite (this is guaranteed by their identification with the CS charges known to be finite) but one can also independently compute their algebra within the metric formalism. We rederive explicitly that the charges fulfill a non-linear $W$-algebra.   

For definiteness, we consider only the spin-3 and spin-4 cases, which illustrate well the general procedure and ideas.  In fact, the central points appear already in the spin-3 case.  The spin-4 case is also covered here to exhibit the technical difficulties encountered in the analysis of higher spins. As we shall see, the computations are indeed rather intricate in the spin-4 case, in contrast to those of the Chern-Simons formulation. We also restrict the analysis to the so-called principal embedding of $sl(2,\mathbb{R})$ into $sl(N,\mathbb{R})$.

Our paper uses the second-order Lagrangian formalism throughout (except the reminder on the pure gravity case given in the next section). A Hamiltonian analysis of the boundary conditions and the charges will be reported elsewhere \cite{InPrep}.

The emergence of the asymptotic $W$-extended conformal structure in higher-spin gravity is a direct generalization of the emergence of the conformal structure in pure gravity.  In the metric approach, this structure emerges through the residual coordinate transformations and higher-spin gauge symmetries that preserve the boundary conditions on the metric and the higher-spin fields.  We show how the conformal Killing vector equations and conformal Killing tensor equations directly arise in this asymptotic analysis.  An asymptotic symmetry of the theory with higher-spin fields up to spin $s$ turns out to be completely parameterised (modulo pure gauge terms) by traceless conformal Killing tensors up to rank $s-1$ of the two-dimensional metric at infinity. 

Our analysis is Lorentzian throughout.  In the Euclidean version of the theory with black hole topology \cite{BH1}, 
the temporal components $g_{0 \mu}$, $h_{0 \mu_1 \cdots \mu_{s-1}}$ of the metric and higher spin fields are related to the inverse temperature and chemical potentials of the angular momentum and higher-spin charges.   They may not take the pre-defined values at infinity given in this paper and for this reason, more flexibility is needed in their asymptotic behavior.
This question was considered in \cite{Henneaux:2013dra,Bunster:2014mua} for the Chern-Simons formulation. Its metric translation is left for future work. 

Our paper is organized as follows.  In Section \ref{sec:spin2}, we recall the metric derivation of the asymptotic symmetries of three-dimensional anti-de Sitter pure gravity \cite{BH}, which preceded in fact the Chern-Simons derivation which was performed later \cite{asympt-CS}.  This is achieved in a manner that prepares the ground for the generalization to higher spins.  We write the boundary conditions and rederive the conformal symmetry at infinity using Schouten brackets and contravariant components, which turns out to simplify the derivation. We point out that the Virasoro generators ${\mathcal L}^{ij}$ appear, from the point of view of the two-dimensional geometry at infinity, as conserved, symmetric, traceless rank-2 tensor densities of weight 2. Next, in Section \ref{sec:spin3}, we consider the coupled spin-2 -- spin-3 system. We provide boundary conditions on the spin-3 field.  We also point out that the boundary conditions on the metric must be strengthened compared with the pure spin-2 case, which can consistently be done.  We show that the boundary conditions are invariant under transformations generated not only by conformal Killing vectors $\e^i$ of the two-dimensional conformal geometry at infinity, but also by rank-2 conformal Killing tensors $\chi^{ij}$. The associated generators ${\mathcal W}^{ijk}$ are conserved, symmetric, traceless rank-3 tensor densities of weight 3.  We compute the algebra, and find the same nonlinear $W_3$-algebra as in the Chern-Simons approach.  The fact that the metric transforms under the spin-3 gauge transformations plays here an essential role.

The need to control an increasing number of subleading terms in the metric and higher spin fields as one increases the maximum spin of the fields involved in the model is a generic phenomenon confirmed in the spin-4 case, to which we turn in Section \ref{sec:spin4}.  The new symmetries are now parametrized by \mbox{rank-3} conformal Killing tensors $\sigma^{ijk}$ of the two-dimensional geometry at infinity, and the associated generators ${\mathcal U}^{ijkl}$ are conserved, symmetric, traceless rank-4 tensor densities of weight 4.  Again, the algebra is found to perfectly match the nonlinear $W_4$-algebra of the Chern-Simons approach.  A new feature appears in the spin-4 case: it is that the self-interactions between the higher spin fields, which come in addition to their gravitational interactions, and the corresponding improvement terms in the gauge transformations, remain relevant asymptotically.  In particular, the coupling constant of the $3-3-4$ vertex enters the asymptotic algebra and can be interpreted as a parameter labeling the different conformal structures that can appear at infinity \cite{triality}.

We then indicate in Section \ref{sec:conclusions} how the analysis generalizes to higher spins and give final comments.  A collection of appendices provide technical information about conventions (Appendix \ref{app:conventions}), isometry algebra of the ``vacuum", i.e., of anti-de Sitter space with zero higher spin field configurations (Appendix \ref{sec:iso}), derivation of the boundary conditions in the metric-like formulation from the boundary conditions in the Chern-Simons formulation (Appendix \ref{app:fields}), more detailed structure of the action and of the gauge transformations for the combined spin-2, spin-3, spin-4 system (Appendices \ref{app:vertex1} and  \ref{app:vertex2}).

\section{Warming up with $AdS_3$ gravity}\label{sec:spin2}

\subsection{Hamiltonian form of the boundary conditions}

The dynamical variables of three-dimensional pure gravity are the spatial metric $g_{ab}$ \mbox{($a,b = 1,2$)} and its conjugate momentum $\pi^{ab}$.  The other components of the metric --  the lapse $N$ and the shift $N^a$ -- are the Lagrange multipliers for the Hamiltonian constraint ${\mathcal H} \approx 0$ and the momentum constraint ${\mathcal H}_a \approx 0$.   

The boundary conditions on the spatial metric and its momentum were given in \cite{BH} and read,
\be
g_{rr} = r^{-2} + \frac{4 \pi}{k}\frac{M_1(\phi)}{r^4} + \cO(r^{-6})\, , \quad g_{r \phi} = \cO(r^{-3})\, , \quad \ g_{\phi \phi} = r^2 + \frac{4 \pi}{k} M_2(\phi) + \cO(r^{-2}) \label{AsMetric01}
\ee
and
\be
\pi^{rr} = \cO(r^{-1})\, , \quad \pi^{r \phi} = \frac{J(\phi)}{2r^2} + \cO(r^{-4})\, , \quad \pi^{\phi \phi} = \cO(r^{-5})
\label{AsMetric02}
\ee
so that
\be
\pi^r{}_{\phi} = \frac{J(\phi)}{2} + \cO(r^{-2}) \, .
\ee
Here, we have set the $AdS$ radius to $\ell = 1$ and we have explicitly written, besides the background  terms (first terms in $g_{rr}$ and $g_{\phi \phi}$), the subleading terms that contribute to the charges (\ref{Charges0}) given below (terms involving $M_1$, $M_2$ and $J$, which are arbitrary functions of $\phi$).  The constant $k$ is a dimensionless constant proportional to the ratio between the $AdS$ radius and Newton's constant, $k = \ell/4G$. The normalization has been chosen so as to simplify the comparison with the discussion of asymptotic symmetries in the Chern-Simons formulation \cite{asympt-CS}, where $k$ denotes the level of the Chern-Simons action.

As shown in \cite{BH}, these boundary conditions are preserved under transformations generated by the constraints,
\be \label{Hgenerator}
H[\xi^\perp, \xi^a] = \int dr d \phi \left(\xi^\perp {\mathcal H} + \xi^a {\mathcal H}_a \right)+ Q[\xi^{\perp,0}, \xi^{\phi, 0}] \, ,
\ee
provided the surface deformation parameters $\xi^\perp$ and $\xi^a$ fulfill
\begin{subequations} \label{AsSymm-tot}
\begin{eqnarray}
\xi^\perp &=& r\, \xi^{\perp, 0}(\phi) + \frac{\alpha(\phi)}{r} + \cO(r^{-3}) \, , \label{AsSymm01} \\[2pt]
\xi^r  &=&   r\, \beta(\phi) + \cO(r^{-1}) \, , \label{AsSymm02} \\[2pt]
\xi^\phi  &=&  \xi^{\phi, 0}(\phi) + \frac{\gamma(\phi)}{r^2} + \cO(r^{-4}) \, ,  \label{AsSymm03}
\end{eqnarray}
\end{subequations}
where (i) $\xi^{\perp,0}(\phi)$ and $\xi^{\phi,0}(\phi)$ are arbitrary functions of $\phi$,  and (ii) the functions $\alpha(\phi)$, $\beta(\phi)$ and $\gamma(\phi)$ are definite functions of the leading orders $\xi^{\perp,0}(\phi)$ and $\xi^{\phi,0}(\phi)$ and also, in the Hamiltonian formalism, of the relevant subleading terms $M_1(\phi)$, $M_2(\phi)$ and $J(\phi)$ appearing in the expansion of the canonical variables.  These functions $\alpha, \beta, \gamma$ are determined by the requirement that the transformation generated by (\ref{Hgenerator}) indeed preserves the boundary conditions (see Appendix of \cite{BH} for a detailed discussion and an explanation of some of the subtleties).  For instance, in the particular case $\xi^{\perp, 0}(\phi) = 1$, $\xi^{\phi, 0}(\phi)= 0$, somewhat tedious but straightforward computations yield
\be
\alpha(\phi) = \frac{2 \pi}{k} M_2(\phi) \, , \; \; \beta(\phi) = 0 \, , \; \; \gamma(\phi) = \frac{2 \pi}{k} J(\phi) \; \; \; \; \; (\hbox{for }\xi^{\perp, 0}(\phi) = 1, \, \xi^{\phi, 0}(\phi)= 0) \, .\label{alphabetagamma}
\ee

The next subleading orders in \eqref{AsSymm-tot} are undetermined but correspond to ``proper gauge transformations" in the terminology of \cite{Benguria:1976in} and so have no physical significance. 

In the generator (\ref{Hgenerator}), the term $Q[\xi^{\perp,0}, \xi^{\phi, 0}]$ is the surface term at infinity that must be added to the bulk piece of $H[\xi^\perp, \xi^a]$ so that $H[\xi^\perp, \xi^a]$ has well-defined functional derivatives \cite{Regge:1974zd}. Explicitly,
\be
Q[\xi^{\perp,0}, \xi^{\phi, 0}] = \int d \phi \left\{\xi^{\perp,0}(\phi)M(\phi)   +  \xi^{\phi,0}(\phi) J(\phi) \right\} , \label{Charges0}
\ee
with $M(\phi) \equiv M_1(\phi)  + 2 M_2(\phi)$. To reach (\ref{Charges0}), we have inserted the asymptotic form of the canonical variables in the formula giving the charges on top of page 222 of \cite{BH} -- where units were chosen so that $16 \pi G = 1$ -- and dropped the constant (first) term, which corresponds to adjusting the charges to be zero for the zero mass black hole.  

The asymptotic symmetries are thus characterized by two arbitrary functions of $\phi$, namely, $\xi^{\perp,0}(\phi)$ and $\xi^{\phi,0}(\phi)$.  As shown in \cite{BH} where the asymptotic algebra is computed, these two arbitrary functions describe the conformal algebra in two dimensions, the two independent Virasoro generators being  $\cL$ and $\tilde{\cL}$, with
\be
M = \cL + \tilde{\cL} \, , \qquad J = \cL - \tilde{\cL} \, .
\ee
The surface integral (\ref{Charges0}) can be rewritten
\be
Q[\xi^+, \xi^-] = \int d \phi \left\{ \xi^+(\phi) \cL(\phi) + \xi^-(\phi) \tilde{\cL}(\phi) \right\} \label{Charges1}
\ee
with $\xi^\pm = \xi^{\perp,0} \pm \xi^{\phi,0}$.  The central charge is $\frac{3 \ell}{2 G}$.

In order to proceed, it is useful to simplify the boundary conditions. As one sees from (\ref{Charges0}), it is only the linear combination $M(\phi) \equiv M_1(\phi) + 2 M_2(\phi)$ that appears in the expression of the charges.  Now, under radial redefinitions 
\be r = r' + \frac{K(\phi)}{r'} \label{RadialRedef}
\ee
which preserves the asymptotic conditions (\ref{AsMetric01}) and (\ref{AsMetric02}), the functions $M_1(\phi)$ and $M_2(\phi)$ are not separately invariant,
$$ M_1 \rightarrow M_1' = M_1 - 4 K \, , \; \; \; M_2 \rightarrow M'_2 = M_2 + 2 K\, , $$ but the charge $M_1(\phi) + 2 M_2(\phi)$ is invariant, as it should.  The radial change of coordinates (\ref{RadialRedef}) is a proper gauge transformation that can be used to set either $M_1(\phi)$ or $M_2(\phi)$ equal to zero.  In standard Schwarzschild coordinates, one sets $M_2(\phi) =0$. For our purposes, it will be more convenient to set instead $M_1(\phi) = 0$.  This fixes the radial coordinate up to order $\cO(r'^{-3})$ (the other coordinates being kept fixed).  We thus use from now on the equivalent set of boundary conditions
\be
g_{rr} = r^{-2}  + \cO(r^{-6})\, , \quad g_{r \phi} = \cO(r^{-3})\, , \quad g_{\phi \phi} = r^2 + \frac{2 \pi}{k} M(\phi) + \cO(r^{-2}) \label{AsMetric1}
\ee
and
\be
\pi^{rr} = \cO(r^{-1})\, , \quad \pi^{r \phi} = \frac{J(\phi)}{2r^{2}} + \cO(r^{-4})\, , \quad \pi^{\phi \phi} = \cO(r^{-5})
\label{AsMetric2}
\ee
for which the surface term at infinity giving the charges reads
\be
Q[\xi^{\perp,0}, \xi^{\phi, 0}] =  \int d \phi \left\{\xi^{\perp,0}(\phi) M(\phi)  +  \xi^{\phi,0}(\phi) J(\phi) \right\} . \label{Charges2}
\ee

\subsection{Covariant form of the boundary conditions}
The boundary conditions were given above in terms of the phase space variables.  To make the generalization to higher spins more direct, it is convenient to rewrite them in terms of the Lagrangian variables, i.e., the spacetime metric $g_{\lambda \mu}$.  This is easy to do if one recalls that phase space can be identified with the space of solutions of the equations of motion. We shall thus integrate the equations of motion asymptotically to get the asymptotic form of the spacetime metric, with (\ref{AsMetric1}) and (\ref{AsMetric2}) as initial conditions.  To that end, we first need to specify the lapse and the shift.

The lapse and the shift, which parametrizes the surface deformation being performed in the actual motion in time, must define asymptotic symmetries, i.e., must belong to the class \eqref{AsSymm-tot}.  In the Minkowskian version of the theory with time ranging from $- \infty$ to $+ \infty$, which we are considering, it is customary to take the functions $\xi^{\perp, 0}(\phi)$ and  $\xi^{\phi, 0}(\phi)$ entering the lapse and the shift as $\xi^{\perp, 0}(\phi) = 1$ and $\xi^{\phi, 0}(\phi)= 0$, so that one marches in time orthogonally to the surfaces $t = const$, in a manner such that $ds= r dt$ (with coefficient one) asymptotically.  This is a particular choice that does not represent the most general motion compatible with the asymptotic symmetry, but it is one that can always be reached within the allowed surface deformation freedom. For definiteness, we shall from now on restrict the motion to that case.

Other choices of lapse and shift might be necessary  in different contexts, e.g., to discuss black hole thermodynamics through the Euclidean continuation \cite{Carlip:1994gc}.  They can easily be covered but this will not be done here.

We thus take (see (\ref{alphabetagamma}))
\begin{subequations}
\begin{eqnarray}
N &=& r  + \frac{2 \pi}{k}\frac{M}{2r} + \cO(r^{-3}) \, , \label{AsSymmN0} \\[2pt]
N^r  &=&    \frac{\delta}{r} + \cO(r^{-3}) \, , \label{AsSymmNr0} \\[2pt]
N^\phi  &=&  \frac{2 \pi}{k}\frac{J}{r^2} + \cO(r^{-4}) \, . \label{AsSymmNphi0}
\end{eqnarray}
\end{subequations}
The  term $\frac{\delta(t,\phi)}{r}$, which was present but  not exhibited in (\ref{AsSymm02}),  is explicitly written here because it corresponds to the definite compensating proper gauge transformation that must accompany the motion in order to maintain the extra gauge condition $M_1(\phi) = 0$ that we have imposed on the radial coordinate.  Given that it generates a proper gauge transformation, its explicit expression is not of great interest and will not be given here.  
This yields for the spacetime metric components at the initial time 
\be
g_{rr} = r^{-2}  + \cO(r^{-6})\, , \quad g_{r \phi} = \cO(r^{-3})\, , \quad  g_{\phi \phi} = r^2 + \frac{2 \pi}{k} M(\phi) + \cO(r^{-2}) \label{AsMetric2a}
\ee
and
\be
g_{tt} = -r^{2}  + \frac{2 \pi}{k} M(\phi) + \cO(r^{-2})\, , \quad g_{r t} = \cO(r^{-3})\, , \quad  g_{t \phi} = \frac{2 \pi}{k}J(\phi)+ \cO(r^{-2}) \, . \label{AsMetric2b}
\ee
Note that one has $h_{tt} = h_{\phi \phi}$ and $g_{t \phi} = N_\phi = 16 \pi G\, \pi^r_{\; \phi}$ to leading order (with $g_{tt} = -r^2 + h_{tt}$, $g_{\phi \phi} = r^2 + h_{\phi \phi}$).  To get the spacetime metric at all times, one needs to determine the time dependence of the two functions $M$ and $J$, or what is the same, $\cL$ and $\tilde{\cL}$.  With $\xi^\perp$ asymptotically equal to $1$ and $\xi^\phi$ asymptotically equal to zero, the generator of time translations is $\int d \phi M = \int d \phi (\cL + \tilde{\cL})$. The time dependence of $\cL$ and $ \tilde{\cL}$ is obtained by taking the bracket with the generator of time translations and follows from the Virasoro algebra, since $\int d \phi (\cL + \tilde{\cL})$ is one of the Virasoro generators. One gets $\dot{\cL} = \cL'$ and $\dot{\tilde{\cL}} = - \tilde{\cL}' $ and therefore
\be
\cL = \cL(x^+) \, , \qquad \tilde{\cL} = \tilde{\cL}(x^-)  \, ,
\ee
where $x^\pm = t \pm \phi$.

Thus, the covariant phase space description of the boundary conditions is
\be
g = g_{rr}{dr^2} + 2\, g_{ri} dr dx^i + g_{ij}(r,x^n) dx^i dx^j \, ,\label{AsMetric3a}
\ee
with $x^{i} = \{t,\phi\}$ and 
\be g_{rr} =  r^{-2}  + \cO(r^{-6})\, , \quad g_{rj} = \cO(r^{-3})  \label{AsMetric3b}
\ee
and
\be
g_{ij} = \frac{r^2}{2}\, \eta_{ij} + \frac{2 \pi}{k} \cL_{ij}  + \cO(r^{-2}) \, .  \label{AsMetric3c}
\ee
Here, $\h_{ij}$ is the \emph{flat} two-dimensional metric $\h_{ij} = \textrm{diag}(-1,1)$, which will be used to raise and lower indices, while $\cL_{ij}$ is a traceless and conserved tensor, 
$$ \partial^i \cL_{ij} = 0 \, , \qquad \h^{ij} \cL_{ij} = 0 \,,$$
i.e., 
\be \label{L_comp}
\cL_{++} = \cL(x^+) \, , \qquad \cL_{--} = \tilde{\cL}(x^-) \, , \qquad \cL_{+-} = 0 \, .
\ee
In (\ref{AsMetric3a}), we have also rescaled the radial coordinate $r$ to conform with the Fefferman-Graham conventions \cite{FG}.

\subsection{Re-derivation of the invariance under the conformal group}\label{sec:conf-metric}

The above boundary conditions contain the known exact solutions to $2+1$ gravity \cite{banados_review,skenderis}, given in the  Fefferman-Graham gauge \cite{FG} by\footnote{When compared with the original Hamiltonian boundary conditions, the Fefferman-Graham gauge involves two additional steps: (i) First, a choice of the radial coordinate, which can be reached by a ``proper" gauge transformation as we explained; (ii) Second, the choice $\xi^{\perp, 0}(\phi) = 1$ and $\xi^{\phi, 0}(\phi)= 0$ for the lapse and the shift, which can be reached by an ``improper" gauge transformation.  This corresponds to a definite choice of the conformal transformation at infinity defined by the motion in time.}
\be
g = \frac{dr^2}{r^2} + g_{ij}(r,x^n) dx^i dx^j \, ,
\ee
with
\be \label{gij}
g_{ij} = \frac{r^2}{2}\, \h_{ij} + \frac{2\p}{k}\, \cL_{ij} + \frac{2\p^2}{k^2\,r^2}\, \cL_{ik} \cL_j{}^k \, .
\ee
In particular,  anti-de Sitter space is recovered by setting
\be
\cL = \tilde{\cL} = -\frac{k}{8\p} \, ,
\ee
while the BTZ black hole ``at rest" has  $\cL = \frac12(M+J)$ and $\tilde{\cL} = \frac12 (M-J)$ with $M$ and $J$ arbitrary constants such that $M \geq 0$ \cite{BTZ,BHTZ}.

These boundary conditions are guaranteed to be invariant under the conformal group in two dimensions, since they are the covariant transcription of the phase space boundary conditions, which have been shown to be so \cite{BH} as we have recalled. It is however instructive to rederive the conformal invariance directly from (\ref{AsMetric3a}), (\ref{AsMetric3b}) and (\ref{AsMetric3c}).  This makes also our discussion self-contained.  The derivation  of asymptotic conformal invariance in the metric formulation was actually also done in \cite{BH}, but we shall repeat it explicitly here in a different way more adapted to the higher spin extension: we shall use the contravariant form of the boundary conditions.  This is because the generalization of the Lie bracket, namely, the Schouten bracket\cite{Nijenhuis}, is naturally defined for contravariant tensors, and this is the geometrical differential operation that  appears when investigating invariance conditions.  Indeed, not only is the variation of any contravariant tensor $T^{\mu_1 \cdots \mu_k}$ under the infinitesimal diffeomorphism generated by the vector field $v^\mu$ given by (minus) its Lie derivative along $v^\mu$, which is equal to its Schouten bracket with $v^\mu$, but the higher spin gauge transformations can also be expressed to leading order in terms of the Schouten bracket of the inverse metric with the higher spin gauge parameters.

More information on the Schouten bracket is given in Appendix \ref{app:conventions}.  Another difference with the treatment of \cite{BH}, which was entirely off-shell, is that the present analysis is performed within the covariant phase space, i.e., dynamical equations of motion can be used when needed.  When higher spins are included, this turns out to be necessary up to some power of $r^{-1}$ that depends on the spin.

In contravariant form, the boundary conditions read
\be \label{bnd2}
g^{\,rr} = r^2 + \cO(r^{-2}) \, , \qquad
g^{\,ri} = \cO(r^{-3}) \, , \qquad
g^{\,ij} = \frac{2}{r^2}\, \h^{ij} - \frac{8\p}{k\,r^4}\, \cL^{ij} + \cO(r^{-6}) \, .
\ee
Asymptotic symmetries correspond to diffeomorphisms which leave the form of \eqref{bnd2} invariant. It is easy to check that the vectors $v^\m$ that generate them, called asymptotic Killing vectors, 
have the same leading dependence on the radial coordinate as the Killing vectors of $AdS_3$, i.e.,
\begin{subequations}
\begin{align} \label{v}
v^r & = r\, \z(x^k) + \frac{\z_1(x^k)}{r} + \cO(r^{-3}) \, , \\[5pt]
v^i & = \e^i(x^k) + \frac{\e_1{}^i(x^k)}{r^2} + \cO(r^{-4}) \, .
\end{align}
\end{subequations} 
From the transformation rule $\delta g^{\mu \nu} = [g, v]^{\mu \nu}$ where $[g, v]^{\mu \nu}$ is the Schouten bracket, $[g, v]^{\mu \nu} = g^{\rho \nu} \partial_\rho v^\mu + g^{\mu \rho} \partial_\rho v^\nu   - v^\rho \partial_\rho g^{\mu \nu} = -\, \cL_v g^{\mu \nu}$, one then gets the variation of the inverse metric as
\begin{subequations} \label{dg_grav}
\begin{align}
\d g^{\,rr} & = -\,4\, \z_1 + \cO(r^{-2}) \, , \\
\d g^{\,ri} & = \frac{2}{r} \left\{ -\, \e_1{}^i + \pr^{\,i} \z \right\} + \cO(r^{-3}) \, ,
\end{align}
\end{subequations}
and\footnote{Indices
between parentheses are meant to be symmetrized with weight one, i.e.\  one divides the symmetrized expression by the number of terms that appears in it, so that symmetrization of a symmetric tensor reproduces the tensor without factor (projector). A fuller account of our conventions is given in Appendix \ref{app:conventions}.}
\be
\begin{split}
\d g^{\,ij} & = \frac{4}{r^2}\! \left\{ \pr^{(i} \e^{\,j)} + \h^{ij} \z \right\} + \frac{4}{r^4}\! \left\{ \pr^{(i} \e_1{}^{j)} + \h^{ij} \z_1 - \frac{2\p}{k} [\, \cL , \e \,]^{ij} - \frac{8\p}{k} \cL^{ij} \z \right\} + \cO(r^{-6}) \, ,
\end{split} \label{232}
\ee
where $[\, \cL , \e \,]^{ij}$ is the Schouten bracket in two dimensions, $[\, \cL , \e \,]^{ij} = \cL^{ik} \partial_k \e^j + \cL^{kj} \partial_k \e^i - \e^k \partial_k \cL^{ij}$. 
The variation preserves the form of $g^{ij}$ only if the terms $\cO(r^{-2})$ vanish and this implies
\be \label{solconf2}
2\,\pr^{(i} \e^{j)} - \h^{ij} \prd \e = 0 \, , \qquad\qquad
\z = -\, \frac{1}{2}\,\prd \e \, . 
\ee
The first condition is the conformal Killing equation for a two-dimensional vector, which is solved by 
\be
\e^{+} = \e(x^+) \, , \qquad\quad \e^{-} = \tilde{\e}(x^-) \, ,
\ee
while the second condition completely determines $\z$ in terms of this conformal Killing vector. 

Imposing now that the terms displayed explicitly in \eqref{dg_grav} vanish leads to
\be \label{solconf2_2}
\e_1{}^i = -\,\frac{1}{2}\,\pr^{\,i}\prd\e \, , \qquad\qquad \z_1 = 0 \, .
\ee
With this information, one can compute how asymptotic symmetries act on the space of solutions, i.e.\ $\d\cL^{ij}$.  One gets from (\ref{232}) 
\be
\d \cL^{ij} = [\, \cL , \e \,]^{ij} - 2\, \cL^{ij} \prd \e + \frac{k}{4\p}\, \pr^{\,i} \pr^{\,j} \prd \e \, , \label{QVariation1}
\ee
or equivalently,
\be
\d\cL = - \left( \e\, \cL' + 2\, \e' \cL \right) + \frac{k}{4\p}\, \e''' \, , \qquad \d\tilde{\cL} =  - \left( \tilde{\e}\, \tilde{\cL}' + 2\, \tilde{\e}^{\,\pe} \tilde{\cL} \right) + \frac{k}{4\p}\, \tilde{\e}^{\,\pe\pe\pe} \, , \label{QVariation2}
\ee
where the prime denotes a derivative with respect to the argument.  This transformation rule for $\cL^{ij}$ is compatible with the tracelessness and transverseness conditions. It is interesting to observe that conversely, imposing that the $\cO(r^{-4})$ piece in (\ref{232}) be traceless and transverse determines $\e_1{}^i $ and $\z_1$.

One thus finds again that the asymptotic symmetries are described by two arbitrary functions ($\e^+$ and $\e^-$) of one argument ($x^+$ or $x^-$).  The commutator of two asymptotic symmetries is equal to the Lie bracket of the corresponding vector fields and is given, up to irrelevant pure gauge subleading terms, by the algebra of the conformal Killing vectors in two dimensions, i.e., the conformal algebra in two dimensions.

\subsection{Comments}

A couple of comments are in order:
\begin{enumerate}
\item \label{FirstPoint}The form of the boundary conditions (\ref{bnd2}) can be characterized as follows:
\begin{itemize}
\item The angular components of the deviation from the background (i.e., $g^{ij} - \frac{2}{r^2}\, \eta^{ij}) $ are such that if one lowers the indices with the background metric, one gets terms of order one, $g_{ij} - \frac{r^2}{2}\, \eta_{ij} = \cO(1)$); these $\cO(1)$-terms are the charges, which obey conservation laws and tracelessness conditions (ensuring that there are only two independent charges).
\item As one replaces one angular index $i$ by one radial index $r$, one increases the order of the background deviation by $r$, i.e., $g^{ri} = r \, \cO\left( g^{ij} - \frac{2}{r^2}\, \eta^{ij} \right)$ and $g^{rr} - r^2 = r \, \cO\left(g^{ri} \right)$.
\end{itemize}  
These rules are consistent, in the sense that (i) they contain the known relevant solutions and (ii) are invariant under asymptotic symmetries which form the conformal group and contain the Killing vectors of the anti-de Sitter background (which are some of the symmetries, forming the so-called ``wedge subalgebra").  These asymptotic symmetries are completely specified, up to irrelevant terms, by 
boundary conformal Killing vectors.

\item These rules for establishing  the boundary conditions are equivalent to the rules that come from the standard Hamiltonian formalism.  The rules can alternatively be derived from the Chern-Simons formulation if one knows the boundary conditions in that formulation, using the map between the metric-like fields and the CS connection given in Appendix \ref{app:fields}.  Actually, for pure gravity, it was the opposite route that was followed, to derive the CS boundary conditions from the metric formulation \cite{asympt-CS}.

\item The set of rules given in point \ref{FirstPoint} are not complete, in that they do not enable one to identify the charges to the $\cO(1)$ terms in the angular components of the background deviation of the (covariant) metric. To do that, one needs to use the action.  It is not sufficient to rely only on symmetry considerations.  However, once one knows  what the charges are, one can read off their algebra from their variations (\ref{QVariation1}) under asymptotic symmetries, since these variations are generated by the charges themselves through the Poisson bracket.  We shall in the sequel borrow the information on what the charges are from the Chern-Simons formulation. The algebra computed within the metric formulation will then be found to coincide with the ($W$-)algebra obtained in the Chern-Simons context, as of course it should.

\item Conformal geometry at infinity: with our choice of boundary conditions, the metric induced on the cylinder at infinity  is the flat metric $\eta_{ij}$ in Minkowskian coordinates.  On could adopt different coordinates at infinity.  Furthermore, it is actually only the conformal class of the metric that is  in fact determined since by the rescaling of $r$, $r \rightarrow r e^{ \Phi(x^i)}$, one may replace $\eta_{ij}$ by $e^{2 \Phi(x^i)} \eta_{ij}$.  Such transformations lead to equivalent descriptions of the boundary conditions.  It is useful to explicitly verify the covariance, under these transformations,  of the quantities and of the equations that characterize the theory at infinity.   If $g_{ij}$ is the metric at infinity ($= \eta_{ij}$ with our choices), we set $\bar{g}_{ij} = \frac{g_{ij}}{\sqrt{-g}} $ and $\bar{g}^{ij} = \sqrt{-g} g^{ij}$. These are, respectively, a  rank-2 covariant tensor density of weight $-1$ and a rank-2 contravariant tensor density of weight $1$, which do not depend on the choice of representative $g_{ij}$ within the conformal class. The conformal Killing equation (\ref{solconf2}) can be rewritten equivalently as $[\bar{g},\epsilon]^{ij}= \lambda \bar{g}^{ij}$ for some $\lambda$, where the Schouten bracket is computed as if $\bar{g}^{ij}$ were an ordinary tensor (without density weight), i.e. $[\bar{g},\epsilon]^{ij}= \bar{g}^{mj} \partial_m \epsilon^i  + \bar{g}^{im} \partial_m \epsilon^j - \epsilon^m \partial_m \bar{g}^{ij}$.  This explicitly displays its invariance under Weyl rescalings of the metric.  Although the Shouten bracket $[\bar{g},\epsilon]^{ij}$ does not transform homogeneously as a tensor density under changes of coordinates, the conformal Killing equation is invariant under changes of coordinates because the terms by which $[\bar{g},\epsilon]^{ij}$ fails to be a tensor density are proportional to $\bar{g}^{ij}$, i.e., have the form of the right-hand side of the conformal Killing equation (so $\lambda$ is not a scalar density and its transformation matches the transformation of the left-hand side).  The easiest way to verify covariance under changes of coordinates is of course to rewrite the conformal Killing equation as $D^j \epsilon^i + D^i \epsilon^j = \mu g^{ij} $ for some $\mu$, which is now a scalar.  Here $D_i$ is the torsionless covariant derivative associated with an arbitrarily chosen two-dimensional metric in the class $\{e^{2 \Phi(x^i)} \eta_{ij} \}$. As we shall see, these direct considerations generalize to higher order conformal Killing tensors. 

The Virasoro generator $\cL^{ij}$ is a rank-2 contravariant tensor density of weight 2.  The traceless condition $\cL^{ij} \bar{g}_{ij} = 0$ is obviously invariant under Weyl rescalings of the metric.  The same property holds for the divergenceless condition $D_j \cL^{ij}= 0$, which may be rewritten as $[\bar{g}, \cL]^{ijk} \bar{g}_{jk} = 0$ (taking into account the tracelessness condition), an expression which is clearly invariant under Weyl rescalings since it involves only the Weyl invariants $\bar{g}^{ij}$ and $\bar{g}_{ij}$.  Here, $[\bar{g}, \cL]^{ijk}$ is again computed without taking into account the density weight of $\bar{g}^{ij}$ and $\cL^{ij}$. Though itself not a tensor density, its contraction $[\bar{g}, \cL]^{ijk} \bar{g}_{jk}$ is. Note that the variation (\ref{QVariation1}) is, apart from the central charge term, (minus) the Lie derivative of a a rank-2 contravariant tensor density of weight 2, as it should.  Note also that the vector density of weight one $j^i[\epsilon] = \cL^{ij} \epsilon_j \equiv \cL^{ij} \epsilon^k \bar{g}_{jk}$ is conserved ($\partial_i j^i[\epsilon] = 0$) for any conformal Killing vector $\epsilon^i$.

\end{enumerate}

\section{Spin-3 field coupled to gravity}\label{sec:spin3}

In this section we consider a rank-3 tensor coupled to three-dimensional gravity as in \cite{metric-like}, assuming that no tensors of higher rank are present. We thus deal with the metric-like counterpart of a $sl(3,\mathbb{R}) \oplus sl(3,\mathbb{R})$ Chern-Simons theory with principal embedding of the gravitational  subalgebra $sl(2,\mathbb{R}) \oplus sl(2,\mathbb{R})$ into $sl(3,\mathbb{R}) \oplus sl(3,\mathbb{R})$.  The aim is to derive the asymptotic symmetries of the coupled spin-2 -- spin-3 system.  The derivation illustrates how the non-linearities that characterize the asymptotic symmetries of three-dimensional higher-spin gauge theories emerge in the metric-like setup.

\subsection{Action \& gauge transformations}\label{sec:action3}

\subsubsection{Action}
At lowest order in an expansion in the spin-3 field the interacting action contains the minimal coupling of Einstein gravity to the free Fronsdal action \cite{Fronsdal}:
\be \label{action3} 
I_{\{3\}} =\int \frac{d^3x\, \sqrt{-g}}{16\p G} \left\{\! \left( R \,+\,
\frac{2}{\ell^2} \right) + \phi^{\,\m\n\r} \left( \cF_{\m\n\r}
- \frac{3}{2} \, g_{(\m\n}\, \cF_{\r)} \right) + \cL_{NM} \right\} +\,
\cO\!\left(\phi^4\right) ,
\ee 
where we have temporarily reinstated $\ell$ for completeness purposes and where $\cF_{\m\n\r}$ is the covariantised Fronsdal tensor,
\be \label{fronsdal3}
\cF_{\m\n\r} =\, \Box\, \phi_{\m\n\r} 
- \frac{3}{2} \left(\, \nabla^\l\nabla_{\!(\m\,} \phi_{\n\r)\l} 
+  \nabla_{\!(\m} \nabla^{\l\,} \phi_{\n\r)\l}\,\right) 
+ 3\, \nabla_{\!(\m}\nabla_{\!\n}\, \phi_{\r)} \, ,
\ee
and $\nabla$ is the Levi-Civita connection.\footnote{The options to introduce gravitational interactions via minimal coupling and to truncate the spectrum to a sole spin-3 field are peculiarities of the three-dimensional setup, which are allowed by the vanishing of the Weyl tensor (see e.g.\ \cite{review_int} for a discussion of higher-spin interactions in $D > 3$ dimensions).} We have also defined $\phi_\m \equiv \phi_{\m\l}{}^\l$ and, likewise, $\cF_\m$ denotes the trace of the Fronsdal tensor. Besides minimal coupling, \eqref{action3} contains all ``non-minimal'' terms involving the Ricci tensor:
\begin{equation} \label{LF}
\begin{split}
\cL_{NM} & = 3\, R_{\rho\sigma} \Big(\, 
k_1\,  \phi^{\,\rho}{}_{\m\n}\, \phi^{\,\sigma\m\n} 
+ k_2\, \phi^{\,\rho\sigma}{}_{\m}\, \phi^{\,\m} 
+ k_3\, \phi^{\,\rho}\, \phi^{\,\sigma} \, \Big)  \\
& + 3\, R\, \Big(\, k_4\, \phi_{\m\n\r}\, \phi^{\,\m\n\r} 
+ k_5\, \phi_{\m}\, \phi^{\,\m} \,\Big) + \frac{1}{\ell^2}\, \Big(\, m_1\, \phi_{\m\n\r}\, \phi^{\,\m\n\r} + m_2\, \phi_{\m}\, \phi^{\,\m} \,\Big) \, .
\end{split}
\end{equation}
One can choose the $k_i$ arbitrarily, while
\begin{equation} \label{masses3}
m_1 \,=\, 6 \left(k_1 + 3 k_4 - 1\right) , \qquad m_2 \,=\, 6 \left(k_2 + k_3 + 3 k_5 + \frac{9}{4} \right) .
\end{equation}
Different $k_i$ do not label inequivalent couplings, but account for the freedom of performing field redefinitions of the metric that are quadratic in the spin-3 field.   For the subsequent analysis, it turns out to be convenient to adopt the choice made in  \cite{metric-like}, to which we refer for more information and motivations.  This choice of the $k_i$ simplifies the gauge transformation of the metric and reads
\begin{equation} \label{numbers}
k_1 = \frac{3}{2} \, , \qquad k_2 = 0 \, , \qquad k_3
= -\, \frac{3}{4} \, , \qquad k_4 = -\, \frac{1}{2} \, , \qquad k_5 = 0 \, .
\end{equation}
It will be assumed from now on.  We shall come back to these ambiguities when discussing the spin-4 case below.

\subsubsection{Gauge transformations}
The action \eqref{action3} is not only invariant under diffeomorphisms, but also  under covariantised Fronsdal gauge transformations,
\be \label{deltaphi}
\d \phi_{\m\n\r} = 3\,\nabla_{\!(\m\,} \x_{\,\n\r)} + \cO(\phi^2) \, , 
\ee
provided that the trace of the gauge parameter vanishes,
\be \label{trace_x}
\x_\l{}^\l =\,0 \, ,
\ee
 and the metric simultaneously transforms as $\d g_{\m\n} \sim \cO(\phi)$. We shall display the precise form of the lowest order in the gauge transformation of the metric in Sect.~\ref{sec:d3g}. The corrections $\cO(\phi^4)$ to the action and the corresponding corrections to the gauge transformations, which are instrumental in preserving the gauge symmetry at all orders, are instead irrelevant for our goals. For more details on the action \eqref{action3} and on its relation with a $sl(3,\mathbb{R}) \oplus sl(3,\mathbb{R})$ Chern-Simons theory we refer to \cite{metric-like}.
 
\subsubsection{Anti-de Sitter solution}\label{sec:AdS}
Anti-de Sitter space $AdS_3$ with zero spin-3 field is a solution of the equations of motion.  This solution is invariant under the diffeomorphisms generated by the 6 independent Killing vectors of anti-de Sitter space, which clearly  leave invariant not only the anti-de Sitter metric but also the zero spin-3 configuration.  

Because the spin-3 field is equal to zero, the spin-3 gauge transformations have no action on the metric while $\d \phi_{\m\n\r} $ reduces to $\d \phi_{\m\n\r} = \nabla^{\textrm{AdS}}_{\!(\m\,} \x_{\,\n\r)}$ where $ \nabla^{\textrm{AdS}}_{\!\m}$ is the covariant derivative in anti-de Sitter space. Invariance  of the spin-3 field under spin-3 gauge transformations, $\d \phi_{\m\n\r} = 0$,  is therefore equivalent to the Killing tensor equation
\be \label{2KTAdS}
\nabla^{\textrm{AdS}}{}_{\hspace{-10pt}(\m\,}\, \x_{\,\n\r)} =0 \, , 
\ee
where the Killing tensor $\x_{\,\n\r}$ should be traceless.

The Killing tensor equations have a long history and it would be out of place to quote here the vast literature referring to that subject. Let us just mention  the works \cite{killing,algebra,kill_tensors_1,kill_tensors_2} related to our purposes.  More information is also provided in Appendix \ref{sec:iso}.

As shown in that Appendix, the equations (\ref{2KTAdS}) possess 10 independent (traceless) solutions. To leading order, the Killing tensors of $AdS_3$ behave as 
$\x^{rr} \sim r^2$, $\x^{ri} \sim r^1$ and $\x^{ij} \sim r^0$ at infinity. With the 6 independent Killing vectors, this gives 16 independent symmetries of anti-de Sitter space.  How the algebra of these symmetries reflects the  $sl(3,\mathbb{R}) \oplus sl(3,\mathbb{R})$ structure underlying the Chern-Simons formulation is discussed also in Appendix \ref{sec:iso}.

Anti-de Sitter space is the solution with the maximum number of symmetries of the theory and is called for that reason ``the vacuum".  Its number of symmetries is finite.  There is an infinite enhancement at infinity of the algebra of exact vacuum symmetries, which generalizes the phenomenon found in the pure gravitational case.  The resulting infinite-dimensional algebra of asymptotic symmetries is $W_3 \oplus W_3$, as we now explicitly exhibit within the metric description.

\subsection{Boundary conditions} \label{sec:bnd3}
In order to develop the asymptotic analysis of the coupled spin-2 -- spin-3 system, we shall proceed in two steps.  First, we shall give the boundary conditions on the fields, motivating them heuristically.  Then, we shall explicitly verify that these boundary conditions fulfill all three conditions outlined in the introduction and so are consistent.

\subsubsection{Boundary conditions on the spin-3 field}

We start by requiring that the angular components $\phi_{ijk}$  with all indices down of the spin-3 field be of $\cO(1)$.  The reason why we demand this property is that it is the analog of the condition $g_{ij} - \frac{r^2}{2}\, \eta_{ij} = \cO(1)$ for the metric. As we shall see, the angular components $\phi_{ijk}$ turn out to be the spin-3 charges.  

The condition $\phi_{ijk} = \cO(1)$ implies $\phi^{ijk} = \cO(r^{-6})$.   The components with radial indices $\phi^{rij}$, $\phi^{rrj}$, $\phi^{rrr}$ then follow the rule that each time one replaces one angular index $i$ by the radial index $r$, the behavior of the leading fall-off term is multiplied by $r$. Furthermore,  we request the leading order of the trace of $\phi^{rij}$ to be zero, as this turns out to be necessary to preserve the boundary conditions on the metric under spin-3 gauge transformations (see discussion below Eq.~(\ref{dg})).

This yields the following boundary conditions  on the spin-3 field:
\begin{subequations} \label{bnd3}
\begin{align}
\phi^{rrr} & = \cO(r^{-3}) \, , \label{phi^rrr} \\[10pt]
\phi^{rri} & = \cO(r^{-4}) \, , \label{phi^rri} \\[10pt]
\phi^{rij} & = r^{-5}\, t^{rij} + \cO(r^{-7}) \, , \; \; \; \h_{ij} t^{rij} = 0 \, , \label{phi^rij} \\[5pt]
\phi^{ijk} & = \frac{6\p C_1}{k\, r^6}\, \cW^{ijk} + \cO(r^{-8}) \, , \label{phi^ijk}
\end{align}
\end{subequations}
where $\cW_{ijk}$ is a symmetric tensor which is both traceless and conserved:
\be
\pr^{\,i} \cW_{ijk} = 0 \, , \qquad\qquad \cW_{ij}{}^j = 0 \, .
\ee
It is the spin-3 analogue of the boundary energy-momentum tensor $\cL_{ij}$   and it admits only two independent chiral components:
\be \label{W_comp}
\cW_{+++} = \cW(x^+) \, , \qquad \cW_{---} = \tilde{\cW}(x^-) \, , \qquad \cW_{++-} = \cW_{+--} = 0 \, .
\ee
The numerical factor $C_1$ in \eqref{phi^ijk} depends on the normalization conventions both for the spin-3 field $\phi^{ijk}$ and for the tensor $\cW^{ijk}$.  Different conventions have been adopted in the literature so that we keep $C_1$ free in our formulas without replacing it by its explicit value.  A definite choice of normalization -- and hence a definite value of $C_1$ -- is given in Appendix \ref{app:fields}.  The choice made there agrees with the standard parameterizations of the exact solutions, as also discussed in that Appendix.  A similar strategy will be adopted below when we introduce spin-4 and higher gauge fields, which also carry normalization-dependent constants.

The tensor $\cW_{ijk}$ has density weight 3.  The trace-free condition is invariant under Weyl rescalings of the metric, and so is the divergence free condition $D_i \cW^{ijk} = 0$ which can equivalently be rewritten as $[\cW,\bar{g}]^{ijkm} \bar{g}_{im}=0$.  

\subsubsection{Boundary conditions on the metric}
The computation of the asymptotic spin-3 symmetries turns out to ``dig deeper" into the asymptotic structure of the metric, because the asymptotic variation of the relevant $\cO(r^{-6})$-term in $\phi^{ijk}$ involves the $\cO(r^{-2})$-term, $\cO(r^{-3})$-term and $\cO(r^{-6})$-term in, respectively, $g^{rr}$, $g^{ri}$ and $g^{ij}$. Thus, we need to specify these terms. This is a novelty of the higher-spin case with respect to pure gravity, which will get amplified as we add further higher spin fields in the sense that even higher order terms in the metric will then have to be specified.

In the covariant description of phase space followed here, the strengthening of the boundary conditions amounts to imposing the equations of motion at the next order.  
This gives explicitly:
\begin{subequations} \label{bnd23}
\begin{align}
g^{rr} & = r^2 + r^{-2}\, h^{rr} + \cO(r^{-4}) \, , \label{g^rr} \\[10pt]
g^{ri} & = r^{-3}\, h^{ri} + \cO(r^{-5}) \, , \label{g^ri} \\[5pt]
g^{ij} & = \frac{2}{r^2}\, \h^{ij} - \frac{8\p}{k\,r^4}\, \cL^{ij} + r^{-6}\, h^{ij} + \cO(r^{-8}) \, , \label{g^ij} 
\end{align}
\end{subequations}
where $h^{rr}$, $h^{ri}$ and $h^{ij}$ are now no longer arbitrary functions of $t$ and $\phi$ but satisfy instead
\be \label{on-shell_h}
h^{ij} = -\, \pr^{(i} h^{j)r} - \frac{1}{2}\, \h^{ij} h^{rr} + \frac{24\p^2}{k^2}\, \cL^i{}_k \cL^{jk} \, .
\ee
At the order where the $h^{\mu\nu}$-coefficients appear, the equations of motion for the metric $G_{\m\n} = 8\p G\, T_{\m\n}$ do not receive spin-3 field back-reaction terms and so reduce to the vacuum field equations $G_{\m\n}= 0$.  Hence the absence in (\ref{on-shell_h}) of the functions appearing in the spin-3 asymptotic expansion.

One could perform a proper gauge transformation to set
\be \label{gauge_fixing3}
h^{rr} = 0 \, , \ h^{ri} = 0 \quad \Rightarrow \quad h^{ij} = \frac{24\p^2}{k^2}\, \cL^i{}_k \cL^{jk} \, .
\ee
This would be the generalization of the radial gauge condition imposed above in the Hamiltonian description. However, we shall refrain from achieving this additional step here as it does not lead to significant simplifications.

The asymptotic form of the fields, i.e.\ \eqref{bnd3} and \eqref{bnd23}, is compatible with the asymptotic form of the known solutions given in  \cite{spin3,metric-like} (see Appendix \ref{app:fields}).

As we stressed already many times, the ultimate justification of the boundary conditions is that they form a set fulfilling all the consistency requirements, as we now show.

\subsection{A first consistency check: asymptotic conformal invariance}\label{sec:conf3}

Besides containing the solutions of \cite{spin3,metric-like}, the boundary conditions can be verified to be compatible with the asymptotic conformal symmetry.    The computations are almost identical to those of the pure metric case and go as follows.  The asymptotic Killing vectors take the form 
\begin{subequations} \label{vBis}
\begin{align} \label{vBis1}
v^r & = r\, \z + \frac{\z_1}{r} +  \frac{\z_2}{r^3} + \cO(r^{-5}) \, , \\[2pt]
v^i & = \e^i + \frac{\e_1{}^i}{r^2} +  \frac{\e_2{}^i}{r^4} + \cO(r^{-6}) \, . \label{vBis2}
\end{align}
\end{subequations} 
One then gets the variation of the inverse metric as
\begin{subequations} \label{dg_gravBis}
\begin{align}
\d g^{\,rr} & = -\,4\, \z_1 + \frac{8}{r^2} \left\{ -\, \z_2 + \frac{1}{2}\, h^{rr} \z - \frac{1}{8}\, \e^i \pr_i h^{rr} + \frac{1}{4}\, h^{ri} \pr_i \z \right\} + \cO(r^{-4}) \, , \\[5pt]
\d g^{\,ri} & = \frac{2}{r} \left\{ -\, \e_1{}^i + \pr^{\,i} \z \right\} \nn \\
& + \frac{4}{r^3}\left\{ -\, \e_2{}^i + \12\, \pr^{\,i} \z_1 - \frac{2 \pi}{k} \cL^{ij} \partial_j \z + \frac{1}{4}\, [\, h^r , \e \,]^i + h^{ri}\z \right\} + \cO(r^{-5}) \, ,
\end{align}
\end{subequations}
and
\be
\begin{split} 
\d g^{\,ij} & = \frac{4}{r^2}\! \left\{ \pr^{(i} \e^{\,j)} + \h^{ij} \z \right\} \\
& + \frac{4}{r^4}\! \left\{ \pr^{(i} \e_1{}^{j)} + \h^{ij} \z_1 - \frac{2\p}{k} [\, \cL , \e \,]^{ij} - \frac{8\p}{k} \cL^{ij} \z \right\} \\
& + \frac{4}{r^6}\! \left\{ \pr^{(i} \e_2{}^{j)} + \h^{ij} \z_2 -  \frac{2\p}{k} [\, \cL , \e_1 \,]^{ij} - \frac{8\p}{k} \cL^{ij} \z_1 + \frac{1}{4}\, [\, h , \e \,]^{ij} + \frac{3}{2}\, h^{ij} \zeta - h^{r(i} \e_1{}^{j)} \right\}\! \\
& + \cO(r^{-8}) \, ,
\end{split} \label{232Bis}
\ee
where, as in the previous section, $[\, \cL , \e \,]^{ij}$ denotes the two-dimensional Schouten bracket at infinity. Preservation of the form of the inverse metric imposes again that $\e^k$ be a conformal Killing vector while $\z$,  $\z_1$ and $\e_1^k$ satisfy again  
\be \label{comp_conf_fix}
\z = -\, \frac12\, \partial \cdot \e \, , \qquad  \z_1 = 0 \, , \qquad  \e_1{}^i = \partial^i \z  
\ee
since the new boundary conditions for the metric are identical to the old ones at the first leading orders. This yields the same variation \eqref{QVariation1} of $\cL^{ij}$ as found above.

The additional terms in  \eqref{dg_gravBis} and \eqref{232Bis} fix the variations $\d h^{rr}$, $\d h^{ri}$ and $\d h^{ij}$ of the subleading terms which made their appearance through the strengthening of the boundary conditions. One can verify that these variations consistently satisfy
\be \label{rel_variations}
\d h^{ij} + \pr^{(i} \d h^{j)r} + \frac{1}{2}\, \h^{ij} \d h^{rr} = \frac{48\p^2}{k^2}\, \cL_k{}^{(i} \d \cL^{j)k} \, ,
\ee
with the $\d \cL^{ij}$ given by \eqref{QVariation1}, provided that one imposes \eqref{on-shell_h} (but without the need to impose any conditions on $\z_2$ and $\e_2{}^i$ which drop from \eqref{rel_variations}). That \eqref{on-shell_h} is preserved is not a surprise, since it is a consequence of the equations of motion and asymptotic symmetries, which are  particular diffeomorphisms,  map solutions of the field equations on solutions. At any rate it is reassuring that both ways to compute $\d \cL^{ij}$, either from the $\cO(r^{-4})$-term in (\ref{232Bis}) or from \eqref{rel_variations}, give identical results.

We now turn to the spin-3 field, which transforms as 
\be
\delta \phi^{\lambda \mu \nu} = [ \phi, v ]^{\lambda \mu \nu} \equiv 3\, \phi^{\rho (\mu \nu} \partial_\rho v^{\lambda)} - v^\rho \partial_\rho \phi^{\lambda \mu \nu} = - \, \cL_v \phi^{\lambda \mu \nu}
\ee
under spacetime diffeomorphisms. Transformations generated by the asymptotic Killing vectors (\ref{vBis}), with coefficients $\zeta, \zeta_1$ and $\e^i, \e^i_1$ determined by the above analysis, are easily verified to preserve the boundary conditions.  Furthermore one finds that the variation of $\cW^{ijk}$ is given by
\be \label{var2W}
\delta \cW^{ijk} = [\, \cW , \e \,]^{ijk} - 3\, \cW^{ijk} \prd \e \, .
\ee
This equation just expresses that $\cW^{ijk}$ is a tensor density of weight 3, in agreement with what was stated above. It preserves therefore the trace-free and divergence-free conditions, and
 implies
\be
\d\cW = - \, \e\, \cW^{\,\pe} - 3\, \e' \, \cW \, , \qquad \d\tilde{\cW} =  -\, \tilde{\e}\, \tilde{\cW}^{\,\pe} - 3\, \tilde{\e}^{\,\pe} \, \tilde{\cW} \, .\label{var2WBis}
\ee
In fact, there is a clear connection between the density weight  of $\cW^{ijk}$ (namely $3$) and the power of $r$ of which it is the coefficient in the expansion of $\phi^{ijk}$ (namely $r^{-6}$).  This connection can be traced to Eq. (\ref{comp_conf_fix}), which states that the radial component $\zeta $ of the infinitesimal three-dimensional diffeomorphism completing to spacetime the infinitesimal two-dimensional diffeomorphism $\epsilon^i$ is $\zeta = - \frac12 \partial \cdot \epsilon$.  The action of $\zeta r \frac{\partial}{\partial r} $ on $\frac{F}{r^{2n}}$, where $F$ is an arbitrary function of $x^i$, is therefore $ n \frac{F}{r^{2n}} \partial \cdot \epsilon $.   This is exactly the variation of a density of weight $n$ under the infinitesimal diffeomorphism $\epsilon^i$.

\subsection{Asymptotic Killing tensors and $W_3$-algebra}\label{sec:killing3}

The analysis just performed of the behavior of the fields under diffeomorphisms that tend to conformal transformations at infinity is a straightforward generalization of what was found for pure gravity and brings no surprise. The emergence of a $W$-algebra is more interesting.  It follows from the study of the transformation of the fields under the spin-3 gauge symmetry.

\subsubsection{Transformation of the spin-3 field}

We start with the spin-3 field because its gauge variation controls the behavior of the spin-3 gauge parameter $\xi^{\m\n}$ at infinity in a neat way.

In the contravariant form more convenient to our purposes, the spin-3 field transforms under spin-3 gauge transformations as
\be \label{d3phi}
\d \phi^{\m\n\r} = [\, g , \xi \,]^{\m \n \r} + \cO(\phi^2) \, .
\ee
The unwritten $\cO\!\left(\phi^2\right)$-terms does not play any role because they are subleading with respect to the significant terms. 

In $AdS_3$ with zero spin-3 field, the variation of $\phi^{\mu \nu \rho}$ reduces to the Killing tensor equation $\nabla_\textrm{AdS}^{\!(\m\,} \, \x^{\,\n\r)} = 0$. Invariance up to lower order terms of the $AdS$ background with zero spin-3 field forces therefore the spin-3 gauge transformations to 
be generated by gauge parameters which have the same leading dependence on the radial coordinate as the Killing tensors\footnote{Conversely, acting with the exact Killing tensors of $AdS_3$ on the known solutions of \cite{spin3,metric-like} generates terms which have the asymptotic behavior given in the text.} of $AdS_3$. One can then derive the additional conditions that they have to satisfy following the same approach as in the pure metric case. We thus consider gauge parameters of the form\footnote{It is in fact easy to see directly that higher powers of $r$, e.g., $r^3$ in $\xi^{rr}$, are in fact not allowed by the asymptotic conditions, so that the behavior assumed in (\ref{xmn}) is not a restriction.}
\begin{subequations} \label{xmn}
\begin{align}
\x^{rr} & = r^2\, \l + \l_1 + r^{-2} \l_2 + \cO(r^{-4}) \, , \\[5pt]
\x^{ri} & = r\, w^i + r^{-1}\, w_1{}^i + r^{-3}\, w_2{}^i + \cO(r^{-5}) \, , \\[5pt]
\x^{ij} & = \chi^{ij} + r^{-2}\, \chi_1{}^{ij} + r^{-4}\, \chi_2{}^{ij} + \cO(r^{-6}) \, ,
\end{align}
\end{subequations}
where the coefficients of the various powers of $r$  are  functions of $t$ and $\phi$. The trace constraint $  g_{\a\b} \x^{\a\b} = 0$ imposes at orders $\cO(r^2)$ and $\cO(1)$%
\be \label{trace3_1}
 \h_{ij} \chi^{ij} = 0
\ee
and 
\be \label{trace3_2}
\l + \12\, \h_{ij}\chi_1{}^{ij} + \frac{2\p}{k}\, \cL_{ij} \chi^{ij} = 0 \,.
\ee

We begin the analysis with the variation of the purely angular components \eqref{phi^ijk}, which contains the physics.   One finds at leading order
\be \label{dphi^ijk_1}
\d \phi^{ijk} = \frac{6}{r^2} \left\{ \pr^{(i} \chi^{jk)} + 2\, \h^{(ij} w^{k)} \right\} + \cO(r^{-4}) \, .
\ee
The $\cO(r^{-2})$-term in the variation has to vanish in order to be compatible with the boundary conditions \eqref{bnd3}. Combining this information with \eqref{trace3_1} one realizes that $\chi^{ij}$ must satisfy 
\be \label{killing3}
\pr^{(i} \chi^{jk)} - \12\, \h^{(ij} \prd \chi^{k)} = 0 \, ,
\ee
while $w^i$ is not independent from $\chi^{ij}$:
\be \label{w^i}
w^i = - \,\frac{1}{4}\, \prd \chi^i \, .
\ee
The condition \eqref{killing3} implies that $\chi^{ij}$ is a \emph{conformal Killing tensor} for the boundary metric (see e.g.\ \cite{algebra}). In terms of $\bar{g}^{ij}$, it can be rewritten as
\be
[\, \bar{g}, \chi \,]^{ijk} = \mu^{(i} \bar{g}^{jk)} 
\ee
for some $\mu^i$, which exhibits its invariance under conformal rescalings of the metric.  In two-dimensions the conformal Killing equation, together with the trace-free condition (\ref{trace3_1}), implies that $\chi^{ij}$ has two independent chiral components:
\be \label{solkilling3}
\chi^{++} = \chi(x^+) \, , \qquad \chi^{--} = \tilde{\chi}(x^-) \, , \qquad \chi^{+-} = 0 \, .
\ee

We now turn to the next order in $r^{-2}$.   A direct computation yields:
\be \label{dphi^ijk_2}
\d \phi^{ijk} = \frac{6}{r^4}\, \Big\{ \pr^{(i} \chi_1{}^{jk)} + 2\, \h^{(ij} w_1{}^{k)} - \frac{4\p}{3k}\, [\, \cL , \chi \,]^{ijk} - \frac{16\p}{k}\, \cL^{(ij} w^{k)} \Big\} + \cO(r^{-6}) \, .
\ee
The terms displayed explicitly in \eqref{dphi^ijk_2} vanish together with the leading orders in
\begin{subequations} \label{dphi_1}
\begin{align}
\d \phi^{rri} & = 4\, \Big\{ - w_1{}^i + \12\, \pr^{\,i} \l \Big\} + \cO(r^{-2}) \, , \\[5pt]
\d \phi^{rij} & = \frac{2}{r}\, \Big\{ - \chi_1{}^{ij} + 2\, \pr^{(i} w^{j)} + 2\, \h^{ij} \l \Big\} + \cO(r^{-3}) \, ,
\end{align}
\end{subequations}
provided that
\begin{subequations} \label{subs3_1}
\begin{align}
\l & = \frac{1}{12}\, \prd\prd \chi - \frac{2\p}{3k}\, \cL_{ij} \chi^{ij} \, , \\[5pt]
w_1{}^i & = \frac{1}{24}\, \pr^{\,i} \Big\{\, \prd\prd \chi - \frac{8\p}{k}\, \cL_{kl} \chi^{kl} \,\Big\} \, , \\[5pt]
\chi_1{}^{ij} & = -\,\12\, \pr^{(i}\prd \chi^{j)} + \frac{1}{6}\, \h^{ij}\, \prd\prd \chi - \frac{4\p}{3k} \, \h^{ij} \cL_{kl} \chi^{kl} \, .
\end{align}
\end{subequations}
These conditions also preserve the trace constraint \eqref{trace3_2}. In \eqref{subs3_1} a peculiarity of higher-spin gauge fields already emerges neatly: the components of the gauge parameters that preserve the boundary conditions depend on the boundary currents. This introduces powers of $\cL^{ij}$ in the variation of $\cW^{ijk}$ and eventually brings the non-linearities in the asymptotic symmetries algebras first observed in \cite{HR,spin3}. 

At the next order the variation of the component $\phi^{ijk}$
reads
\be \label{dphi^ijk_3}
\begin{split}
\d \phi^{ijk} = \frac{6}{r^6}\, \Big\{& \pr^{(i} \chi_2{}^{jk)} + 2\, \h^{(ij} w_2{}^{k)} - \frac{4\p}{3k}\, [\, \cL , \chi_1 \,]^{ijk} - \frac{16\p}{k}\, \cL^{(ij} w_1{}^{k)} \\
& + \frac{1}{6} \left[\, h , \chi \,\right]^{ijk} + 3\, h^{(ij} w^{k)} - h^{r(i} \chi_1{}^{j)} \Big\} + \cO(r^{-8}) \, ,
\end{split}
\ee
where the subleading components of the metric $h^{rr}$, $h^{ri}$ and $h^{ij}$ appear for the first time.  One also finds
\begin{subequations} \label{final_var3}
\begin{align}
\d \phi^{rrr} & = -\, 6\,r\, \l_1 + \, \frac{12}{r}\, \Big\{ -\, \l_2 + h^{rr} \l - \frac{1}{4}\, w^i \pr_i h^{rr} + \frac{1}{4}\, h^{ri} \pr_i \l \Big\} + \cO(r^{-3}) \, , \label{dphi^rrr} \\[10pt]
\d \phi^{rri} & = \frac{8}{r^2}\, \Big\{ - w_2{}^i + \frac{1}{4}\, \pr^{\,i} \l_1 - \frac{\p}{k}\, \cL^{ij} \pr_j \l \nn \\
& \phantom{\frac{8}{r^2}\qquad\ } + \frac{1}{4}\, [\, h^r , w \,]^i + h^{ri} \l - \frac{1}{8}\, \chi^{ij} \pr_j h^{rr} + \frac{1}{2}\, h^{rr} w^i \Big\} + \cO(r^{-4}) \, , \\[10pt]
\d \phi^{rij} & = \frac{4}{r^3}\, \Big\{ - \chi_2{}^{ij} + \pr^{(i} w_1{}^{j)} + \h^{ij} \l_1 - \frac{2\p}{k}\, [\,\cL , w\,]^{ij} - \frac{8\p}{k}\, \cL^{ij} \l \nn \\
& \phantom{\frac{8}{r^2}\qquad\ } + \frac{1}{4}\, [\, h^r , \chi \,]^{ij} + 2\, h^{r(i} w^{j)} \Big\} + \cO(r^{-5}) \, . 
\end{align}
\end{subequations}

From the variations of the components with at least one radial index, one gets $\l_1$,  $\l_2$, $w_2{}^i$ and $\chi_2{}^{ij}$.  Substituting the resulting expressions into \eqref{dphi^ijk_3} gives then $\d \cW^{ijk}$,
\begin{align}
& \d \cW^{\,ijk} = -\, \frac{1}{6 C_1} \bigg\{\, 2\,\pr^{\,i}\pr^{\,j}\pr^{\,k}\!\left( \cL_{mn}\chi^{mn}\right) + 3\, \pr_m \pr^{(i}\cL^{jk)} \prd \chi^m + 9\, \pr_m \cL^{(ij}\pr^{k)} \prd \chi^m \nn \\[2pt]
& + 6\left( \pr^{m}\cL^{(ij}\pr_m\prd\chi^{k)} - \pr^{(i}\cL^{j|m}\pr_m\prd\chi^{|k)} \right) + \left( 8\, \cL^{(ij}\pr^{k)} + 5\, \h^{(ij}\cL^{k)m}\pr_m \right) \prd\prd \chi \nn \\[5pt]
& - 18\,\cL^{m(i}\pr_m\pr^{\,j}\prd\chi^{k)} - \frac{k}{4\p}\,\pr^{\,i}\pr^{\,j}\pr^{\,k}\prd\prd\chi - \frac{8\p}{k} \Big[ \left( 8\, \cL^{(ij}\pr^{k)} + 5\, \h^{(ij}\cL^{k)p}\pr_p \right) \!\left(\cL_{mn}\chi^{mn}\right) \nn \\
& - 9\, \chi^{m(i|}\pr_m \!\left( \cL_n{}^{|j}\cL^{k)n} \right)  - \frac{27}{2}\, \cL^{m(i}\cL_m{}^{j}\prd\chi^{k)} + 9\, \cL^m{}_n \cL^{n(i}\pr_m \chi^{jk)} \Big] \bigg\} \, , \label{VarW21}
\end{align}
where we have explicitly used the relation \eqref{on-shell_h} on the metric fluctuations $h^{rr}$, $h^{ri}$ and $h^{ij}$.  It is important to realize that if this relation did not hold, the resulting  $\d \cW^{ijk}$ would not have been a traceless, conserved tensor density of weight 3. 
This is the reason why we  imposed this relation, which is, as we have indicated, a consequence of the equations of motion.  These must therefore hold up to some appropriate order in $1/r$. 
In the metric-like formulation the conformal invariance at the boundary is thus achieved only \emph{on shell}, consistently with the on-shell closure of the algebra of metric-like gauge transformations (see e.g.\ \cite{metric-like,metric-like2}). %

From (\ref{VarW21}), one can derive the variation of the two independent components of $\cW_{ijk}$ and obtain
\be
\d \cW = \frac{1}{6 C_1} \Big\{ 2\,\chi\cL''' + 9\,\chi'\cL'' + 15\,\chi''\cL' + 10\,\chi'''\cL - \frac{k}{4\p}\, \chi^{(5)} - \frac{64\p}{k} \left( \chi\cL\cL' + \chi'\cL^2 \right)\! \Big\} \label{varW21Bis}
\ee
in perfect agreement e.g.\ with Eq.~(4.20b) of \cite{spin3} apart from a flip in the sign of $\cL$ due to a different choice of conventions.  A similar expression holds for $\delta \tilde{\cW}$. 

\subsubsection{Transformation of the metric}\label{sec:d3g}

The asymptotic transformation of the spin-2 field  $\cL_{ij}$ under the asymptotic spin-3 symmetries is strictly speaking not needed since it follows from the asymptotic transformation of the spin-3 field $\cW^{ijk}$ under the asymptotic spin-2 symmetries, which we have already computed in \eqref{var2W}. This is because these fields are the generators of the corresponding transformations.  Since the  Poisson bracket is antisymmetric, knowledge of $\delta \cW \sim \{\,\cW, \cL\,\}$ determines completely $\delta \cL \sim \{\,\cL, \cW\,\} = -\, \{\,\cW, \cL\,\}$.

However, as a consistency check, it is useful to derive  $\delta_\chi \cL_{ij}$ directly from the variation of the metric under spin-3 gauge transformations. This is an interesting computation because it tests the terms linear in $\phi^{\mu \nu \rho}$ in $\delta_3 g_{\mu \nu}$. For instance, $\delta_3 g_{\mu \nu}$ is sensitive to the trace of the tensor $t^{rij}$ which we displayed in the boundary conditions \eqref{bnd3} for the spin-3 field, while $\d_2 \phi_{\m\n\r}$ is not. As a result, the knowledge of the variation of the metric is instrumental in fixing the precise boundary conditions on the fields.

The spin-3 gauge transformation of the metric that compensates the variation of the Fronsdal Lagrangian under the transformation \eqref{d3phi} of the spin-3 field is 
\begin{equation} \label{dg}
\begin{split}
\d g^{\m\n} = -\, 3 & \, \bigg\{ 
 \a\, \phi_{\rho\sigma}{}^{(\m} \nabla^{\n)} \x^{\rho\sigma} 
+ \b\, \phi_{\rho} \nabla^{(\m} \x^{\n)}{}^{\rho} 
+ 4\, \x^{\rho\sigma}\, \nabla_{\!\rho}\, \phi^{\m\n}{}_{\sigma} 
+ (\a-8)\, \x^{\rho\sigma} \nabla^{(\m} \phi^{\n)}{}_{\rho\sigma}  \\
& 
+ 8\, \x^{\rho(\m} \nabla\cdot\phi^{\n)}{}_{\rho} - 8\, \x^{\rho\,(\m} \nabla_{\!\rho}\, \phi^{\n)}
+ (\b-8)\, \x^{\rho(\m} \nabla^{\n)} \phi_{\rho} 
+ 2\, \x^{\m\n\,} \nabla\cdot \phi \\
& - g^{\m\n}\, \Big[\,  4\, \x^{\rho\sigma}\, \nabla\!\cdot \phi_{\rho\sigma} - 8\, \x^{\rho\sigma}\, \nabla_{\!\rho}\, \phi_{\sigma} \, \Big] \bigg\} + \cO(\phi^3) \, ,
\end{split}
\end{equation}
an expression computed with the choice (\ref{numbers}) of the coefficients $k_i$ entering the action \eqref{LF}.  Different choices would have led to more involved expression for $\d g^{\m\n}$ \cite{metric-like}.  The variation  (\ref{dg}) contains two free parameters, $\a$ and $\b$, which parametrize field dependent diffeomorphisms generated by
\be \label{field_dep_diffeo}
v^\m = \a\, \x^{\r\s} \phi^\m{}_{\r\s} + \b\, \x^{\m\r} \phi_\r \, .
\ee
Computing (\ref{dg}) near the boundary, one finds, using \eqref{bnd3} and $t^{r}{}_k{}^k = 0$, that the metric transforms as 
\begin{subequations} \label{d3g_lead}
\begin{align}
\d g^{rr} & = -\, \frac{3}{2}(12 - \a)\, t^{rij} \chi_{ij} + \cO(r^{-2}) \, , \\[5pt]
\d g^{ri} & = -\, \frac{9\p C_1}{2k\, r} (12- \a)\, \cW^{\,i}{}_{jk} \chi^{jk} + \cO(r^{-3}) \, , \\[5pt]
\d g^{ij} & = \frac{3}{2\,r^4} \left\{ 2(12 - \a)\, t^{rk(i} \chi^{j)}{}_k  \right\} -\, \frac{9\p C_1}{k\, r^4}\, \big\{ 4\, \chi^{kl} \pr_k \cW_l{}^{ij} + \a\, \cW_{kl}{}^{(i} \pr^{j)} \chi^{kl} \nn \\
& - (8 - \a)\, \chi^{kl} \pr^{(i} \cW^{j)}{}_{kl} - 2\, \cW_k{}^{l(i} \pr_l \chi^{j)k} \big\}
+ \cO(r^{-6}) \, , \label{d3g^ij_lead}
\end{align}
\end{subequations}
where, as anticipated, one can notice a dependence on $t^{rij}$ in $\d g^{rr}$ and $\d g^{ij}$.

Making the choice $\a = 12$ eliminates the leading orders in $\d g^{rr}$ and $\d g^{ri}$, while $\d g^{ij}$ becomes
\be \label{dgbnd1}
\d g^{ij} =  - \frac{36\p C_1}{k\, r^4} \left\{\, \chi^{kl}\! \left( \pr_k \cW_l{}^{ij} + \pr^{(i} \cW^{j)}{}_{kl} \right) + 3\, \cW_{kl}{}^{(i} \pr^{j)} \chi^{kl} \,\right\} + \cO(r^{-6}) \, .
\ee
One immediately extract from \eqref{dgbnd1} the variation of the Virasoro charges $\cL^{ij}$ under the spin-3 gauge transformations,
\be \label{d3L}
\d_\chi \cL^{ij} = \frac{9C_1}{2} \left\{\, \chi^{kl}\! \left( \pr_k \cW_l{}^{ij} + \pr^{(i} \cW^{j)}{}_{kl} \right) + 3\, \cW_{kl}{}^{(i} \pr^{j)} \chi^{kl} \,\right\} ,
\ee
an expression which is a traceless, transverse tensor density of weight 2  provided that $\chi^{ij}$ is a conformal Killing tensor.\footnote{Conversely, the trace of \eqref{d3L} vanishes only if $\pr_- \chi^{++} = \pr_+ \chi^{--} = 0$. These two conditions, combined with $\chi^{+-} = 0$ required by \eqref{trace3_1}, imply that $\chi^{ij}$ is a conformal Killing tensor.} Taking into account \eqref{solkilling3}, the variation of the components is finally found to be
\be \label{d3L_comp}
\d \cL = \frac{9C_1}{2} \left(\, 2\, \chi\,\cW^{\,\pe} + 3\, \chi'\,\cW \,\right) , \qquad 
\d \tilde{\cL} = \frac{9C_1}{2} \left(\, 2\, \tilde{\chi}\,\tilde{\cW}^{\pe} + 3\, \tilde{\chi}'\,\tilde{\cW} \,\right) .
\ee

To conclude this section, a comment is in order: had we taken different values of the $k_i$ coefficients in the action, we would have found the same final value for $\delta \cL^{ij}$, but a compensating diffeomorphism might have been needed.  A  judicious choice of the free parameters in the action is  helpful to simplify the computation of the asymptotic symmetries, but it does not affect the variation of the charges. This is because the field redefinitions under consideration do not affect our boundary conditions to leading order.  As we shall see in the next section, this state of affairs becomes more intricate when fields of spin higher than 3 are included.

\subsection{Charges \& asymptotic symmetries}\label{sec:charges3}

One can verify that our boundary conditions are equivalent to the Chern-Simons boundary conditions given in \cite{HR,spin3}, in the sense that if one computes the metric and spin-3 field from the Chern-Simons connection using the formulas given in Appendix \ref{app:fields}, one gets fields that obey the boundary conditions given here.

In addition to checking agreement, this computation reveals that the coefficients $\cL^{ij}$ and $\cW^{ijk}$ appearing in the angular components of the metric and the spin-3 fields are indeed the charges generating the $W$-symmetry.  This is of course not surprising given that $\cL^{ij}$ and $\cW^{ijk}$ are conserved and traceless, and was anticipated in our terminology.  Much in the same way as the conserved current associated with the conformal Killing vector $\epsilon^i$ is the vector density of weight one $j^i[\epsilon] = \cL^{ij} \epsilon_j \equiv \cL^{ij} \epsilon^k \bar{g}_{jk}$, the conserved current associated with the conformal Killing tensor $\chi^{ij}$ is the vector density of weight one $j^i[\chi] = \cW^{ijk} \chi_{jk}\equiv \cW^{ijk} \chi^{lm} \bar{g}_{jl}\bar{g}_{km}$.

Alternatively, one can also identify $\cL^{ij}$ and $\cW^{ijk}$ with the charges by using the Hamiltonian formalism, ``\`a la Regge-Teitelboim" \cite{Regge:1974zd}.  This will be done in a forthcoming work  \cite{InPrep}.  Yet another method is provided by the covariant approach of  \cite{Barnich:2001jy}.

Once one knows that the charges are $\cL^{ij}$ and $\cW^{ijk}$, one can read off their algebra from their variation under the $W$-transformations through the formula $\delta_B Q_A = \{Q_A, Q_B\}$ where $Q_A$ stands for a generic charge and $\delta_B Q_A$ is the known variation of $Q_A$ under the transformation generated by $Q_B$.  The variations of all the charges were computed above and given in formulas (\ref{QVariation2}), (\ref{var2WBis}), (\ref{varW21Bis})  and (\ref{d3L_comp}), from which one infers
\begin{subequations}\label{W3}
\begin{align}
\left\{ \cL(u) , \cL(v) \right\} & = \d(u-v) \cL^\pe(u) \,+\, 2\,\d^\pe(u-v)\cL(u) - \frac{k}{4\pi}\, \d^{\pe\pe\pe}(u-v)\, , \label{AlgLW1} \\[5pt]
\left\{ \cL(u) , \cW(v) \right\} & = 2\,\d(u-v) \cW^\pe(u) \,+\, 3\,\d^\pe(u-v)\cW(u) \, , \label{AlgLW2} \\[5pt]
\left\{ \cW(u) , \cW(v) \right\} & = \frac{1}{6 C_1} \,\bigg( 2\,\d(u-v)\cL^{\pe\pe\pe}(\theta) + 9\,\d^\pe(u-v)\cL^{\pe\pe}(u) + 15\,\d^{\pe\pe}(u-v)\cL^\pe(u) \\
& \hspace{-2.6cm} + 10\,\d^{\pe\pe\pe}(u-v)\cL(u) + \frac{64\pi}{k} \left( \d(\th-\th^\pe)\cL(u)\cL^\pe(u) + \d^\pe(u-v)\cL^2(u) \right) + \frac{k}{4\pi}\, \d^{(5)}(u-v) \bigg) \, . \nn \label{AlgLW3}
\end{align}
\end{subequations}
Note that, as it should,  the expression obtained for the bracket $\{ \cL(u), \cW(w) \}$ is the same whether one computes it from $\delta_\cW \cL$ or from $\delta_\cL \cW$. The formulas (\ref{W3}) are in complete agreement with those of \cite{HR,spin3} and give the same nonlinear classical $W_3$ algebra with central charge $c = \frac{3 \ell}{2 G}$ identical to that of pure gravity.

In contrast to the finite-dimensional algebra of exact symmetries of the vacuum, the asymptotic symmetry algebra is infinite-dimensional.  This is exactly as in the pure gravity case.  Just as in that case, the exact symmetry algebra of the vacuum corresponds to the first Fourier modes of $\cL$ and $\cW$, namely $\cL_0$, $\cL_{\pm 1}$, $\cW_0$, $\cW_{\pm 1}$ and $\cW_{\pm 2}$ on each chiral side (``wedge algebra").

\section{Fields of spin 3 and 4 coupled to gravity}\label{sec:spin4}

We now add to the previous setup a single rank-4 symmetric tensor, thus moving to the metric-like counterpart of a $sl(4,\mathbb{R})\oplus sl(4,\mathbb{R})$ Chern-Simons theory with principal embedding of the gravitational subsector. The aim is to illustrate another novelty introduced by higher-spin gauge fields, namely the influence of  interacting vertices on the structure of asymptotic symmetries. This brings into the action some parameters which cannot be absorbed by redefinitions of the fields, and that are the seeds of the different extensions of the conformal algebra that one can realize asymptotically.

\subsection{Action \& gauge transformations}\label{sec:action4}

At lowest order in an expansion in the higher-spin fields the action contains the minimal coupling of Einstein gravity to the spin-3 and spin-4 free Fronsdal actions. As in Sect.~\ref{sec:action3} one can also add ``non-minimal'' terms which are quadratic in the fields, but they can be always eliminated by a field redefinition of the metric. The action is invariant under the infinitesimal gauge transformations
\begin{subequations} \label{free_gauge}
\begin{align}
\d g_{\m\n} & = 2\,\nabla_{\!(\m} v_{\n)} + \cO\!\left(\phi,\vf\right) , \\[5pt]
\d \phi_{\m\n\r} & = 3\,\nabla_{\!(\m\,} \x_{\,\n\r)} + \cO\!\left(\phi,\vf\right) , \\[5pt]
\d \vf_{\m\n\r\s} & = 4\,\nabla_{\!(\m\,} \k_{\n\r\s)} + \cO\!\left(\phi,\vf\right) ,
\end{align}
\end{subequations}
provided that all gauge parameters are traceless and the double-trace of the spin-4 field vanishes:\footnote{One can actually weaken this condition: at the interacting level the constraint \eqref{double-trace} of the free theory is compatible with a constraint of the form $\vf_\l{}^\l{}_\r{}^\r \sim \cO(\vf^2)$. One can however always eliminate the non-linear terms with a field redefinition and go back to the constraint \eqref{double-trace}. This is our choice, while we shall comment more on the double-trace constraint in Appendix \ref{app:fields}.}
\be \label{double-trace}
\vf_\l{}^\l{}_\r{}^\r = 0 \, .
\ee
The schematic form \eqref{free_gauge} of the gauge transformations does not provide sufficient information, however,  to identify completely the asymptotic symmetries of the model. With hindsight this is not surprising since a similar phenomenon was encountered already in the coupled spin-2 -- spin-3 case, where the higher spin corrections to the gauge transformations of the metric were needed.  Here, one also needs the corrections to the spin-3 gauge transformations -- which were not necessary in the previous section --,  because the additional contributions that should appear in $\d \cW_{ijk}$ in order to reproduce the Chern-Simons result call for extra terms in the spin-3 gauge transformations. Within the current setup we shall indeed see that some of the omitted contributions in \eqref{free_gauge}  have to be worked out because they do affect the variation of relevant terms in the boundary conditions.  We must therefore keep in mind that we must keep control of the first of these terms in the expansion.

We are thus led to consider the action
\be \label{I4}
I_{\{3,4\}} = \int \frac{d^3x \sqrt{-g}}{16\p G} \left( \cL_{EH} + \cL_{3} + \cL_{4} + \cL_{3-3-4} + \cL_{4-4-4} \right) + \cdots \, ,
\ee
where $\cL_{EH}$ is the Einstein-Hilbert Lagrangian, while $\cL_{3}$ and $\cL_{4}$ are the covariantised free Fronsdal Lagrangians for a spin-3,
\be \label{L3quad}
\begin{split}
\cL_{3} & = \phi^{\,\m\n\r} \left( \cF_{\m\n\r}
- \frac{3}{2} \, g_{(\m\n} \cF_{\r)} \right) + 3 R_{\a\b} \Big(\, 
k_1\,  \phi^{\,\a}{}_{\m\n\,} \phi^{\,\b\m\n} 
+ k_2\, \phi^{\,\a\b}{}_{\!\m\,} \phi^{\,\m} 
+ k_3\, \phi^{\,\a} \phi^{\,\b} \, \Big) \\
& + 3 R\, \Big(\, k_4\, \phi_{\m\n\r} \phi^{\,\m\n\r} 
+ k_5\, \phi_{\m} \phi^{\,\m} \,\Big) + \frac{1}{\ell^2} \Big(\, m_1\, \phi_{\m\n\r} \phi^{\,\m\n\r} + m_2\, \phi_{\m} \phi^{\,\m} \,\Big) \, ,
\end{split}
\ee
and a spin-4 field,
\be \label{L4quad}
\begin{split}
\cL_{4} & = \vf^{\m\n\r\s}\! \left( \cF_{\!\m\n\r\s}
\!- 3\, g_{(\m\n} \cF_{\!\r\s)} \right) + 6 R_{\a\b} \Big(\, 
l_1\,  \vf^{\,\a}{}_{\m\n\r} \vf^{\,\b\,\m\n\r} 
\!+ l_2\, \vf^{\,\a\b}{}_{\!\m\n\,} \vf^{\,\m\n} 
\!+ l_3\, \vf^{\,\a}{}_{\!\m} \vf^{\,\b\m} \, \Big) \\
& + 6 R\, \Big(\, l_4\, \vf_{\m\n\r\s} \vf^{\,\m\n\r\s} 
+ l_5\, \vf_{\m\n} \vf^{\,\m\n} \,\Big) + \frac{1}{\ell^2} \Big(\, n_1\, \vf_{\m\n\r\s} \vf^{\,\m\n\r\s} + n_2\, \vf_{\m\n} \vf^{\,\m\n} \,\Big) \, .
\end{split}
\ee
In both cases $\cF$ denotes the covariantised Fronsdal tensor, with a symmetric ordering for the covariant derivatives as in \eqref{fronsdal3}.  At lowest order the presence of $\cL_4$ does not affect the conditions for the gauge invariance of $\cL_3$. The $m_i$ are therefore fixed as in \eqref{masses3}, while
\be \label{masses4}
n_1 = 2 \left( 6 l_1 + 18 l_4 - 7 \right) , \qquad n_2 = 12 \left( l_2 + 2l_3 + 3l_5 + \frac{19}{4} \right) .
\ee

The terms $\cL_{3-3-4}$ and $\cL_{4-4-4}$ denote instead 
the cubic vertices with at most two derivatives that one can build with $\phi_{\m\n\r}$ and $\vf_{\m\n\r\s}$. The action \eqref{I4} displays the same number of gauge symmetries as in \eqref{free_gauge} only if cubic vertices are fixed --  up to an overall coupling constant and up to field redefinitions. Their detailed structure is shown in appendices \ref{app:vertex1} and \ref{app:vertex2}. The corresponding gauge transformations are also given explicitly there and we shall only reproduce here their schematic form\footnote{In \eqref{gauge4} fields and gauge parameters are meant to carry the same indices as in \eqref{free_gauge} and one has to distribute the derivative over all tensors and consider all possible contractions of indices (see appendices \ref{app:vertex1} and \ref{app:vertex2} for details).}
\begin{subequations} \label{gauge4}
\begin{align}
\d g & = \nabla v + \phi \nabla \x + \vf \nabla \k + \phi \vf \nabla \x + \phi^2 \nabla \k + \vf^2 \nabla \k + \cdots \, , \label{dg4} \\[5pt]
\d \phi & = \nabla \x + \vf \nabla \x + \phi \nabla \k + \phi \nabla v + \cdots \, , \label{dphi4} \\[5pt]
\d \vf & = \nabla \k + \vf \nabla \k + \phi \nabla \x + \vf \nabla v + \cdots \, . \label{dvf4}
\end{align}
\end{subequations}
These transformations leave the action invariant up to terms of quadratic order in the higher-spin fields. Invariance up to that order does not impose any restriction on the coupling constants in the cubic vertices, but these can be fixed either by demanding invariance up to the cubic order, or equivalently, by asking for the closure of the algebra of asymptotic symmetries (see discussion below \eqref{var3Wext_comp}). As in Sect.~\ref{sec:action3} we omitted higher-order corrections in both the action and the gauge transformations. They are instrumental to secure the gauge symmetry, but irrelevant to determine the asymptotic symmetries of the model.

\subsection{Boundary conditions}\label{sec:bnd4}

In analogy with our treatment of the asymptotic analysis of the coupled spin-2 -- spin-3 system, we shall first give the boundary conditions on the fields, motivating them heuristically. Then, we shall verify their consistency, i.e.\ that they fulfill the three conditions outlined in the introduction.

\subsubsection{Boundary conditions on the spin-4 field}

The rationale behind the boundary conditions on the spin-4 field is the same as in Sect.~\ref{sec:bnd3}: we require that the angular components with all indices down are $\cO(1)$, which implies $\vf^{ijkl} = \cO(r^{-8})$. The components with radial indices then follow the rule that each time one replaces one angular index $i$ by the radial index $r$, the behavior of the leading fall-off term is multiplied by $r$. In analogy with the spin-3 case, the independent angular components turn out to be the two independent spin-4 charges.

The existence of a self-interacting cubic vertex for the spin-4 field requires however an additional important specification: the first subleading components of the field must satisfy the asymptotic equations of motion, since they enter the computation of asymptotic symmetries even in the absence of tensors of spin $>4$. This yields the following boundary conditions on the spin-4 field:
\begin{subequations} \label{bnd4}
\begin{align}
\vf^{rrrr} & = \cO(r^{-4}) \, , \label{vf^rrrr} \\[10pt]
\vf^{rrri} & = \cO(r^{-7}) \, , \label{vf^rrri} \\[10pt]
\vf^{rrij} & = \cO(r^{-8}) \, , \label{vf^rrij} \\[10pt]
\vf^{rijk} & = r^{-9}\, u^{rijk} + \cO(r^{-11}) \, , \qquad \h_{ij} u^{rijk} = 0 \, , \label{vf^rijk} \\[5pt]
\vf^{ijkl} & = \frac{8\p C_2}{k} \left( \frac{1}{r^8}\, \cU^{ijkl} - \frac{10\p}{k\,r^{10}}\, \cL_m{}^{(i} \cU^{jkl)m} \right) + \cO(r^{-12}) \, , \label{vf^ijkl}
\end{align}
\end{subequations}
where $\cU^{ijkl}$ is a symmetric tensor which is both traceless and conserved:
\be
\pr_{\,i} \cU^{ijkl} = 0 \, , \qquad\qquad \cU^{ijk}{}_k = 0 \, .
\ee
It is the spin-4 analogue of the boundary currents $\cL^{ij}$ and $\cW^{ijk}$, and it admits only two independent chiral components:
\be \label{U_comp}
\cU_{++++} = \cU(x^+) \, , \qquad \cU_{----} = \tilde{\cU}(x^-) \, , \qquad \cU_{+++-} = \cU_{++--} = \cU_{+---} = 0 \, .
\ee
The strengthening of the boundary conditions on almost all radial components with respect to the rule recalled above is forced by the asymptotic equations of motion: the leading terms in $\vf^{rrri}$, $\vf^{rrij}$ and $\vf^{rijk}$ vanish on shell if one fixes the $\cO(r^{-10})$ term in $\vf^{ijkl}$ as in \eqref{vf^ijkl}.  One could also proceed without setting them to zero and taking into account the relations among components imposed by the equations of motion, but this will complicate the already intricate computation of asymptotic symmetries. The overall factor in (\ref{vf^ijkl}) is instead a matter of conventions. It has been chosen so as to agree with the parameterisation of the exact solutions discussed in Appendix \ref{app:fields}. 

The tensor $\cU^{ijkl}$ has density weight 4. As we discussed above, this is because it is the $\cO(r^{-8})$-term in the expansion of $\varphi^{ijkl}$.  The trace-free condition and the divergence-free condition $D_i\, \cU^{ijkl} = \frac{1}{2} [\, \cU , \bar{g} \,]^{ijklm} \bar{g}_{im} = 0$ are invariant under Weyl rescalings of the metric.

\subsubsection{Boundary conditions on the spin-3 field and on the metric}

Increasing the spin of the charges increases their density weight and decreases the power of $r$ at which they appear in contravariant tensors (compare \eqref{bnd2}, \eqref{phi^ijk} and \eqref{vf^ijkl}). Now, the asymptotic variation of the relevant $\cO(r^{-8})$ terms in $\vf^{ijkl}$ naturally involves subleading terms of higher orders than the ones written so far in both the metric and the spin-3 field.  One must therefore, as we already found in the spin-3 case, ``dig deeper" and specify these higher order terms in $g^{ij}$ and $\phi^{ijk}$. The additional higher-order contributions, which were present but unwritten above, must of course be compatible with the asymptotic equations of motion.  We thus consider the following boundary conditions on the spin-3 field:
\begin{subequations} \label{bnd3Bis}
\begin{align}
\phi^{rrr} & = \cO(r^{-5}) \, , \label{phi2^rrr} \\[10pt]
\phi^{rri} & = \cO(r^{-6}) \, , \label{phi2^rri} \\[10pt]
\phi^{rij} & = r^{-7}\, t_2^{rij} + \cO(r^{-9}) \, , \qquad \h_{ij} t^{rij}_2 = 0 \, , \label{phi2^rij} \\[5pt]
\phi^{ijk} & = \frac{6\p C_1}{k} \left( r^{-6}\, \cW^{ijk} + r^{-8}\, t^{ijk} \right) +\, \cO(r^{-10}) \, . \label{phi2^ijk}
\end{align}
\end{subequations}
To simplify computations, we have fixed the gauge   $t^{rrr} = t^{rri} = t^{rij} = 0$ with respect to \eqref{bnd3}, and the field equations correspondingly fix the subleading correction in $\phi^{ijk}$ as
\be \label{t^ijk}
t^{ijk} = -\,\frac{8\p}{k}\, \cL_m{}^{(i} \cW^{jk)m} \, .
\ee
For the metric we consider the following boundary conditions:
\begin{subequations} \label{bnd23Bis}
\begin{align}
g^{rr} & = r^2 + \cO(r^{-6}) \, , \label{g2^rr} \\[10pt]
g^{ri} & = \cO(r^{-7}) \, , \label{g2^ri} \\[5pt]
g^{ij} & = \frac{2}{r^2}\, \h^{ij} - \frac{8\p}{k\, r^4}\, \cL^{ij} + r^{-6}\, h^{ij} + r^{-8}\, h_2{}^{ij} + \cO(r^{-10}) \, . \label{g2^ij} 
\end{align}
\end{subequations}
We have imposed $h^{rr} = h^{ri} = 0$ as suggested by \eqref{gauge_fixing3} as well as similar gauge conditions on the subleading components, thus obtaining  the conditions on $h^{ij}$ and $h_2^{\; ij}$ corresponding to \eqref{on-shell_h} in the form, 
\begin{subequations}
\begin{align}
h^{ij} & = \frac{24\p^2}{k^2}\, \cL^i{}_k \cL^{jk} \, , \label{h^ij} \\
h_2{}^{ij} & = - \left( \frac{64\p^3}{k^3}\, \cL^i{}_k \cL^j{}_l \cL^{kl} + \frac{10\p^2}{3k^2}\, \cW^i{}_{kl} \cW^{jkl} \right) . \label{h2^ij}
\end{align}
\end{subequations}
Note that there is a back-reaction of the spin-3 field on  the $\cO(r^{-8})$-order in $g^{ij}$. 

In \eqref{bnd3Bis} and \eqref{bnd23Bis} we wrote explicitly the terms that one needs to compute the variation of all charges. To check that asymptotic symmetries preserve our boundary conditions on the spin-4 field at, e.g., order $\cO(r^{-10})$ in $\vf^{ijkl}$ one should also impose that the next subleading corrections satisfy the equations of motion.

\subsection{Asymptotic conformal invariance}\label{sec:conf4}

Besides containing the solutions that one can derive from the Chern-Simons formulation, the boundary conditions can be verified to be compatible with the asymptotic conformal symmetry. The check is however slightly subtler in this case, since compensating higher-spin gauge transformations become relevant.

The check that the new boundary conditions for the metric are compatible with the asymptotic conformal invariance proceeds in full analogy with Sect.~\ref{sec:conf3}. Since we consider more subleading contributions in $g^{\m\n}$, one has to consider extra contributions in the gauge parameters too:
\begin{subequations} \label{vTris}
\begin{align} 
v^r & = r\, \z + \frac{\z_1}{r} + \frac{\z_2}{r^3} + \frac{\z_3}{r^5} + \cO(r^{-7}) \, , \\[2pt]
v^i & = \e^i + \frac{\e_1{}^i}{r^2} + \frac{\e_2{}^i}{r^4} + \frac{\e_3{}^i}{r^6} + \cO(r^{-8}) \, . 
\end{align}
\end{subequations}
The variation of the metric agrees with that displayed in Sect.~\ref{sec:conf3} up to the corresponding orders (recall however that now we fixed the gauge $h^{rr} = h^{ri} = 0$). The new contributions are instead
\begin{subequations} \label{var3g_diffeo}
\begin{align}
\d g^{rr} & = \cdots - \frac{12}{r^4}\, \z_3 + \cO(r^{-6}) \, , \label{var3g1} \\
\d g^{ri} & = \cdots + \frac{6}{r^5} \left\{ -\, \e_3{}^i + \frac{1}{3}\, \pr^i \z_3 - \frac{4\p}{3k} \cL^{ij} \pr_j \z_2 + \frac{1}{6}\, h^{ij} \pr_j \z \right\} + \cO(r^{-7}) \, , \label{var3g2}\\
\d g^{ij} & = \cdots + \frac{4}{r^8} \left\{ \pr^{(i} \e_3{}^{j)} + \h^{ij} \z_3 - \frac{2\p}{k}\, [\, \cL , \e_2 \,]^{ij} - \frac{8\p}{k}\, \cL^{ij} \z_2 + \frac{1}{4}\, [\, h , \e_1 \,]^{ij} + \frac{3}{2}\, h^{ij} \z_1   \right. \nn \\
& \left. \phantom{\cdots + \frac{4}{r^8} \qquad} + \frac{1}{4} \, [\, h_2 , \e \,]^{ij} + 2\, h_2{}^{ij} \z \right\} + \cO(r^{-10}) \, . \label{var3g3}
\end{align}
\end{subequations}
The pattern that already emerged in the spin-3 case repeats itself exactly along the same lines here: from \eqref{var3g1} and \eqref{var3g2} one fixes $\z_3$ and $\e_3{}^i$. Substituting the result in \eqref{var3g3} gives a $\d h_2{}^{ij}$ which is consistent with its definition in terms of $\cL_{ij}$ and $\cW_{ijk}$ (i.e.\ the variation computed from \eqref{var3g_diffeo} and that computed from the variations \eqref{QVariation1} and \eqref{var2W} agree).

Preservation of the boundary conditions imposes again that $\e^i$ be a conformal Killing vector and $\z$ takes the same form as in \eqref{comp_conf_fix}. As a result the variation of $\cW_{ijk}$ remains the same as in \eqref{var2W}. In the subleading orders, however, preserving the new boundary conditions requires to dispose of the variations induced by diffeomorphisms with a compensating higher-spin gauge transformation. For instance:
\be \label{var3diff_sub}
\d \phi^{rij} = \frac{6\p C_1}{k\,r^5}\, \cW^{ijk} \pr_k \z + \cO(r^{-7}) \, ,
\ee
so that preserving $\phi^{rij} = \cO(r^{-7})$ requires the combination of the asymptotic diffeomorphism with another gauge transformation. One can easily achieve this result using the component $\chi_3{}^{ij}$ of the gauge parameter, which enters algebraically the variation $\d \phi^{rij}$ at the order $\cO(r^{-5})$. Since it does not play any role in determining $\d \cW^{ijk}$, one can use $\chi_3{}^{ij}$ to absorb the variation \eqref{var3diff_sub} without spoiling the discussion of Sect.~\ref{sec:killing3}. Moreover, the same compensating gauge transformation is instrumental in obtaining the correct transformation for the $t^{ijk}$ in \eqref{phi2^ijk}, i.e.\ $\d t^{ijk} = - \frac{8\p}{k} ( \d\cL_m{}^{(i}\cW^{jk)m} + \cL_m{}^{(i}\d\cW^{jk)m} )$.

In the case of the spin-4 field, a diffeomorphism
\be
\d \vf^{\m\n\r\s} = [\, \vf , v \,]^{\m\n\r\s} \equiv 4\, \vf^{\a(\m\n\r} \pr_\a v^{\s)} - v^\a \pr_\a \vf^{\m\n\r\s} = -\, \cL_v \vf^{\m\n\r\s}  
\ee
generated by the asymptotic Killing vectors \eqref{vTris} induces the variations $\d \vf^{rrrr} = \cO(r^{-4})$, $\d \vf^{rrrr} = \cO(r^{-7})$ and $\d \vf^{rrij} = \cO(r^{-8})$ which are consistent with the boundary conditions \eqref{bnd4}. Preserving the condition on the remaining component with radial indices requires instead a compensating gauge transformation since
\be \label{var4_diff1}
\d \vf^{rijk} = \frac{8\p C_2}{k\,r^7}\, \cU^{ijkl} \pr_l \z + \cO(r^{-9}) \, .
\ee
It is direct to see, without specifying explicitly the compensating spin-4 gauge transformation, that such a compensating transformation does exist.  This is all that is required for our purposes. The mechanism is the same as the one that we have already seen at work in \eqref{var3diff_sub}. Some components of the spin-4 gauge parameter $\k^{\m\n\r}$ enter algebraically the contribution from the Schouten bracket at this order, and they do not contribute to the variation of the charges. Therefore, one can safely use them to cancel the variation \eqref{var4_diff1}. 

From the variation of the component with only angular indices one obtains finally
\be \label{var2U}
\delta \cU^{ijkl} = [\, \cU , \e \,]^{ijkl} - 4\, \cU^{ijkl} \prd \e \, .
\ee
This equation confirms that $\cU^{ijkl}$ is a tensor density of weight 4. It preserves the trace-free and divergence-free conditions and implies
\be
\d\cU = - \, \e\, \cU^{\,\pe} - 4\, \e' \, \cU \, , \qquad \d\tilde{\cU} =  -\, \tilde{\e}\, \tilde{\cU}^{\,\pe} - 4\, \tilde{\e}^{\,\pe} \, \tilde{\cU} \, .
\ee
%

\subsection{Asymptotic Killing tensors and $W_4$-algebra}\label{sec:killing4}

To complete the analysis of the asymptotic symmetries of the model, one has to consider also the remaining gauge transformations. In the present setup the first corrections to the quadratic action become relevant. It is thus convenient to organize the higher-spin gauge transformations as 
\be \label{var-schem_3}
\d_3 g = \d^{(0)}_3 g + \d^{(1)}_3 g + \cdots \, , \quad 
\d_3 \phi = \d^{(0)}_3 \phi + \d^{(1)}_3 \phi + \cdots \, , \quad
\d_3 \vf = \d^{(1)}_3 \vf + \cdots \, ,
\ee
and
\be \label{var-schem_4}
\d_4 g = \d^{(0)}_4 g + \d^{(1)}_4 g + \cdots \, , \quad 
\d_4 \phi = \d^{(1)}_4 \phi + \cdots \, , \quad
\d_4 \vf = \d^{(0)}_4 \vf + \d^{(1)}_4 \vf + \cdots \, .
\ee
The terms of lowest order in the variations of the higher-spin fields are the Schouten brackets of the inverse metric with the higher-spin gauge parameters,
\be \label{lowest_delta}
\d^{(0)}_3 \phi^{\m\n\r} = [\, g , \x \,]^{\m\n\r} \, , \qquad
\d^{(0)}_4 \vf^{\m\n\r\s} = [\, g , \k \,]^{\m\n\r\s} \, ,
\ee
and they are accompanied by corresponding variations of the metric (see e.g.\ \eqref{dg} for the spin-3 case).\footnote{At each order of the expansion in the higher-spin fields we consider the full non-linear coupling with the metric. For this reason the covariantised lowest-order gauge transformations \eqref{lowest_delta} are accompanied by a transformations of the metric, which is not present if one considers all fields as linearised fluctuations around an $AdS_3$ background. The detailed form of the lowest order in a generic higher-spin transformation of the metric can be found in Appendix C of \cite{metric-like}.} The next to leading orders have the schematic form already recalled in \eqref{gauge4}:
\begin{alignat}{5}
\d^{(1)}_3 g & = \phi \vf \nabla \x \, , \qquad 
& \d^{(1)}_3 \phi & = \vf \nabla \x \, , \qquad
& \d^{(1)}_3 \vf & =  \phi \nabla \x \, , \\[5pt]
\d^{(1)}_4 g & = \phi^2 \nabla \k + \vf^2 \nabla \k \, , \qquad 
& \d^{(1)}_4 \phi & = \phi \nabla \k \, , \qquad
& \d^{(1)}_4 \vf & = \vf \nabla \k \, .
\end{alignat}
Precise (but rather lengthy\ldots) expressions for these gauge transformations are given in appendices \ref{app:vertex1} and \ref{app:vertex2}.

One easily convinces oneself that asymptotic Killing tensors must continue to have the same leading dependence on the radial coordinate as the exact Killing tensors of $AdS_3$. We thus consider spin-3 gauge parameters of the form
\begin{subequations} \label{xmn4}
\begin{align}
\x^{rr} & = r^2\, \l + \l_1 + r^{-2} \l_2 + r^{-4} \l_3 + \cO(r^{-4}) \, , \\[5pt]
\x^{ri} & = r\, w^i + r^{-1}\, w_1{}^i + r^{-3}\, w_2{}^i + r^{-5}\, w_3{}^i + \cO(r^{-5}) \, , \\[5pt]
\x^{ij} & = \chi^{ij} + r^{-2}\, \chi_1{}^{ij} + r^{-4}\, \chi_2{}^{ij} + r^{-6}\, \chi_3{}^{ij} + \cO(r^{-6}) \, ,
\end{align}
\end{subequations}
(we now have to control an additional subleading order with respect to \eqref{xmn}) and spin-4 gauge parameters of the form
\begin{subequations} \label{kmnr}
\begin{align}
\k^{rrr} & = r^3\, \a + r\, \a_1 + r^{-1}\, \a_2 + r^{-3}\, \a_3 + \cO(r^{-3}) \, , \\[5pt]
\k^{rri} & = r^2\, \b^i + \b_1{}^i + r^{-2}\, \b_2{}^i + r^{-4}\, \b_3{}^i + \cO(r^{-4}) \, , \\[5pt]
\k^{rij} & = r\, \g^{ij} + r^{-1}\, \g_1{}^{ij} + r^{-3}\, \g_2{}^{ij} + r^{-5}\, \g_3{}^{ij} + r^{-7}\, \g_4{}^{ij} + \cO(r^{-7}) \, , \\[5pt]
\k^{ijk} & = \s^{ijk} + r^{-2}\, \s_1{}^{ijk} + r^{-4}\, \s_2{}^{ijk} + r^{-6}\, \s_3{}^{ijk} + r^{-8}\, \s_4{}^{ijk} + \cO(r^{-8}) \, .
\end{align}
\end{subequations}
In both cases one also has to take into account the trace constraints $\x_\l{}^\l = \k^{\m\l}{}_\l = 0$, which impose algebraic relations on the components and allow, e.g., to eliminate $\x^{rr}$, $\k^{rrr}$ and $\k^{rri}$ in terms of the other components.

\subsubsection{Boundary conformal Killing tensors}

We begin the asymptotic analysis by looking at the leading behavior of each contribution to the purely angular components coming from \eqref{var-schem_3} and \eqref{var-schem_4}: 
\begin{subequations}
\begin{alignat}{5}
\d^{(0)}_3 \phi^{ijk} & = \cO(r^{-2}) \, , \qquad\quad 
& \d^{(1)}_3 \phi^{ijk} & = \cO(r^{-6}) \, , \qquad\quad 
& \d^{(1)}_4 \phi^{ijk} & = \cO(r^{-4}) \, , \\[5pt]
\d^{(0)}_4 \vf^{ijkl} & = \cO(r^{-2}) \, , \qquad\quad 
& \d^{(1)}_4 \vf^{ijkl} & = \cO(r^{-6}) \, , \qquad\quad 
& \d^{(1)}_3 \vf^{ijkl} & = \cO(r^{-6}) \, .
\end{alignat}
\end{subequations}
The first important observation is that, for both $\phi^{ijk}$ and $\vf^{ijkl}$, the leading term comes from the Schouten bracket, i.e.
\be \label{dphi^ijk_Bis}
\d \phi^{ijk} = \frac{6}{r^2} \left\{ \pr^{(i} \chi^{jk)} + 2\, \h^{(ij} w^{k)} \right\} + \cO(r^{-4})
\ee
and
\be \label{dvf^ijkl}
\d \vf^{ijkl} = \frac{8}{r^2} \left\{ \pr^{(i} \s^{jkl)} + 3\, \h^{(ij} \g^{kl)} \right\} + \cO(r^{-4}) \, .
\ee
One can thus repeat the first step in the analysis of Sect.~\ref{sec:killing3} verbatim: cancellation of the $\cO(r^{-2})$ contribution in \eqref{dphi^ijk_Bis} is required by consistency with the boundary conditions and imposes that $\chi^{ij}$ be a conformal Killing tensor. In full analogy, $\s^{ijk}$ and $\g^{ij}$ must be traceless because
\be
g_{\a\b} \k^{\a\b r} = \frac{r^3}{2}\, \h_{ij} \g^{ij} + \cO(r) \, , \qquad
g_{\a\b} \k^{\a\b i} = \frac{r^2}{2}\, \h_{jk} \s^{ijk} + \cO(1) \, . 
\ee
Combining this information with \eqref{dvf^ijkl}, one concludes that $\s^{ijk}$ must satisfy the conformal Killing tensor equation  
\be \label{killing4}
\pr^{(i} \s^{jkl)} - \12\, \h^{(ij} \prd \s^{kl)} = 0 \, ,
\ee
while
\be \label{gammaij}
\g^{ij} = -\, \frac{1}{6}\, \prd \s^{ij} \, .
\ee
In terms of $\bar{g}^{ij}$, \eqref{killing4} can be rewritten as
\be
[\, \bar{g}, \s \,]^{ijkl} = \mu^{(ij} \bar{g}^{kl)} 
\ee
for some $\mu^{ij}$, which exhibits its invariance under conformal rescalings of the metric. In two dimensions the conformal Killing equation, together with the tracefree condition, implies that $\s^{ijk}$ has two independent chiral components:
\be \label{solkilling4}
\s^{+++} = \s(x^+) \, , \qquad \s^{---} = \tilde{\s}(x^-) \, , \qquad \s^{++-} = \s^{+--} = 0 \, .
\ee

In conclusion, asymptotic symmetries continue to be generated by conformal Killing tensors of the flat boundary metric. This information is already encoded in the linearised gauge transformations, and we just explicitly verified that higher-spin interactions do not spoil it. However, as we shall see below, the higher-spin terms do modify the transformations of the charges. 

One should now study the behavior at $r \to \infty$ of the spin-3 and spin-4 gauge transformations of all fields with two goals:
\begin{itemize}
\item derive the transformations of the charges $\cL^{ij}$, $\cW^{ijk}$, $\cU^{ijkl}$ generated by the boundary conformal Killing tensors $\e^i$, $\chi^{ij}$, $\s^{ijk}$;
\item check the consistency of our boundary conditions, i.e.\ control also the variation of the radial components and of the subleading orders that have been specified in \eqref{phi2^ijk} and \eqref{g2^ij}.
\end{itemize}
We shall proceed by first examining the spin-3 transformations and then moving to the spin-4 ones. We shall however mainly focus on the variations $\d_\chi \cW^{ijk}$ and $\d_\s \cU^{ijkl}$ that suffice to display all novelties of the spin-4 case without loosing one's way in technicalities.

\subsubsection{Spin-3 gauge transformations}\label{sec:var3spin4}

We begin with reconsidering the variation of $\cW^{ijk}$ under spin-3 transformations. The aim is to show how interactions influence asymptotic symmetries and, viceversa, how the Jacobi identities of the asymptotic symmetry algebra constrain the coupling constants of the metric-like theory.

We already examined $\d^{(0)}_3 \phi^{\m\n\r}$ in Sect.~\ref{sec:killing3}, but when one adds a rank-4 tensor the interactions between higher-spin fields require the additional $\d^{(1)}_3 \phi^{\m\n\r}$ detailed in \eqref{dx-phi}. Asymptotically the extra term in the gauge variation gives
\begin{subequations} \label{var3(1)phi3}
\begin{align}
\d^{(1)}_3 \phi^{rrr} & = \cO(r^{-1}) \, , \\[5pt]
\d^{(1)}_3 \phi^{rri} & = \frac{2C_2\p}{3k\,r^2}\, c_1\, \cU^{ij}{}_{kl} \pr_j \chi^{kl} + \cO(r^{-4}) \, , \\[5pt]
\d^{(1)}_3 \phi^{rij} & = \frac{2C_2\p}{3k\,r^3} (2a_2-a_3-3b_1-6b_2)\, \cU^{ij}{}_{kl} \chi^{kl} + \cO(r^{-5}) \, , \\[5pt]
\d^{(1)}_3 \phi^{ijk} & = \frac{4C_2\p}{k\,r^6} \Big\{ a_1\, \cU^{ijkl} \prd \chi_{l} + b_1\, \chi^{lm} \pr_{l\,} \cU^{ijk}{}_m + a_2\, \cU_{\,lm}{}^{(ij} \pr^{k)} \chi^{lm} \nn \\
& + b_2\, \chi^{lm} \pr^{(i} \cU^{jk)}{}_{lm} + a_3\, \cU^{lm(ij} \pr_{l\,} \chi^{k)}{}_m + c_1\, \h^{(ij} \cU^{k)}{}_{lmn} \pr^l \chi^{mn} \nn \\
& + (4a_1 + 2a_2 + a_3 - 5b_1 - 2b_2 )\, \cU^{ijkl} w_l \Big\} + \cO(r^{-8}) \, ,
\end{align}
\end{subequations}
where we displayed only the contributions that influence $\d_\chi \cW^{ijk}$. The coefficients are fixed as in \eqref{d3phi3_par} and \eqref{d3phi3_field}; they depend on the coupling constant $\g$ of the 3--3--4 vertex and on a set of free coefficients, denoted by $r_i$, which parameterise field redefinitions (see \eqref{redef3->34} and \eqref{redef4->33}). One can simplify \eqref{var3(1)phi3} using the strategy adopted in Sect.~\ref{sec:d3g} to study the gauge variation of the metric: one can (i) fix conveniently the parameters\footnote{This is allowed because the field redefinitions \eqref{redef3->34} do not affect our boundary conditions \eqref{bnd3Bis}.} $r_i$ as in \eqref{rnumbers}, and (ii) take into account that $\chi^{ij}$ is an asymptotic Killing vector while $w^i$ satisfies \eqref{w^i}. If one also fixes $a_2 = \frac{9\g}{2}$, the variations of the radial components then become
\be
\d^{(1)}_3 \phi^{rrr} = \cO(r^{-1}) \, , \qquad
\d^{(1)}_3 \phi^{rri} = \cO(r^{-4}) \, , \qquad
\d^{(1)}_3 \phi^{rij} = \cO(r^{-5}) \, .
\ee
They are subleading with respect to \eqref{final_var3}, so that $\l_1$, $w_1{}^i$, $w_2{}^i$, $\chi_1{}^{ij}$ and $\chi_2{}^{ij}$ are the same as in the coupled spin-2 -- spin-3 system. 

The variation of the angular components receives instead the following correction at order $\cO(r^{-6})$:
\be \label{var3fin^ijk}
\d^{(1)}_3 \phi^{ijk} = \frac{3C_2\p\g}{k\,r^6}\, \Big\{ \chi^{lm}\! \left(\, 2\, \pr_{l\,} \cU_m{}^{ijk} + \pr^{(i} \cU^{jk)}{}_{lm} \right) + 6\, \cU_{\,lm}{}^{(ij} \pr^{k)} \chi^{lm} \Big\} + \cO(r^{-8}) \, .
\ee
One has therefore to add to the variation $\d_\chi \cW^{ijk}$ given by \eqref{VarW21} the terms
\be \label{var3Wext}
\d_\chi \cW^{ijk} = \cdots + \frac{C_2}{2 C_1}\,\g\, \Big\{ \chi^{lm}\! \left(\, 2\, \pr_{l\,} \cU_m{}^{ijk} + \pr^{(i} \cU^{jk)}{}_{lm} \right) + 6\, \cU_{\,lm}{}^{(ij} \pr^{k)} \chi^{lm} \Big\} \, ,
\ee
which preserve the trace-free and divergence-free conditions on $\cW^{ijk}$ and imply
\be \label{var3Wext_comp}
\begin{split}
\d_\chi \cW = \frac{3C_2}{2 C_1}\,\g \left( \chi\, \cU' + 2\, \chi' \cU \right) - \frac{1}{6 C_1} \Big\{ & 2\,\chi\cL''' + 9\,\chi'\cL'' + 15\,\chi''\cL' + 10\,\chi'''\cL \\
& - \frac{64\p}{k} \left( \chi\cL\cL' + \chi'\cL^2 \right) - \frac{k}{4\p}\, \chi^{(5)} \Big\}
\end{split}
\ee
in agreement with the result in the Chern-Simons formulation (see e.g.\ \cite{GH,Wlambda,Tan:2011tj}).
We thus see, as announced,  that although the transformations of the charges are always generated by conformal Killing tensors, their precise form depends on the spectrum of the theory.

A few comments are in order: the first, more technical, is that we could have obtained the same $\d_\chi \cW^{ijk}$ working with arbitrary $r_i$. The corrections to $w_2{}^i$ and $\chi_2{}^{ij}$ would have been compensated by the different structure of \eqref{var3fin^ijk}. This is the analogue of what we discussed at the end of Sect.~\ref{sec:d3g}: to detect the influence of field redefinitions on asymptotic symmetries one has to deal with spin-4 charges as we shall do in the next subsection.

The second comment concerns instead the structure of the model: the variation \eqref{var3Wext} does contain the coupling constant $\g$ of $\cL_{3-3-4}$, but the overall coefficient $C_2 \g$ cannot be freely taken once one has fixed the normalization of all charges. Demanding that asymptotic symmetries satisfy the Jacobi identities without the need for additional generators fixes $C_2 \g$ and hence the coupling constant $\g$ of $\cL_{3-3-4}$ in terms of $C_2$. We can compare the value of $\g$  e.g.\ with Eq.~(3.27) of \cite{GH}. To this end one has to rescale $\chi$, obtaining
\be \label{shift_chi}
\chi \to \frac{N_3 C_1}{2}\, \chi \quad \Rightarrow  \quad \g =  \frac{8}{3N_3C_2} \, ,
\ee
where $N_3$ is the function defined in \eqref{N_l}, that for the present model reads $N_3 = \frac{12}{5}$. In general, i.e.\ in the presence also of symmetric tensors of higher rank, the coupling constant $\g$ corresponds to the parameter introduced in \cite{triality} to label the conformal structures that can appear at infinity. This comparison also implies that models involving one symmetric tensor of each rank from 2 to $\infty$ -- corresponding to the Chern-Simons theories with $hs[\l]$ gauge algebra briefly recalled in Appendix \ref{app:fields} -- should admit only a single independent dimensionless coupling constant besides Newton's constant.

Let us now turn to the spin-3 variation of the metric: the first correction to the gauge transformation, i.e.\ the $\d_3^{(1)} g^{\m\n}$ induced by the 3--3--4 vertex, is subleading with respect to the terms that we considered in \eqref{d3g_lead}. As a result, the variation of the spin-2 charges is not affected as it should, and $\d_\chi \cL^{ij}$ remains the same as in \eqref{d3L}. 

The spin-4 field varies as well under spin-3 gauge transformations. Preservation of our boundary conditions requires a compensating spin-4 transformation, in the spirit of what we have already seen e.g.\ in \eqref{var4_diff1} when we discussed diffeomorphisms. To complete the calculation of $\d_\chi \cU^{ijkl}$ one thus has to control also the spin-4 gauge transformations which we discuss below. At any rate, the covariant calculation is rather involved and, with our present understanding, not particularly illuminating. For this reason, we confine ourselves to report the variation of the left-moving component of $\cU^{ijkl}$,
\be \label{d3U}
\begin{split}
\d_\chi \cU = & - \frac{3C_1\g}{80C_2} \Big\{ \chi \cW''' + 6\, \chi' \cW'' + 14\, \chi'' \cW' + 14\, \chi''' \cW \\
& - \frac{4\p}{k} \left( 25\, \chi \cL' \cW + 18\, \chi \cL \cW' + 52\, \chi' \cL \cW \right) \Big\} \, ,
\end{split}
\ee
which agrees with the outcome of the computation in the Chern-Simons setup. One can obtain this result working with arbitrary $r_i$, so that all spin-3 variations of the charges are not affected by field redefinitions.  Notice that all constants which enter the variation $\d_\chi \cU$ of (\ref{d3U}) have been  fixed by our prior analysis. One can verify, for instance, that if one rescales the gauge parameter as in \eqref{shift_chi} one reproduces the correct ratio between the normalization of the higher-spin fields in \eqref{C2}:
\be
\frac{C_1^2}{C_2^2} = - \frac{4 N_4}{3 N_3} = 2 \, . 
\ee

In the boundary conditions \eqref{bnd4}, \eqref{bnd3Bis} and \eqref{bnd23Bis} we have specified also some terms in the expansion in powers of $r^{-2}$ that are subleading with respect to the ones which accommodate the charges. Besides computing the variations of the charges, one should also verify that the variations of the subleading components agree with our boundary conditions. This is a formidable task due to the intricate structure of the gauge transformations, in particular $\d_3^{(1)} g^{\m\n}$.  But this is however only a consistency check guaranteed to hold since, as we already recalled in Sect.~\ref{sec:conf4}, asymptotic symmetries map solutions of the equations of motion into other solutions. 

\subsubsection{Spin-4 gauge transformations}\label{sec:var4spin4}

As in the analysis of the coupled spin-2 -- spin-3 system of Sect.~\ref{sec:killing3},  it is convenient to first study the variation of the spin-4 field in order to control the structure of the allowed asymptotic spin-4 gauge transformations. We already discussed in \eqref{dvf^ijkl} the leading order in the variation of the purely angular components, and we noticed that its cancellation implies that $\s^{ijk}$ -- the leading order in the purely angular component of (\ref{kmnr}) -- is a conformal Killing tensor. Furthermore, we also fixed $\g^{ij}$ in terms of $\s^{ijk}$ in \eqref{gammaij}. These conditions,  however, do not suffice to guarantee the preservation of our boundary conditions and, again in analogy with what we have seen in the spin-2 -- spin-3 system, one has to constrain the other parameters  that appear in the expansion \eqref{kmnr} of the asymptotic symmetries. 

To elucidate the procedure we can look at the variations
\begin{subequations} \label{first_delta_vf}
\begin{align}
\d_4^{(0)} \vf^{rrij} & = 4 \left\{ -\, \g_1{}^{ij} + \pr^{(i} \b^{j)} + \h^{ij} \a \right\} + \cO(r^{-2}) \, , \\[5pt]
\d_4^{(0)} \vf^{rijk} & = \frac{2}{r} \left\{ -\, \s_1{}^{ijk} + 3\, \pr^{(i} \g^{jk)} + 6\, \h^{(ij} \b^{k)} \right\} + \cO(r^{-3}) \, ,
\end{align}
\end{subequations}
which are the counterparts of \eqref{dphi_1}. Preservation of our boundary conditions requires to express $\s_1{}^{ijk}$ and $\g_1{}^{ij}$ in terms of the boundary conformal Killing tensor $\s^{ijk}$. The trace constraint on the gauge parameter fixes indeed also $\a$ and $\b^i$: the condition $\k^{r\l}{}_\l = 0$ at order $\cO(r)$ implies 
\be
\a + \frac{1}{2}\, \h_{ij} \g_1{}^{ij} + \frac{2\p}{k}\, \cL_{ij} \g^{ij} = 0 \, ,
\ee
while the condition $\k^{i\l}{}_\l = 0$ at order $\cO(1)$ implies
\be
\b^i + \frac{1}{2}\, \h_{jk} \s_1{}^{ijk} + \frac{2\p}{k}\, \cL_{jk} \s^{ijk} = 0 \, .
\ee
In this case the deformation of the linearised gauge transformations is irrelevant since both $\d_s^{(1)} \vf^{rijk}$ and $\d_s^{(1)} \vf^{rrij}$ are subleading with respect to \eqref{first_delta_vf} for $s = 3,4$. 

This structure clearly repeats itself at each order in the expansion in powers of $r^{-2}$. From the variations of the components $\vf^{rijk}$ and $\vf^{rrij}$ one fixes $\s_n{}^{ijk}$ and $\g_n{}^{ij}$ and the trace constraint fixes accordingly $\b_{n-1}{}^i$ and $\a_{n-1}$ (one can express the components $\k^{rrr}$ and $\k^{rri}$ of the gauge parameter in terms of the others using the trace constraint). At this point the variations of the remaining components of the field are also fixed up to a certain order, and one only has to verify that they are consistent with our boundary conditions. The only difference with respect to Sect.~\ref{sec:killing3} is that the deformations of the linearised gauge transformations start to play a role, and they have to be taken into account when one expresses $\s_n{}^{ijk}$ and $\g_n{}^{ij}$ in terms of the boundary conformal Killing tensors. For instance, at the next order in $r^{-2}$ one obtains
\begin{subequations} \label{2nd_delta_vf}
\begin{align}
\d_4^{(0)} \vf^{rrij} & = \frac{8}{r^2}\, \Big\{ - \g_2{}^{ij} + \frac{1}{2}\, \pr^{(i} \b_1{}^{j)} + \frac{1}{2}\, \h^{ij} \a_1 - \frac{\p}{k} [\, \cL , \b \,]^{ij} - \frac{4\p}{k}\, \cL^{ij} \a \Big\} + \cO(r^{-4}) \, , \\[5pt]
\d_4^{(0)} \vf^{rijk} & = \frac{4}{r^3}\, \Big\{ - \s_2{}^{ijk} + \frac{3}{2}\, \pr^{(i} \g_1{}^{jk)} + 3\, \h^{(ij} \b_1{}^{k)} - \frac{24\p}{k}\, \cL^{(ij} \b^{k)} - \frac{2\p}{k} [\, \cL , \g \,]^{ijk} \Big\} \nn \\
& + \cO(r^{-5}) \, ,
\end{align}
\end{subequations}
but the deformations of the gauge transformations begin to contribute. The spin-4 gauge transformation is deformed as follows:
\begin{subequations} \label{var41spin4_expl}
\begin{align}
\d_4^{(1)} \vf^{rrij} & = -\, \frac{\p C_2}{180k\,r^2} \Big\{ 3(5\r -8\,\tilde{r}_1 + 8\,\tilde{p_2})\, \s^{klm} \pr^{(i} \cU^{j)}{}_{klm} \nn \\
& - 2(23\r -8\,\tilde{r}_1 + 8\,\tilde{p}_2)\, \cU_{klm}{}^{(i} \pr^{j)} \s^{klm} \Big\} + \cO(r^{-4}) \, , \\[5pt]
\d_4^{(1)} \vf^{rijk} & = -\, \frac{3\p C_2}{20k\,r^{3}}(11\r - 8\,\tilde{r}_1 + 8\,\tilde{p}_2)\, \cU_{lm}{}^{(ij} \s^{k)lm} + \cO(r^{-5}) \, .
\end{align}
\end{subequations}
Spin-3 transformations also contribute at this order since
\begin{subequations} \label{var31spin4_expl}
\begin{align}
\d_3^{(1)} \vf^{rrij} & = \frac{\p C_1}{20k\,r^2} \Big\{ 2(9\g -2\,r_1 - 2\,\tilde{a}_2)\, \chi^{kl} \pr^{(i} \cW^{j)}{}_{kl} \nn \\
& - 3(15\g -2\,r_1 - 2\,\tilde{a}_2)\, \cW_{kl}{}^{(i} \pr^{j)} \chi^{kl} \Big\} + \cO(r^{-4}) \, , \\[5pt]
\d_3^{(1)} \vf^{rijk} & = \frac{9\p C_1}{5k\,r^{3}}(6\g-r_1-\tilde{a}_2)\, \cW_l{}^{(ij} \chi^{k)l} + \cO(r^{-5}) \, .
\end{align}
\end{subequations}
In both \eqref{var41spin4_expl} and \eqref{var31spin4_expl} we have not fixed the free coefficients in the gauge transformations since there are no preferred choices that cancel the deformations. 

We can now make more precise the comment on compensating gauge transformations made in the paragraph right above Eq.~(\ref{d3U}). In order to preserve the boundary conditions, an asymptotic variation generated by $\chi^{ij}$ must be accompanied by a compensating spin-4 gauge transformation with $\s^{ijk} = 0$ and
\begin{subequations}
\begin{align}
\g_2{}^{ij} & = \frac{\p C_1}{160k} \Big\{ 2(9\g -2\,r_1)\, \chi^{kl} \pr^{(i} \cW^{j)}{}_{kl}  - 3(15\g -2\,r_1)\, \cW_{kl}{}^{(i} \pr^{j)} \chi^{kl} \Big\} \, , \\[5pt]
\s_2{}^{ijk} & = \frac{9\p C_1}{20k}(6\g-r_1)\, \cW_l{}^{(ij} \chi^{k)l} \, ,
\end{align}
\end{subequations}
where we set $\tilde{a}_2 = 0$ for brevity.

Once one has fixed $\g_2{}^{ij}$, $\s_2{}^{ijk}$ and $\b_1{}^i$, $\a_1$ one can verify that the gauge variations of the remaining components satisfy
\be
\d \vf^{rrrr} = \cO(1) \, , \qquad \d \vf^{rrri} = \cO(r^{-1}) \, , \qquad \d \vf^{ijkl} = \cO(r^{-8}) \, ,
\ee 
where we consider here the sum of all variations. To continue one should approach in the same way the next subleading order: preservation of our boundary conditions imposes $\d \vf^{rrij} = \cO(r^{-6})$ and $\d \vf^{rijk} = \cO(r^{-7})$. From these conditions one fixes $\g_3{}^{ij}$ and $\s_3{}^{ijk}$ (whose expression is influenced by the deformations of the linearised gauge transformations). As a result one obtains also $\d \vf^{rrrr} = \cO(r^{-2})$ and $\d \vf^{rrri} = \cO(r^{-3})$ for any value of the coefficients $r_i$ and $\tilde{r}_i$ which parameterise redefinitions of the higher-spin fields.

So far, the computations of the subleading terms in the asymptotic symmetries are somewhat tedious but straightforward and without new feature.  When moving to the purely angular components one encounters instead another qualitative difference with respect to what we discussed in the previous sections: the trace-free and divergence-free conditions are not preserved for arbitrary values of the free parameters associated to field redefinitions. For instance, in light-cone coordinates one obtains
\be
\d \vf^{+++-} = F_1(r_i,p_2)\, \s\cW' + F_2(r_i,p_2)\, \s' \cW \, ,
\ee
where the functions $F_i$ vanish if one chooses e.g.
\be \label{fix-r_i}
p_2 \to r_1 - r_4 + 6\g \, , \qquad r_6 \to -\, \frac{1}{7} \left( 10\,r_1 + 3\,r_2 + \frac{7\,r_7}{2} - \frac{15\,\g}{2} - \frac{200}{\g\,C_1^2} \right) .
\ee
For generic values of the $r_i$ one would instead violate our boundary conditions. This result can be interpreted as follows: field redefinitions of the form $\vf \to \phi^2$ (see \eqref{redef4->33}) do affect our boundary conditions and, as a result, preserving them requires a tuning of the free parameters in the action.

Having fixed the parameters as in \eqref{fix-r_i} one can obtain the variation of the charges from the variations of the components $\vf^{++++}$ and $\vf^{----}$:
\be
\begin{split}
& \d_\s \cU = \frac{\r}{80}\, \Big\{ \s\,\cU''' + 5\,\s'\,\cU'' + 9\,\s''\,\cU' + 6\,\s'''\,\cU - \frac{56\p}{k} \left( \s\cL\,\cU' + \s\cL'\cU + 2\,\s'\cL\,\cU \right) \Big\} \\
& + \frac{1}{180C_2}\, \Big\{ 3\,\s\cL^{(5)} + 20\,\s'\cL^{(4)} + 56\,\s''\cL''' + 84\,\s'''\cL'' + 70\,\s^{(4)}\cL' + 28\,\s^{(5)}\cL \\
& - \frac{4\p}{k} \left( 78\,\s\cL\cL''' + 177\,\s\cL'\cL'' + 352\,\s'\cL\cL'' + 295\,\s'\cL'\cL' + 588\,\s''\cL\cL' + 196\,\s'''\cL^2 \right) \\ 
& + \frac{672\p}{k} \left( \s\cW\cW' + \s'\cW^2 \right) + \frac{2304\p^2}{k^2} \left( 3\,\s\cL^2\cL' + 2\,\s'\cL^3 \right) - \frac{k}{4\p}\, \s^{(7)} \Big\} \, ,
\end{split}
\ee 
while a similar expression holds for $\d_\s \tilde{\cU}$. This time we cannot compare anymore with \cite{GH}, but all relative coefficients agree with Eq.~(3.27e) of \cite{Wlambda}. Even if the normalizations of the latter paper are different from our present ones, it is clear that the coupling constant $\r$ of the 4--4--4 vertex cannot be independent. In general -- if one decides, mimicking \cite{triality}, to parameterise different models with the coupling constant $\g$ of the 3--3--4 vertex -- $\r$ will be a function of $\g$.

The analysis of the gauge variation of the metric is completely analogous to the discussion of Sect.~\ref{sec:d3g} up to the computation of the variation of the spin-2 charges. We therefore refrain from showing the details and we simply report the final result:
\be \label{d4L}
\d_\s \cL^{ij} = \frac{4}{C_2} \left\{\, 3\,\s^{klm}\! \left( \pr_{k\,} \cU_{lm}{}^{ij} + \pr^{(i} \cU^{j)}{}_{klm} \right) + 8\, \cU_{klm}{}^{(i} \pr^{j)} \s^{klm} \,\right\} ,
\ee
which implies
\be
\d_\s \cL = \frac{8}{C_2} \left( 3\,\s\,\cU' + 4\,\s'\,\cU \right) .
\ee

The variation of the spin-3 field under spin-4 transformations is the counterpart of the variation of the spin-4 field under spin-3 transformations since $\d_\s \cW \sim \{\, \cU , \cW \,\} = -\, \{\, \cW ,\cU \,\}$ and $\d_\chi \cU \sim \{\, \cW , \cU \,\}$. The covariant computation of $\d_\s \cW^{ijk}$ is intricate as the covariant computation of $\d_\chi \cU^{ijkl}$, but we verified that the resulting $\d_\s \cW$ is compatible with the $\d_\chi \cU$ displayed in \eqref{d3U} and reads
\be
\begin{split}
\d_\s \cW = & - \frac{\g}{20} \Big\{ 5\, \s \cW''' + 20\, \s' \cW'' + 28\, \s'' \cW' + 14\, \s''' \cW \\
& - \frac{4\p}{k} \left( 27\, \s \cL' \cW + 34\, \s \cL\, \cW' + 52\, \s' \cL\, \cW \right) \Big\} \, . 
\end{split}
\ee

\subsection{Charges \& asymptotic symmetries}\label{sec:charges4}

One can verify also in this case that our boundary conditions are equivalent to the Chern-Simons boundary conditions given in \cite{HR,spin3}. The matching is however more laborious and involves several field redefinitions, as we discuss in Appendix \ref{app:fields}. At any rate, one eventually gets fields that obey the boundary conditions given here, and the importance of field redefinitions in getting the final result explains why we had to fix some of the free parameters in the gauge transformations in the computations of asymptotic symmetries.

In addition to checking agreement, this computation reveals that the coefficients $\cU^{ijkl}$ appearing in the angular components of the spin-4 field are indeed the charges generating the $W$-symmetry. The discussion of Sect.~\ref{sec:charges3} thus extends smoothly to the spin-4 case in spite of the significant increase in the complications involved in the computation of the variation of the charges. Much in the same way as the conserved current associated with the conformal Killing vector $\epsilon^i$ is the vector density of weight one $j^i[\epsilon] = \cL^{ij} \epsilon_j \equiv \cL^{ij} \epsilon^k \bar{g}_{jk}$, the conserved current associated with the conformal Killing tensor $\s^{ijk}$ is the vector density of weight one $j^i[\s] = \cU^{ijkl} \s_{jkl} \equiv \cU^{ijkl} \s^{mnp} \bar{g}_{jm}\bar{g}_{kn} \bar{g}_{lp}$.

Also in this case, once one knows that the charges are $\cL^{ij}$, $\cW^{ijk}$ and $\cU^{ijkl}$, one can read off their algebra from their variation under the $W$-transformations through the formula $\delta_B Q_A = \{Q_A, Q_B\}$ where $Q_A$ stands for a generic charge and $\delta_B Q_A$ is the known variation of $Q_A$ under the transformation generated by $Q_B$.  The variations of all the charges were computed above and given in formulas (\ref{QVariation2}), (\ref{var2WBis}), (\ref{varW21Bis})  and (\ref{d3L_comp}), and we refer to Eq.~(C.3) of \cite{Wlambda} for an explicit expression of the Poisson brackets.

\section{Summary and conclusions}\label{sec:conclusions}
In this paper, we have investigated the asymptotic symmetries of the system consisting of anti-de Sitter gravity coupled to higher spin gauge fields, described in lowest order by the sum of the Einstein-Hilbert action and the Fronsdal action for each higher spin field.  We have focused on the spin-3 and spin-4 cases but the procedure for even higher spins follows the same pattern.  We summarize it here in the general case. 

The crux of the boundary conditions can be synthesized as follows:

\begin{enumerate}
\item The metric behaves asymptotically as
\be \label{metric00}
g  = \frac{dr^2}{r^2} + \left\{ \frac{r^2}{2}\, \h_{ij} + \frac{2\p}{k}\, \cL_{ij}  \right\} dx^i dx^j  + \textrm{``subleading terms"}
\ee
where the subleading terms, although not contributing to the charges because at least of order $\cO(r^{-4})$ with respect to the background metric, cannot, however, be dropped.  Their role will be commented upon below.   The physical information about the gravitational field is contained in the $(i,j)$ (``angular") components of the metric, with $x^i \equiv (t, \phi$).  More specifically, the gravitational charges are the $\cL^{ij}$'s, which are transverse and traceless.  This means in particular that, as discussed in Section \ref{sec:spin2}, the mass is completely shifted to the angular components of the metric through our coordinate choices.  There is no contribution to it from the $g_{rr}$-components, contrary to what happens in standard Schwarzschild coordinates, which therefore do not obey the requested asymptotic behavior and must be transformed to (\ref{metric00}) by a coordinate transformation.  In contravariant form, the asymptotic behavior of the metric reads
\be \label{metric}
g^{-1} = r^2 \theta_r^2 + \left\{ \frac{2}{r^2}\, \h^{ij} - \frac{8\p}{kr^4}\, \cL^{ij}  \right\} \theta_i \theta_j  + \textrm{``subleading terms"}
\ee
where the $\theta_\mu \equiv \frac{\partial}{\partial x^\mu}$'s are dual to the $dx^\mu$'s.
\item The physical information about the spin-$s$ field is contained in the ``angular" components, which behave asymptotically as
\be
\vf_s^{\textrm{cov}} \sim \cW_{i_1 \cdots\, i_s}\, dx^{i_1} \cdots dx^{i_s} + \textrm{``subleading terms"} \, ,
\ee
or in contravariant form
\be \label{spin-s}
\vf_s^{\textrm{contr}} \equiv \vf_s \sim \frac{1}{r^{2s}}\, \cW^{i_1 \cdots\, i_s}\, \theta_{i_1} \cdots \theta_{i_s} + \textrm{``subleading terms"} \, ,
\ee
where again the subleading terms, although not contributing to the charges because at least of order $\cO(r^{-2})$ with respect to the written leading behavior of the field, cannot, however, be dropped.
The spin-$s$ charges are the $\cW^{i_1 \cdots i_s}$'s, which are transverse and traceless. 

\item The asymptotic symmetries are completely determined, up to irrelevant pure gauge terms, by the leading part of their angular components, which is of order $\cO(1)$,
\begin{eqnarray}
v^i (r, x^j)&=& \epsilon^i (x^j)+ \cO(r^{-2}) \, , \\[5pt]
\xi^{i_1 \cdots i_{s-1}}(r,x^j) &=& \lambda^{i_1 \cdots i_{s-1}}(x^j) + \cO(r^{-2}) \, .
\end{eqnarray}
Here, $\epsilon^i$ and $\lambda^{i_1 \cdots i_{s-1}}$ are respectively conformal Killing vectors and traceless conformal Killing tensors of the flat two-dimensional metric $\eta_{ij}$.  The conformal Killing vector and tensor equations,
\begin{eqnarray}
\partial^{(i} \epsilon^{j) } &=& \frac12\, \eta^{ij} \partial \cdot \epsilon \, , \\
\pr^{(i_1} \l^{i_2 \cdots\, i_s)} &=& \frac{1}{2}\, \h^{(i_1i_2} \prd \l^{i_3 \cdots\, i_s)} \, , \label{conf-killConcl}
\end{eqnarray} 
emerge when requesting the vanishing of the $\cO(r^{-2})$-terms in the variation of the contravariant angular components  of the metric and higher spin fields. The space of solutions of these equations is infinite-dimensional and spanned by independent chiral functions, $\epsilon^+ = \epsilon(x^+)$, $\epsilon^- = \tilde{\epsilon}(x^-)$, $\lambda^{+ \cdots +} = \lambda(x^+)$, $\lambda^{- \cdots -} = \tilde{\lambda}(x^-)$, with any mixed component equal to zero, $\lambda^{+ -\cdots }=0$.  The subleading terms of the angular components and the components with at least one radial index are not equal to zero but are completely determined up to irrelevant pure gauge terms by the above $\cO(1)$-terms of the angular components.  Accordingly,  they do not carry independent information.   

The asymptotic symmetries contain the exact vacuum symmetries for low values of the Fourier modes (wedge algebra, see Appendix \ref{sec:iso}) but enlarge them considerably, to the infinite-dimensional $W$-algebras.

\item Requesting the vanishing of the subsequent terms of orders $\cO(r^{-4})$, $\cO(r^{-6})$ up to $\cO(r^{-2s+2})$ in the variation of the contravariant angular components of the higher spin fields, as well as the vanishing of similar unwanted terms in the asymptotic form of the components of the metric and the higher spin fields with at least one radial index, determines successively the subleading terms in the asymptotic symmetries in terms of the leading terms $\epsilon^i$, $\lambda^{i_1 \cdots i_{s-1}}$ and of the  generators $\cL^{ij}$, $\cW^{i_1 \cdots i_s}$.  The polynomial order in the generators generically increases at each successive step.  Finally, the computation of the $\cO(r^{-4})$ terms in the variation of the inverse metric and of the $\cO(r^{-2s})$-terms in the variation of $\varphi_s$, provides the variations of the generators $\cL^{ij}$ and $\cW^{i_1 \cdots i_s}$.  

\item This recursive procedure is identical in spirit to the procedure followed in the Chern-Simons formulation where the successive steps correspond to increasing values of the level.  [Note that on the Chern-Simons side, the level actually also counts the inverse powers of $r$ prior to performing the gauge transformation of \cite{asympt-CS} that eliminates the asymptotic dependence on $r$.] However, the recursive procedure is more intricate in the metric formulation because it involves also the unwritten subleading terms in (\ref{metric}) and (\ref{spin-s}).   For instance, the $\cO(r^{-2k})$ terms in the variation of a higher spin field involve the $\cO(r^{-2j})$-terms ($j\leq k$) of the metric, and if $j>2$, these terms are subleading in (\ref{metric}).   This is the reason why one must specify the subleading terms in (\ref{metric}) and (\ref{spin-s}).  Within the context of the covariant phase space approach adopted here, this is achieved by solving the equations of motion asymptotically. This resolution produces a unique expression for the subleading terms. One needs actually to solve the equations of motion only up to some finite power of $r^{-1}$ that depends on the spin of the fields coupled to gravity. It is interesting to note that as one increases the spin, one must dig deeper into the expansion - being fully on-shell in the limit of infinite spin. 

\item The obtained variations of the generators $\cL^{ij}$, $\cW^{i_1 \cdots i_s}$ are compatible with the transverse and tracelessness conditions and in complete agreement with the nonlinear $W$-algebras derived in the Chern-Simons formulation. The conserved current associated with a boundary conformal Killing tensor $\lambda^{i_1 \cdots i_{s-1}}$ is  $j^k[\lambda] = \cW^{k}_{\; \; i_1 \cdots i_{s-1}} \lambda^{i_1 \cdots i_{s-1}}$ and the corresponding charge is obtained by integration of $j^0$ over a spacelike slice. 

\end{enumerate}

The analysis of the asymptotic structure of higher spin anti-de Sitter gravity provides insight on the emergence of the conformal structure at infinity and is interesting from this point of view.  However, it is quite intricate.  Besides the difficulties already mentioned concerning the necessity to control the subleading terms in the solutions of the equations of motion, the computation is complicated by the fact that one must know in detailed form not only the leading terms, but also higher-order terms in the gauge transformations and the action in an expansion in powers of the higher spin fields, according to the rule that everything that can contribute does actually contribute. Increasing the spin somehow digs deeper in the non-linear structure of the theory since the polynomial terms in the algebra can and indeed do receive corrections from interaction vertices. The first higher-order terms are non negligible at infinity, where they play in fact a crucial role.  They are generically not known in closed form. Determining them, even only up to the needed relevant orders,  is a formidable technical task.  A further complication is that one must take into account the possibility to make field redefinitions, which have a non trivial incidence on the form of the boundary conditions at infinity. 

By contrast, the Chern-Simons approach is much more direct and powerful.\footnote{We should stress, however, that the Chern-Simons formulation is not always available, when dealing with matter couplings or attempting to add a topological mass.  The ``pedestrian" approach analyzed in this paper is at present the only available option for dealing with such cases.} Is there a lesson to be drawn from this difference in complexity?  Perhaps the lesson is again that one must de-emphasize the metric.  While the metric definitely plays a special role in dealing with lower spin fields, which can be treated as fields propagating in a dynamical geometry, the separation into ``geometry" and ``matter" is artificial when higher spin gauge fields are present. This has been already repeatedly emphasized and follows from the fact that the metric transforms under the higher spin gauge symmetries.  This paper shows that the mixing of the metric with the higher spin fields remains relevant at infinity, even though the higher spin fields are ``weak" there.  The simplicity of the Chern-Simons approach follows in great part from the fact that all fields, including the metric, are packaged into a single connection. How to perform the packaging in the metric-like approach -- in three and higher dimensions -- deserves further study.  

In order to overcome at least partially the technical difficulties recalled above, it would be interesting to identify the metric-like counterpart of the so called $u$-gauge in the Chern-Simons formulation (see e.g.\ \cite{Wlambda} and references therein). This setup gives the  algebra of asymptotic symmetries in a basis that contains at most quadratic terms, and this could correspond to a clever choice of field redefinitions that ``neutralize'' asymptotically the contributions coming from higher-order interactions of the higher-spin fields.

Finally, it would be interesting to extend the analysis to include half-integer spin fields. This can in principle be done along the lines of \cite{spinor}, which involves suitable projections of the spinor fields at infinity.  One expects the appropriate conformal Killing spinor-tensor equations to emerge through the preservation of the boundary conditions. Covering half-integer fields would automatically allow one to treat non-principal embeddings.

\subsection*{Acknowledgments}

We would like to thank the Albert Einstein Institute in Potsdam for hospitality during the course of this work. We thank G.~Barnich, X.~Bekaert, S.~Fredenhagen, E.~Joung, T.~Nutma, M.~Taronna and S.~Theisen for helpful discussions. M.H.\ thanks the Alexander von Humboldt Foundation for a Humboldt Research Award. Our research was partially supported by the ERC Advanced Grant ``SyDuGraM'', by FNRS-Belgium (convention FRFC PDR T.1025.14 and  convention IISN 4.4514.08) and by the ``Communaut\'e Fran\c{c}aise de Belgique" through the ARC program.


\begin{appendix}

\section{Conventions}\label{app:conventions}

Greek letters denote indices which take values on all coordinates $x^\m = \{t,r,\phi\}$ of the three-dimensional spacetime, that we assume to have the topology of a cylinder (whose boundary is approached at $r \to \infty$). Latin letters denote instead indices associated to surfaces that are transverse to the radial coordinate, i.e.\ $x^i = \{t,\phi\}$.

A pair of parentheses denotes the symmetrization of the indices it encloses with weight one, i.e.\ one has to divide by the number of terms that enter the symmetrization as, for instance, in
\be
A_{(\m} B_{\n)} = \frac{1}{2} \left( A_\m B_\n + A_\n B_\m \right) .
\ee

We define the Schouten bracket \cite{Nijenhuis} for the symmetric contravariant tensors $v$ (of rank~$p$) and $w$ (of rank~$q$) as the following symmetric contravariant tensor of rank $p+q-1$: 
\be \label{schouten}
[ v , w ]^{\m_1 \cdots \m_{p+q-1}} = \frac{(p+q-1)!}{p!\,q!} \left(\, p\, v^{\a(\m_1 \cdots} \pr_\a w^{\cdots \m_{p+q-1})} - q\, w^{\a(\m_1 \cdots} \pr_\a v^{\cdots \m_{p+q-1})} \,\right) .
\ee
One can equivalently define the previous bracket by associating to the symmetric tensor $v^{\m_1 \cdots \m_p}(x)$ the phase polynomial $v(x,p) = \frac{1}{p!}\, v^{\m_1 \cdots \m_p}(x)\,p_{\m_1} \!\cdots p_{\m_p}$. The bracket \eqref{schouten} is induced by the standard Poisson bracket
\be
\{ v, w \} = \frac{\pr v}{\pr x^\a} \frac{\pr w}{\pr p_\a} - \frac{\pr w}{\pr x^\a} \frac{\pr v}{\pr p_\a}
\ee
as
\be
\{ v(x,p), w(x,p) \} = -\,\frac{1}{(p+q-1)!}\, [ v , w ]^{\m_1 \cdots \m_{p+q-1}}(x)\, p_{\m_1} \!\cdots p_{\m_{p+q-1}} \, .
\ee
The Schouten bracket obeys the Jacobi identity, and one can substitute the ordinary derivative in \eqref{schouten} with any torsionless connection.  

\section{``Isometry'' algebra of the vacuum}\label{sec:iso}

In this Appendix we first briefly recall how one can build all Killing tensors of $AdS_3$ and then we show how one can naturally associate to them a one-parameter family of Lie structures, which corresponds to the one-parameter family of Lie algebras $hs[\lambda]$. These are identified with the wedge algebras of the asymptotic symmetries of models involving a single symmetric tensor of each rank from $2$ to $\infty$ that, for particular values of the parameter $\l$, truncate to the theories with a finite number of symmetric tensors on which we focussed in the main body of the paper. We will therefore explain in which sense the wedge algebras of asymptotic symmetries can be considered as generalized ``isometries'' of the vacuum even in three spacetime dimensions, where several consistent interacting theories are available.

\subsection{Killing tensors of $AdS_3$}\label{app:killing}

In the light-cone coordinates that we often use in the paper, in which the $AdS_3$ space is parameterised as
\be
ds^2 = \ell^2 \left\{ \frac{dr^2}{r^2} - \frac{1}{4} \left( \left( \frac{r^2}{\ell^2} + \frac{\ell^2}{r^2} \right)  dx^+ dx^- + (dx^+)^2 + (dx^-)^2 \right) \right\} ,
\ee
the 6 Killing vectors of $AdS_3$ read
\begin{subequations} \label{killing_vectors}
\begin{align}
J^\pm_0 & = \mp\, \frac{\pr}{\pr x^{\pm}} \, , \\[4pt]
J^\pm_1 & = -\, \frac{r}{2} \sin x^\pm \frac{\pr}{\pr r} - \frac{r^4 + \ell^4}{r^4 - \ell^4} \cos x^\pm \frac{\pr}{\pr x^\pm} + \frac{2\ell^2r^2}{r^4 - \ell^4} \cos x^\pm \frac{\pr}{\pr x^\mp} \, , \\[4pt]
J^\pm_2 & = \mp\, \frac{r}{2} \cos x^\pm \frac{\pr}{\pr r} \pm \frac{r^4 + \ell^4}{r^4 - \ell^4} \sin x^\pm \frac{\pr}{\pr x^\pm} \mp \frac{2\ell^2r^2}{r^4 - \ell^4} \sin x^\pm \frac{\pr}{\pr x^\mp} \, ,
\end{align}
\end{subequations}
where we reinstated the dependence on the $AdS$ radius $\ell$ for clarity. Notice that the components in each set are chiral functions and that this presentations of the Killing vectors makes manifest the isomorphism $so(2,2) \simeq sl(2,\mathbb{R}) \oplus sl(2,\mathbb{R})$ since they satisfy
\be \label{J_basis}
\left[\, J^\pm_a \comma J^\pm_b \,\right] = \e_{ab}{}^c J^\pm_c \, , \qquad  \left[\, J^+_a \comma J^-_b \,\right] = 0 \, .
\ee
The other relevant information is that the components of a generic Killing vector $v^\m$ behave as $v^r = \cO(r)$ and $v^i = \cO(1)$. 

As discussed in Sect.~\ref{sec:AdS}, the anti-de Sitter solution is left invariant by higher-spin transformations generated by \emph{traceless} Killing tensors since, when higher-spin fields vanish, their gauge transformations reduce to
\be
\d \vf_{\m_1 \cdots \m_s} = \nabla^{\textrm{AdS}\,}{}_{\hspace{-15pt}(\m_1\,} \x_{\,\m_2 \cdots \m_s)} =0 \, . 
\ee 
Every Killing tensor of a space of constant curvature is a linear combination of symmetrised products of Killing vectors \cite{killing}. However, generic products are not traceless. Efficient ambient space techniques have been developed to build the traceless Killing tensors of the anti-de Sitter solutions in arbitrary spacetime dimensions (see e.g.\ \cite{algebra,kill_tensors_1,kill_tensors_2}), but in the case of $AdS_3$ one can also take advantage of the chiral splitting. Indeed, for each chiral copy one can introduce the basis
\be
L^\pm_1 = J^\pm_0 + J^\pm_1 \, , \qquad L^\pm_0 = J^\pm_2 \, , \qquad L^\pm_{-1} = J^\pm_0 - J^\pm_1 
\ee 
and take advantage of the following facts:
\begin{enumerate}
\item The Killing tensors 
\be
(W^l_l)^{\m_1 \cdots \m_l} = (L_1)^{\m_1} \cdots (L_1)^{\m_l} \, , \qquad (W^l_{-l})^{\m_1 \cdots \m_l} = (L_{-1})^{\m_1} \cdots (L_{-1})^{\m_l}
\ee
are traceless because the contraction of two Killing tensors is proportional to the Killing metric of $sl(2,\mathbb{R})$,
\be \label{killing-form}
g_{\m\n} (L_i)^\m (L_j)^\n = \frac{\ell^2}{2}\, \k_{ij} \, , \qquad \textrm{with} \quad
\k = \begin{pmatrix}
0 & 0 & -1 \\
0 & \12 & 0 \\
-1 & 0 & 0	
\end{pmatrix} .
\ee

\item The Lie derivative of a traceless Killing tensor along a Killing vector is again a traceless Killing tensor.
\end{enumerate}
Therefore one can build $2l + 1$ traceless Killing tensors of rank $l$ using one of the following and equivalent recursion relations
\be
(W^l_{m\pm1})^{\m_1 \cdots \m_l} = \frac{1}{\pm\,l-m}\, [\, L_{\pm 1} , W^l_m \,]^{\m_1 \cdots \m_l} \, , \label{recursion}
\ee
which are just the translation in the present context of the usual recursion relations that one uses to define the generators of $hs[\l]$ in terms of the generators of its $sl(2,\mathbb{R})$ subalgebra. The resulting tensors satisfy by construction the properties
\be
\nabla^{(\m_1} (W^l_m)^{\m_2 \cdots \m_l)} = 0 \, , \qquad  
g_{\a\b} (W^l_m)^{\a\b\m_1 \cdots \m_{l-2}} = 0 \, , \label{primary}
\ee
and one can easily prove that they also satisfy
\be
\nabla_{\!\a} (W^l_m)^{\a\m_1 \cdots \m_{l-1}} = 0 \, , \qquad 
\left[\, \Box - \frac{l(l+1)}{\ell^2} \,\right] (W^l_m)^{\m_1 \cdots \m_{l}} = 0 \, , \label{wave}
\ee
where here and in the rest of this Appendix $\nabla_\m$ denotes the anti-de Sitter covariant derivative. 
One can repeat the construction for each chiral copy and obtain in total $2(2l+1)$ independent traceless Killing tensors of rank $l$.

With this construction one can easily realize that, in the coordinates we used in \eqref{killing_vectors}, the traceless Killing vectors of $AdS_3$ satisfy $\x^{\,r\cdots\, r\, i_1\cdots\, i_{s-n}} = \cO(r^{n})$ i.e.\ their components behave as $r^{n}$ at $r \to \infty$, where $n$ is the number of radial indices.

\subsection{Algebra of Killing tensors}\label{app:kill-algebra}

Thanks to the construction depicted in the previous subsection we have a one-to-one correspondence between the traceless Killing tensors of $AdS_3$ and the generators of $hs[\lambda]$, but it is still unclear how to recover a one-parameter family of algebras starting from the Killing tensors.

The Schouten bracket provides a natural Lie structure on the previous vector space: even if the bracket of two tensors is in general not traceless, all its traceless components do satisfy the Killing equation. Therefore the algebra of traceless Killing tensors closes if one adds to it the inverse metric, that becomes a central element since $g^{\m\n}$ commutes with all Killing tensors.\footnote{One can actually equivalently define Killing tensors by imposing that their Schouten bracket with the metric vanishes.} For instance, the traceless part of the Schouten bracket $[\,W^2_m , W^2_n \,]^{\m\n\r}$ is proportional to the rank-3 tensor $(W^3_{m+n})^{\m\n\r}$, while its trace satisfies
\be
[\,W^2_m , W^2_n \,]^{\m}{}_\a{}^\a = -\, \frac{\ell^2}{36}(m-n)(2m^2+2n^2-mn-8) (L_{m+n})^\m \, .
\ee
As a result, the decomposition into traceless components gives
\be \label{[W2,W2]}
[\,W^2_m , W^2_n \,]^{\m\n\r} = 2(m-n)(W^3_{m+n})^{\m\n\r} - \frac{\ell^2}{60}(m-n)(2m^2+2n^2-mn-8)\, g^{(\m\n} (L_{m+n})^{\r)} .
\ee
The analogue commutator for $hs[\l]$ reads
\be \label{[W2,W2]_hs}
[\,W^2_m , W^2_n \,] = 2(m-n)\, W^3_{m+n} - \frac{\l^2-4}{60}\,(m-n)(2m^2+2n^2-mn-8)\, L_{m+n} \, .
\ee
Rescaling the generators as
\be
W^\ell_m \rightarrow \l^{l-1}\, W^l_m \, ,
\ee
and taking the limit $\l \to \infty$ one matches \eqref{[W2,W2]} with the identification $\ell^2 g^{\m\n} \sim I$.\footnote{By comparing only \eqref{[W2,W2]} and \eqref{[W2,W2]_hs} one could think to introduce the $\l$ dependence in \eqref{[W2,W2]} with a rescaling of the inverse metric. However, this is excluded by the comparison with another $\l$ dependent set of commutators.}

In conclusion, one can identify the space of traceless Killing tensors of $AdS_3$ supplemented by the Schouten bracket with the algebra $hs[\infty]$. One can also make this identification more precise realizing $hs[\infty]$ as the algebra of area preserving diffeomorphisms of a two-dimensional hyperboloid \cite{Blencowe2}.

One could recover other $hs[\lambda]$ algebras constructing the Lie bracket from the cubic interacting vertices as discussed in \cite{class_algebras,cubic_symmetries}. On the other hand, one can also introduce a Lie structure in another natural way that do not require any information on the structure of higher-spin interactions: it suffices to associate to each Killing vector the differential operator 
\be
L_i \equiv (L_i)^\m\, \nabla^{\textrm{AdS}}_{\!\m} \, .
\ee
The composition of operators defines a non-commutative product and the commutator of two $L_i$ reads
\be \label{comm_diff}
\begin{split}
[\, L_i , L_j \,] & = L_i{}^\m L_j{}^\n [\, \nabla_{\!\m} , \nabla_{\!\n} \,] \,+\, \left( L_i{}^\m \pr_\m L_j{}^\n - L_j{}^\m \pr_\m L_i{}^\n \right) \nabla_{\!\n} \\[5pt]
& = L_i{}^\m L_j{}^\n [\, \nabla_{\!\m} , \nabla_{\!\n} \,] \,+\, (i-j)L_{i+j}{}^\m\, \nabla_{\!\m} \, .
\end{split}
\ee
In general the first term on the right-hand side does not vanish, but if one acts with $[\,L_i,L_j\,]$ on a scalar function it does. This suggests to introduce the operators
\be \label{W_op}
W^l_m \equiv (-1)^{l-m}\,
\frac{(l+m)!}{(2l)!} \,
 \Bigl[ \underbrace{L_{-1}, \dots [\,L_{-1}, [\,L_{-1}}_{\hbox{\footnotesize{$l-m$ terms}}}, (L_{1})^{l}\,]]\Bigr]
\ee
and to act with them on scalar functions. The symmetrised product of two $sl(2,\mathbb{R})$ generators coincides with the differential operator build out of the symmetrised product of their components since
\be
L_{(i}L_{j)} = L_i{}^{(\m} L_j{}^{\n)}\, \nabla_{\!\m} \nabla_{\!\n} + \left( L_i{}^\m \pr_\m L_j{}^\n + L_j{}^\m \pr_\m L_i{}^\n \right) \nabla_{\!\n}
\ee
and the Killing equation, combined with \eqref{killing-form}, implies
\be
L_i{}^\m \pr_\m L_j{}^\n + L_j{}^\m \pr_\m L_i{}^\n = -\, \pr^{\,\n}\! \left( g_{\a\b}\, L_i^\a L_j^\b \right) = 0 \, .
\ee
As a result, the operators $W^l_m$ defined in \eqref{W_op} are in one-to-one correspondence with the traceless Killing tensors defined in \eqref{recursion}: $W^l_m = (W^l_m)^{\m_1 \cdots \m_l}\,\nabla_{\!\m_1} \cdots \nabla_{\!\m_l}$.

One can then compute the commutators of the $W^l_m$ in two ways: either using the definition \eqref{W_op} and the basic commutator \eqref{comm_diff} or using the definition \eqref{recursion} and distributing the derivatives with the Leibniz rule. The first approach gives for instance
\be \label{[W2,W2]_cas}
[\,W^2_m , W^2_n \,] = 2(m-n)\, W^3_{m+n} - \frac{1}{60}(m-n)(2m^2+2n^2-mn-8) \left( 4\,C - 3 \right) L_{m+n} \, ,
\ee
where
\be \label{casimir}
C \equiv L_0^2 - \12 \left( L_{-1}L_{1} - L_1L_{-1} \right)
\ee
is the Casimir operator of $sl(2,\mathbb{R})$.
The second approach gives instead
\be \label{[W2,W2]_f}
\begin{split}
[\,W^2_m , W^2_n \,] f & = 2(m-n)(W^3_{m+n})^{\m\n\r}\, \nabla_{\!\m}\nabla_{\!\n}\nabla_{\!\r} f \\
& - \frac{l^2}{60}(m-n)(2m^2+2n^2-mn-8)\, (L_{m+n})^\m\, \nabla_{\!\m} \! \left( \Box - \frac{3}{l^2} \right)\! f \, ,
\end{split}
\ee
in agreement with the explicit expression of the Casimir operator in this context:
\be
C f \equiv \left( L_0{}^\m L_0{}^\n - L_{-1}{}^{(\m} L_{1}{}^{\n)} \right) \nabla_{\!\m}\nabla_{\!\n} f = \frac{\ell^2}{4}\, \Box f \, ,
\ee
where $f$ is a scalar function.

In $hs[\l]$ the commutator \eqref{[W2,W2]_hs} is recovered via the identification
\be \label{quotient}
C \equiv \frac{1}{4} \left(\l^2 - 1\right) I \, ,
\ee
i.e.\ by choosing a representation for the $sl(2,\mathbb{R})$
algebra and building the generators of $hs[\l]$ as products of the representatives of the $L_i$. In the present context the same identification is possible, but it amounts to restrict the space of functions on which the differential operators acts to the kernel of the wave equation
\be \label{scalar_mass}
\left( \Box - \frac{\l^2-1}{\ell^2} \right)\! f = 0 \, .
\ee 
The mass in \eqref{scalar_mass} is the same as that of the scalars that enter the Vasiliev equations in $D=3$. This confirms the consistency of the whole procedure, that somehow revisits the construction of higher-spin algebras of \cite{algebra}. Even in the absence of matter couplings, one can use this bulk construction to relate the algebra $hs[\l]$ to the ``isometries'' of the vacuum.

\section{Metric-like fields from Chern-Simons}\label{app:fields}

Higher-spin gauge theories can be alternatively described in a frame-like language, where the symmetric tensors used in this paper are substituted by two differential forms that generalize the dreibein and the spin connection of the frame formulation of Einstein gravity \cite{frame1,frame2}. In three spacetime dimensions one can consider the fields
\be \label{frame-fields}
e = e_\m{}^A\, T_A\, dx^\m \, , \qquad \o = \o_\m{}^A\, T_A\, dx^\m \, ,
\ee
(where $T_A$ collects the generators of a suitable gauge algebra) and the action
\be \label{CS}
I = \frac{1}{16\p G} \int_{\cM_3}\! \tr \left( e \ww R + \frac{1}{3\ell^2}\, e \ww e \ww e \right) \qquad \mathrm{with} \quad R = d\o + \o \ww \o \, .
\ee
In sects.~\ref{sec:spin3} and \ref{sec:spin4} we discussed e.g.\ the metric counterparts of the models based on the algebras $sl(3,\mathbb{R})$ and $sl(4,\mathbb{R})$.\footnote{One should also specify the embedding of the Lorentz $so(1,2) \sim sl(2,\mathbb{R})$ subalgebra in the full gauge algebra. In this paper we only deal with the \emph{principal embedding}.} These are examples of a more general setup where one deals with the infinite-dimensional gauge algebra $hs[\l]$, which can be considered as a continuation of $sl(N)$ under $N \to \l$ (see e.g.\ \cite{Wlambda,hslambda} and references therein). For generic values of $\l$ the action \eqref{CS} describes fields with spin ranging from $2$ to $\infty$, while for $\l = N$ the trace becomes degenerate and the theory describes fields with spin ranging from $2$ to $N$. For $\ell > 0$ one can also rewrite \eqref{CS} as the difference of two $hs[\l]$ (or $sl(N)$) Chern-Simons actions \cite{Blencowe1,Blencowe2}. 

A map between the fields \eqref{frame-fields} and their metric-like peers has been proposed in \cite{spin3,Wlambda}: for the fields considered in sects.~\ref{sec:spin3} and \ref{sec:spin4} it reads e.g.
\begin{subequations} \label{metric-like_ansatz}
\begin{align}
g & = 2\, \tr(e_\m e_\n) dx^\m dx^\n \, , \\[6pt]
\phi & = C_1\, \tr(e_\m e_\n e_\r) dx^\m dx^\n dx^\r \, , \label{metric-like_3} \\
\vf & = C_2 \left\{ \tr(e_\m e_\n e_\r e_\s) - \frac{3\l^2-7}{10}\, \tr(e_\m e_\n) \tr(e_\r e_\s) \right\} dx^\m dx^\n dx^\r dx^\s \, , \label{metric-like_4} 
\end{align}
\end{subequations}
where the trace is normalised such that on the Lorentz $sl(2,\mathbb{R})$ subalgebra it corresponds to the matrix trace in the representation of dimension two.\footnote{Our normalization agrees with e.g.\ \cite{spin3,GH,Wlambda,hslambda} but, for $\l = N$, our trace \emph{does not} agree with the matrix trace in the fundamental of $sl(N,\mathbb{R})$.} Overall normalizations depend on the normalization of kinetic terms in the metric-like action and are discussed below. One has to trace over fiber indices because the action \eqref{CS} is invariant under the transformations
\be \label{lorentz-like}
\d e = [\, e , \L \,] \, , \qquad \d \o = d \L + [\, \o , \L \,] \, , 
\ee
which generalize Lorentz transformations and have no counterparts in the metric formulation (see \cite{spin3,Wlambda,metric-like} for details). The relative coefficients between multiple traces are instead not fixed by first principles, and indeed one can shift them with field redefinitions. The relative coefficient in \eqref{metric-like_4} has been however fixed in \cite{Wlambda} by requiring that the spin-4 field vanishes when the connections \eqref{frame-fields} take their vacuum value.  

In this Appendix we first recall how one can extract from \eqref{metric-like_ansatz} exact solutions of the metric-like models which we discuss in Sections~\ref{sec:spin3} and \ref{sec:spin4}. Then we show how one can build solutions which fit into the general discussion of Section~\ref{sec:conclusions} by fixing the relative coefficients between multiple traces. Let us stress that with this procedure one fully controls the space of solutions of the equations of motion in three dimensions.  This is a crucial ingredient in the AdS/CFT correspondence, where one aims at computing the on-shell action, but in arbitrary spacetime dimensions only solutions of the free Fronsdal equations have been studied in detail (see e.g.\ \cite{JM_AdS-CFT}).

\subsection{Spin-3 and spin-4 fields}

The boundary conditions displayed in the main body of the paper are the metric counterpart of the ``highest-weight'' boundary conditions in the Chern-Simons formulation \cite{HR,spin3}. There one defines the gauge connections $A = \o + e$ and $\tilde{A} = \o - e$, and imposes the following boundary conditions:
\be \label{pre-bnd_CS}
A = b^{-1} a_i\,  b\, dx^i + b^{-1} d b \, , \qquad 
\tilde{A} = -\, b\, \tilde{a}_i\,  b^{-1} dx^i + b d b^{-1} \, ,
\ee
where $b = e^{\log r\, W^1_0}$ while the $hs[\l]$-valued connections $a$ and $\tilde{a}$ read
\begin{subequations} \label{bnd_CS}
\begin{align}
a & = \left( W^1_1 - \frac{2\p}{k} \cL(x^+) W^1_{-1} + \frac{2\p}{kN_3} \cW(x^+) W^2_{-2} + \frac{2\p}{kN_4}\, \cU(x^+) W^3_{-3} + \cdots \right) dx^+ \, , \\
\tilde{a} & = \left( W^1_{-1} - \frac{2\p}{k} \tilde{\cL}(x^-) W^1_{1} + \frac{2\p}{kN_3} \tilde{\cW}(x^-) W^2_{2} + \frac{2\p}{kN_4}\, \tilde{\cU}(x^-) W^3_{3} + \cdots \right) dx^- \, .
\end{align}
\end{subequations}
The $W^l_m$ (with $l \geq 1$ and $-l \leq m \leq l$) form a basis of $hs[\l]$ such that
\begin{subequations}
\begin{align}
[\, W^1_m , W^l_n \,] & = (lm-n) W^l_{m+n} \, , \\[5pt]
\tr(W^k_m W^l_n) & = (-1)^{l-m} N_{l+1}(\l) \frac{(l+m)!(l-m)!}{(2l)!}\, \d^{k,l} \d_{0,m+n} \, ,
\end{align}
\end{subequations}
where the coefficients $N_{l+1}$ are defined as
\be
N_{l+1}(\l) = - \frac{6\,(l !)^{2}}{(2l +1)!}\, \prod_{i\,=\,2}^l\, (i-\l)(i+\l) \, , \label{N_l} 
\ee
so that $\tr(W^1_{-1}W^1_{1}) = N_2(\l) = -1$.\footnote{For $\l = N$ the trace in the fundamental of $sl(N,\mathbb{R})$ would instead give $\mathrm{Tr}_{N \times N}(W^1_{-1}W^1_{1}) = - \frac{N(N^2-1)}{6}$.}
The functions $\cL$, $\cW$, $\cU$ have to be identified with the left-moving components of the currents $\cL_{ij}$, $\cW_{ijk}$, $\cU_{ijkl}$ and normalisations are fixed as in \cite{GH}. Note that \eqref{pre-bnd_CS} and \eqref{bnd_CS} define flat connections: the boundary conditions of \cite{HR,spin3} therefore put the system on shell (at least asymptotically).

If one fixes $\l = 3$, then one can ignore in \eqref{bnd_CS} all $W^l_m$ with $l \geq 3$. Substituting the field $e$ defined as above in \eqref{metric-like_ansatz}, one obtains that the spin-4 field vanishes, while the metric and the spin-3 field solve the equations of motion derived from the action \eqref{action3} (with free coefficients fixed as in \eqref{numbers}) provided that\footnote{Although for $\l=3$ we have $N_3 = 1$, we display explicitly the factor $N_3$ in both \eqref{C1} and \eqref{sol3CS} to stress that, in general, the precise on-shell expression for the fields obtained from \eqref{metric-like_ansatz} depends on $\l$.} 
\be \label{C1}
C_1 = \frac{2}{3\sqrt{|N_3|}} = \frac{2}{3} \, .
\ee 
Taking advantage of the relations \eqref{L_comp} and \eqref{W_comp} between $\cL$, $\cW$ and the boundary currents $\cL_{ij}$, $\cW_{ijk}$, the resulting metric and spin-3 field read \cite{spin3}
\begin{subequations} \label{sol3CS}
\begin{align}
g & = \frac{dr^2}{r^2} + \left\{ \frac{r^2}{2}\, \h_{ij} + \frac{2\p}{k}\, \cL_{ij} + \frac{2\p^2}{k^2 r^2}\, \cL_{ik}\cL_j{}^k + \frac{2\p^2}{k^2 N_3\, r^4}\, \cW_{ikl} \cW_j{}^{kl} \right\} dx^i dx^j \, , \\
\phi & = \frac{3\p C_1}{4k} \left\{ \cW_{ijk} + \frac{4\p}{kr^2}\, \cL_{im} \cW_{jk}{}^m + \frac{4\p^2}{k^2 r^4}\, \cL_{im} \cL_{jn} \cW_k{}^{mn} \right\} dx^i dx^j dx^k \, .
\end{align}
\end{subequations}
Their contravariant correlatives satisfy our boundary conditions \eqref{bnd3} and \eqref{bnd23} with $h^{rr} = h^{ri} = 0$.

We can now repeat the same steps for $\l = 4$: the metric and the spin-3 field receive corrections in $\cU_{ijkl}$ at the orders, respectively, $\cO(r^{-6})$ and $\cO(r^{-4})$. In analogy with \eqref{sol3CS}, the spin-4 field satisfies
\be
\vf_{ijkl} = \frac{\p C_2}{2k}\, \cU_{ijkl} + \cO(r^{-2}) \, ,
\ee
and solves the equations of motion derived from the action \eqref{I4} provided that 
\be \label{C2}
C_1 = \frac{2}{3\sqrt{|N_3|}} = \frac{\sqrt{5}}{3\sqrt{3}} \, , \qquad C_2 = \frac{1}{\sqrt{3|N_4|}} = \frac{\sqrt{5}}{3\sqrt{6}} \, ,
\ee 
and one appropriately tunes the free coefficients. Two complications emerge however starting from this example. The first is that not all radial components vanish:
\be
\vf_{rrij} = \frac{\p^2 C_2 N_4}{15 k^2 N_3^2\,r^6}\, \cW_{ikl} \cW_j{}^{kl} + \cO(r^{-8}) \, .
\ee
Moreover the rank-4 tensor defined by \eqref{metric-like_4} is \emph{not} doubly traceless, as one can verify using the explicit on-shell expressions for the metric and $\vf$. For $\l = 4$ it satisfies instead the constraint
\be
\begin{split}
g^{\m\n} g^{\r\s} \vf_{\m\n\r\s} & = \frac{5C_2}{18C_1^2}\, g^{\m\n} g^{\r\s} g^{\a\b} \left( 2\, \phi_{\m\r\a}\phi_{\n\s\b} + 3\, \phi_{\m\r\s}\phi_{\n\a\b} \right) \\
& - \frac{5}{27C_2}\, g^{\m\n} g^{\r\s} g^{\a\b} g^{\g\d} \left( \vf_{\m\r\a\g} \vf_{\n\s\b\d} + 3\, \vf_{\m\r\a\b} \vf_{\n\s\g\d} \right) + \cdots \, ,
\end{split}
\ee
where omitted terms are at least cubic in the fields or contain double traces of $\vf$.\footnote{In general the coefficients depend on $\l$ and the omitted terms also involve fields of higher spin.} Changing the relative coefficient in \eqref{metric-like_4} does not help, and even worsen the fall-off of the double trace for $r \to \infty$. Therefore we cannot compare directly \eqref{metric-like_4} with our boundary conditions, which have been given for tensors satisfying $g^{\m\n} g^{\r\s} \vf_{\m\n\r\s} = 0$. One can nevertheless easily recover a doubly traceless field. It suffices to introduce a projector by a field redefinition:
\be \label{proj_spin4}
\vf_{\m\n\r\s} \rightarrow \left( \d^\a_\m \d^\b_\n \d^\g_\r \d^\d_\s - \frac{1}{5}\, g_{(\m\n}g_{\r\s)} g^{\a\b} g^{\g\d} \right) \vf_{\a\b\g\d} \, .
\ee
The price to pay is that the result does not have a finite expansion in powers of $r^{-1}$ like the metric, the spin-3 field and the spin-4 field defined by \eqref{metric-like_4}.

We can now compare the exact solution built from \eqref{bnd_CS} with our boundary conditions. The inverse metric satisfies $g^{rr} = r^2$, $g^{ri} = 0$, while $g^{ij}$ has the same form as in \eqref{g2^ij}. The spin-3 field satisfies $\phi^{rrr} = \phi^{rri} = \phi^{rij} = 0$, while $\phi^{ijk}$ has the same form as in \eqref{phi2^ijk}. The doubly-traceless spin-4 field \eqref{proj_spin4} satisfies instead $\vf^{rrri} = \vf^{rijk} = 0$, while
\be \label{radial4}
\vf^{rrrr} = \frac{4\p^2 C_2 N_4}{5k^2 N_3^2\, r^2}\, \cW_{ijk} \cW^{ijk} + \cO(r^{-4}) \, , \quad 
\vf^{rrij} = \frac{4\p^2 C_2 N_4}{3k^2 N_3^2\, r^6}\, \cW^i{}_{kl} \cW^{jkl} + \cO(r^{-8}) \, , 
\ee
and $\vf^{ijkl}$ has the same form as in \eqref{vf^ijkl}. Due to \eqref{radial4}, the rank-4 tensor still does not have the desired form, but one can eliminate the unwanted terms in $\vf^{rrrr}$ and $\vf^{rrij}$ by performing an additional field redefinition of the type discussed in \eqref{redef4->33}:
\be \label{redef_fin}
\vf^{\m\n\r\s}\! \to \frac{8 C_2 N_4}{45 C_1^2 N_3^2}\! \left\{ 10\, \phi_\a{}^{(\m\n} \phi^{\r\s)\a} - g^{(\m\n|}\! \left[ \phi_{\a\b}{}^{|\r} \phi^{\s)\a\b} + g^{|\r\s)}\! \left( \phi_{\a\b\g} \phi^{\a\b\g} + 6\, \phi_\a \phi^\a \right) \right] \!\right\}\! .
\ee
All field redefinitions in \eqref{redef4->33} preserve the double trace constraint by construction, so that the resulting \mbox{rank-4} tensor eventually fits into our boundary conditions \eqref{bnd4}.\footnote{It would be interesting to understand if \mbox{-- by} fixing appropriately the free coefficients in the \mbox{Lagrangian --} one can find an exact solution of the equations of motion where all components with radial indices vanish as in \eqref{sol3CS}, although the issue goes beyond the scope of this paper.}

Note once again that, even if field redefinitions cannot influence the physics, they do influence the boundary conditions and the presentation of asymptotic symmetries. They can thus hide or manifest possible geometric structures. This is not a surprise: if one expands the Einstein-Hilbert Lagrangian around a given background one obtains a specific non-polynomial action. Redefining the fluctuations one can modify its form, but this generically obscures the relation with the Ricci scalar.  

\subsection{Fields of spin $s > 4$}

In the introductory remarks of this Appendix we recalled that the relative coefficients in the map between frame and metric-like fields are not fixed a priori. It could be anyway useful to identify a ``canonical'' map, like the one that we already encountered in the definition of the spin-4 field in  \eqref{metric-like_4}. The vanishing of all fields but the metric on the vacuum is a desirable property that however does not suffice to fix all relative coefficients for $s > 4$. For instance, the term $\tr(e^2)\tr(e^3)$ that appears in the most general ansatz for a spin-5 field vanishes identically when $e$ takes its background value. Nevertheless, for arbitrary $s$, one can fix completely the ansatz by requiring that, if one starts from a ``highest-weight'' connection in the Chern-Simons theory, one obtains
\be \label{asympspins}
\vf_s \sim \cW_{i_1 \cdots\, i_s} dx^{i_1} \cdots dx^{i_s} + \cO(r^{-2}) \, .
\ee
This is the fall-off which fits into the boundary conditions that we discuss in Section~\ref{sec:conclusions}! Before showing that matching \eqref{asympspins} fully fixes the ansatz, let us recall that the freedom in the relative coefficients between multiple traces does not parameterise all possible field redefinitions, but only those which do not contain the inverse metric. Some of the latter play an important role in this paper, since they are required to match our complete boundary conditions as in \eqref{proj_spin4} and \eqref{redef_fin}. These redefinitions, however, only affect \eqref{asympspins} at subleading orders and thus do not affect the following discussion.

Let us consider the spin-5 example to begin with. Suppose for simplicity that only left-moving components are switched on: then \eqref{pre-bnd_CS} and \eqref{bnd_CS} imply that
\be
\tr(e^5) = \left\{ a_1\, \cZ\, \tr\left((W^1_1)^4W^4_{-4}\right) + a_2\, \cL\,\cW\,  \tr\left((W^1_1)^3W^1_{-1}W^2_{-2}\right) \right\} (dx^+)^5 + \cO(r^{-2}) \, ,
\ee
where we denoted by $\cZ$ the spin-5 charge. The only way to fulfill the condition \eqref{asympspins} is to cancel the term with $\cL\cW$ by properly combining $\tr(e^5)$ with $\tr(e^2)\tr(e^3)$ as in
\be
\vf_5 \sim \tr(e^5) - \frac{5(3\l^2-13)}{21}\, \tr(e^2)\tr(e^3) \, . 
\ee 
In general one has an equal number of unwanted combinations of the generators of $hs[\l]$ that appear in the highest-weight connection and multiple traces in the ansatz for the fields. For instance, for $s=6$ one obtains contributions of order greater or equal to $\cO(1)$ from $\tr((W^1_1)^5W^5_{-5})$, $\tr((W^1_1)^4W^1_{-1}W^3_{-3})$, $\tr((W^1_1)^4(W^2_{-2})^2)$ and $\tr((W^1_1)^3(W^1_{-1})^3)$, while the ansatz for the field, besides $\tr(e^6)$, contains also $\tr(e^2)\tr(e^4)$, $\tr(e^3)^2$ and $\tr(e^2)^3$. Compatibility with the asymptotic expansion \eqref{asympspins} fixes the relative coefficients as\footnote{The factor $\l^2-4$ in the denominator does not signal any pathology of the theory for $\l = 2$, since it is cancelled by an identical factor coming from the traces that multiply it ($\tr(e^3)$ vanishes in pure gravity, i.e.\ it is proportional to $\l^2-4$). For $\l = 2$ the final outcome is actually $\vf_6 = 0$ for any $e$.}
\be
\begin{split}
\vf_6 \sim\ & \tr(e^6) -\frac{5(5\l^4-65\l^2+264)}{63(\l^2-4)}\, \tr(e^3)^2 - \frac{5(\l^2-7)}{6}\, \tr(e^2)\tr(e^4) \nn \\
& + \frac{6\l^4-71\l^2+125}{42}\, \tr(e^2)^3 \, .
\end{split}
\ee
Using e.g.\ the $\star$-product realization of the trace of $hs[\l]$ first introduced in \cite{sphere}, one can easily continue along these lines. The rank-7 symmetric tensor which complies with the asymptotic expansion \eqref{asympspins} is e.g.
\be
\begin{split}
\vf_7 \sim\ & \tr(e^7) - \frac{35(5\l^4-95\l^2+636)}{198(\l^2-4)}\, \tr(e^3)\tr(e^4) - \frac{7(3 \l^2 - 31)}{22}\, \tr(e^2)\tr(e^5) \nn \\
& + \frac{35(3\l^6 -71 \l^4 + 488 \l^2 -840)}{198(\l^2-4)}\, \tr(e^2)^2\tr(e^3) \, .
\end{split}
\ee
As a side remark, note that with the same procedure one can express the charges $\cL$, $\cW$, etc.\ in terms of traces of powers of the connection $a$ defined in \eqref{bnd_CS}. This is a useful way to compute the $\cW$-charges starting from other gauges, that has been exploited in the study of smooth solutions in the Chern-Simons formulation \cite{conical1,hslambda,conical2}, although to our knowledge explicit expressions for the $\cW$-charges were given only up to spin 4.

\section{3--3--4 cubic vertex}\label{app:vertex1}

Thanks to the vanishing of the Weyl tensor, the higher-spin gauge transformations of the metric contain a single derivative (see e.g.\ \eqref{dg}). As a result, the interacting vertices needed to restore the gauge invariance lost after covariantisation by the quadratic actions \eqref{L3quad} and \eqref{L4quad} do not contain more than two derivatives as in $D > 3$. Furthermore, the frame-like action \eqref{CS} is of first order, and the generalized spin connection can be expressed in terms of the generalized vielbein and its first derivative through its equation of motion \cite{metric-like,metric-like2}. For these reasons in the present and in the following Appendix we only consider interacting vertices with at most two derivatives.

Efficient techniques to classify and build cubic vertices for higher-spin particles have been developed over the last few years. Since three spacetime dimensions are blessed by the absence of higher derivatives, we follow instead a very pragmatic approach: we display the ugly but still controllable general ansatz and the values of the coefficients in the action and in the gauge transformations which guarantee gauge invariance. Computations have been performed using xAct packages for Mathematica \cite{xAct}, and in particular the package xTras \cite{xtras}.

\subsection{Action}\label{app:action3}

One cannot build vertices with two derivatives and an odd number of tensors of odd rank, while one can build a vertex with two rank-3 tensors and one rank-4 tensor (which we assume to have vanishing double trace). The general ansatz can be conveniently decomposed as
\be \label{ansatz334}
\cL_{3-3-4} = \vf^{\m\n\r\s} \left( J_1 + J_2 + J_3 + J_4 \right)_{\m\n\r\s} \, ,
\ee
where we do not allow derivatives on the rank-4 tensor in order to eliminate the redundancies induced by integrations by parts. The quadratic currents are defined as follows: $J_4$ contains the terms that one can set to zero in three spacetime dimensions thanks to the identities which follow from the vanishing of antisymmetrizations over more than three indices.\footnote{A systematic way to construct all identities satisfied by a set of tensors in a given dimension is described e.g.\ in \cite{xtras}.} $J_2$ and $J_3$ collect the terms that can be independently shifted by field redefinitions, respectively, of the higher-rank tensors and of the metric. All coefficients in $J_2$ and $J_3$ are therefore free, in analogy with the $k_i$ which appear e.g.\ in the quadratic spin-3 Lagrangian \eqref{L3quad}. $J_1$ contains instead the non-trivial part of the vertex, which is fixed up to an overall coupling constant if one imposes that the action \eqref{I4} be gauge invariant up to quadratic order in the higher-spin fields.

Before displaying explicitly the ansatz \eqref{ansatz334}, let us stress that one can fix all coefficients in $J_1$ by asking for gauge invariance on an $AdS$ background. The terms in the Ricci tensor that one has to add to restore gauge invariance on an arbitrary background can always be absorbed by a field redefinition of the metric. As we have discussed in the main body of the paper, this choice is however not necessarily the best one to compute asymptotic symmetries. In the following we will thus work with generic $\cL_2$ and $\cL_3$.

The portion of the vertex which is non trivial in our parameterisation is
\begin{align}
& (J_1)_{\m\n\r\s} = A_1 \nabla_{\!\a} \phi_{\b\m\n} \nabla^\b \phi^\a{}_{\r\s} + A_2 \nabla_{\!\m} \phi_{\n\a\b} \nabla^\a \phi^\b{}_{\r\s} + A_3 \nabla_{\!\m} \phi_{\n\a\b} \nabla_{\!\r} \phi_\s{}^{\a\b} + B_1 \nabla\cdot\phi_{\m\n} \nabla\cdot\phi_{\r\s} \nn \\[3pt]
& + B_2 \nabla_{\!\m} \phi_{\n\r}{}^\a \nabla\cdot\phi_{\a\s} + B_3 \nabla_{\!\a} \phi_{\m\n\r} \nabla_{\!\s} \phi^\a + \nabla_{\!\m} \phi_{\n\r}{}^\a \big( B_4 \nabla_{\!\s} \phi_\a + B_5 \nabla_{\!\a} \phi_\s \big) + B_6 \nabla_{\!\m} \phi_\n \nabla_{\!\r} \phi_\s \nn \\[3pt]
& + B_7 \nabla\cdot \phi_{\m\n} \nabla_{\!\r} \phi_\s + B_8 \nabla_{\!\m} \phi_{\n\r\s} \nabla\cdot\phi + \phi_{\m}{}^{\a\b} \big( A_4 \nabla_{\!\n} \nabla_{\!\r} \phi_{\s\a\b} + A_5 \nabla_{\!(\n} \nabla_{\!\a)} \phi_{\b\r\s} \,\big) \nn \\[3pt]
& + \phi_{\m\n\r} \big( B_{9} \nabla\!\cdot\!\nabla\!\cdot \phi_{\s} + B_{10} \nabla_{\!(\s} \nabla_{\!\a)} \phi^\a \big) + \phi_{\m\n}{}^\a \big( B_{11} \nabla_{\!(\a} \nabla_{\!\b)} \phi_{\r\s}{}^\b + B_{12} \nabla_{\!(\r} \nabla_{\!\b)} \phi_{\s\a}{}^\b \nn \\[3pt]
& + B_{13} \nabla_{\!(\a} \nabla_{\!\r)} \phi_\s + B_{14} \nabla_{\!\r} \nabla_{\!\s} \phi_\a \big) + \phi_\m \big( B_{15} \nabla_{\!(\n} \nabla_{\!\a)} \phi_{\r\s}{}^\a \nn + B_{16} \nabla_{\!\n} \nabla_{\!\r} \phi_{\s} \big) + \phi^\a \big( B_{17} \nabla_{\!(\a} \nabla_{\!\m)} \phi_{\n\r\s} \nn \\
& + B_{18} \nabla_{\!\m} \nabla_{\!\n} \phi_{\r\s\a} \big) + g_{\m\n} \Big\{ C_1 \nabla_{\!\a} \phi_{\b\g\r} \nabla^{\b} \phi_{\s}{}^{\a\g} + C_2 \nabla_{\!\r} \phi_{\a\b\g} \nabla^{\a} \phi_{\s}{}^{\b\g} + C_3 \nabla_{\!\r} \phi_{\a\b\g} \nabla_{\!\s} \phi^{\a\b\g} \nn \\
& + \nabla\cdot \phi^{\a\b} \big( C_4 \nabla_{\!\a} \phi_{\b\r\s} + C_5 \nabla_{\!\r} \phi_{\s\a\b} \big) + C_6 \nabla\cdot \phi_{\r\s} \nabla\cdot \phi + \big( C_7 \nabla\cdot \phi_\r{}^\a + C_8 \nabla^\a \phi_\r \big) \nabla\cdot\phi_{\a\s} \nn \\[3pt]
& + C_{9} \nabla_{\!\b} \phi_\a \nabla^\a \phi_{\r\s}{}^\b + C_{10} \nabla_{\!\r} \phi_{\s\a\b} \nabla^\a \phi^\b \nn + \nabla_{\!\r} \phi^\a\big( C_{11} \nabla\cdot \phi_{\a\s} + C_{12} \nabla_\a \phi_\s \big) + C_{13} \nabla_{\!\r} \phi_\s \nabla\cdot\phi \nn \\[3pt]
& + \phi_{\r\s}{}^\a \big( C_{14} \nabla_{\!(\a} \nabla_{\!\b)} \phi^\b  + C_{15} \nabla\cdot \nabla\cdot \phi_\a \big) + \phi_\r{}^{\a\b} \big( C_{16} \nabla_{\!(\s} \nabla_{\!\g)} \phi_{\a\b}{}^\g + C_{17} \nabla_{\!(\a} \nabla_{\!\g)} \phi_{\s\b}{}^\g \nn \\[3pt]
& + C_{18} \nabla_{\!(\s} \nabla_{\!\a)} \phi_{\b} + C_{19} \nabla_{\!\a} \nabla_{\!\b} \phi_\s \big) + \phi^{\a\b\g} \big( C_{20} \nabla_{\!(\r} \nabla_{\!\a)} \phi_{\s\b\g} + C_{21} \nabla_{\!\a} \nabla_{\!\b} \phi_{\g\r\s} \big) \nn \\[3pt]
& + \phi_\r \big( C_{22} \nabla_{\!(\s} \nabla_{\!\a)} \phi^\a + C_{23} \nabla\cdot \nabla\cdot \phi_\s \big) + \phi^\a \big( C_{24} \nabla_{\!(\a} \nabla_{\!\b)} \phi_{\r\s}{}^\b + C_{25} \nabla_{\!(\r} \nabla_{\!\b)} \phi_{\s\a}{}^\b \nn \\
& + C_{26} \nabla_{\!(\r} \nabla_{\!\a)} \phi_\s + C_{27}\, \phi_\a \nabla_{\!\r} \nabla_{\!\s} \phi^\a \big) \Big\} \nn \\
& + \ell^{-2} \Big\{ D_1\, \phi_{\m\n\a} \phi_{\r\s}{}^\a + D_2\, \phi_{\m\n\r} \phi_\s + g_{\m\n} \big( D_3\, \phi_{\r\a\b} \phi_{\s}{}^{\a\b} + D_4\, \phi_{\r\s\a} \phi^\a + D_5\, \phi_\r \phi_\s \big) \Big\} \, ,
\end{align}
where we labelled with $A_i$ the terms that would appear also in the traceless and transverse gauge reviewed  in \cite{review_noether}. A generic field redefinition of the form $\phi \to \phi\vf$ contains 7 terms,
\be \label{redef3->34}
\begin{split}
\phi^{\m\n\r} \to &\, f_1\, \phi_{\a\b}{}^{(\m} \vf^{\n\r)\a\b} + f_2\,\phi_{\a}{}^{(\m\n} \vf^{\r)\a} + f_3\, \phi_\a \vf^{\a\m\n\r} + f_4\, \phi^{(\m} \vf^{\n\r)} \\[3pt]
& + g^{(\m\n|}\! \left\{ f_5\, \phi_{\a\b\g} \vf^{|\r)\a\b\g} + f_6\, \phi^{|\r)\a\b} \vf_{\a\b} + f_7\, \phi_{\a} \vf^{|\r)\a} \right\} ,
\end{split}
\ee 
while a generic field redefinition of the form $\vf \to \phi^2$ contains again 7 terms, but only 5 independent coefficients if one wants to preserve the double trace constraint:
\begin{align}
\vf^{\m\n\r\s} \to & \, f_8\, \phi_\a{}^{(\m\n} \phi^{\r\s)\a} + f_9\, \phi^{(\m} \phi^{\n\r\s)} + g^{(\m\n|}\! \left\{ f_{10}\, \phi_{\a\b}{}^{|\r} \phi^{\s)\a\b} + f_{11}\, \phi^{|\r\s)\a} \phi_\a + f_{12}\, \phi^{|\r} \phi^{\s)} \right\} \nn \\[3pt]
& - \frac{1}{15}\, g^{(\m\n} g^{\r\s)}\! \left\{ (2f_8 + 5 f_{10})\, \phi_{\a\b\g} \phi^{\a\b\g} + (f_8 + 3f_9 + 5(f_{11}+f_{12}))\, \phi_\a \phi^\a \right\} . \label{redef4->33}
\end{align}
Therefore $J_2$ must contain 12 terms which, following  \cite{review_noether}, we choose as
\be \label{J2_3}
\begin{split}
& (J_2)_{\m\n\r\s} = r_1 \nabla_{\!\a} \phi_{\b\m\n} \nabla^\a \phi^\b{}_{\r\s} + r_2 \nabla_{\!\a} \phi_{\m\n\r} \nabla^\a \phi_\s + r_3\, \phi_{\m\n\r} \Box \phi_\s + r_4\, \phi_{\m\n}{}^\a \Box \phi_{\a\r\s} \\[3pt]
& + r_5\, \phi_\m \Box \phi_{\n\r\s} + g_{\m\n} \big( r_6 \nabla_{\!\a} \phi_{\b\g\r} \nabla^{\a} \phi_{\s}{}^{\b\g} + r_7 \nabla_{\!\a} \phi_\b \nabla^\a \phi_{\r\s}{}^\b + r_8 \nabla_{\!\a} \phi_\r \nabla^\a \phi_\s  \\[3pt]
& + r_9\, \phi_\r{}^{\a\b} \Box \phi_{\a\b\s} + r_{10}\, \phi_{\r\s}{}^\a \Box \phi_\a + r_{11}\, \phi^\a \Box \phi_{\a\r\s} + r_{12}\, \phi_\r \Box \phi_\s \big) \, .
\end{split}
\ee
There are instead 22 field redefinitions of the metric that affect the vertex, but only 20 of them are independent. Correspondingly $J_3$ contains all terms with the Ricci tensors but two:
\begin{align}
& (J_3)_{\m\n\r\s} = R_{\a\b} \big( t_1\, \phi_{\m\n}{}^\a \phi_{\r\s}{}^\b + t_2\, \phi_{\m\n\r} \phi_\s{}^{\a\b} \big) + R_{\m\a} \big( t_3\, \phi_{\n\r\b} \phi_\s{}^{\a\b} + t_4\, \phi_{\n\r\s} \phi^\a + t_5\, \phi^\a{}_{\n\r} \phi_\s \big) \nn \\[3pt]
& + R_{\m\n} \big( t_6\, \phi_{\r\a\b} \phi_\s{}^{\a\b} + t_7\, \phi_{\r\s\a} \phi^\a \big) + R\, \big( t_8\, \phi_{\m\n\a} \phi_{\r\s}{}^\a + t_9\, \phi_{\m\n\r} \phi_\s \big) \label{J3_3} \\
& + g_{\m\n} \Big\{ R_{\a\b} \big( t_{10}\, \phi_{\r\g}{}^{\a} \phi_{\s}{}^{\b\g} + t_{11}\, \phi_{\r\s\g} \phi^{\a\b\g} + t_{12}\, \phi_{\r\s}{}^\a \phi^\b + t_{13}\, \phi_\r{}^{\a\b} \phi_\s \big) + t_{14}\, R_{\r\s} \phi_{\a\b\g} \phi^{\a\b\g} \nn \\
& + R_{\r\a} \big( t_{15}\,\phi_{\s\b\g} \phi^{\a\b\g} + t_{16}\, \phi_{\s\b}{}^{\a} \phi^\b + t_{17}\, \phi_\r \phi^\a \big) + R \big( t_{18}\, \phi_{\r\a\b} \phi_\s{}^{\a\b} + t_{19}\, \phi_{\r\s\a} \phi^\a + t_{20}\, \phi_\r \phi_\s \big) \Big\} . \nn
\end{align}
Finally, there are 6 independent identities that involve two covariant derivatives and two tensors of rank-3. One can thus eliminate from the general ansatz the following terms: 
\be \label{J4_3}
\begin{split}
& (J_4)_{\m\n\r\s} = z_1 \nabla_{\!\a} \phi_{\m\n\r} \nabla\cdot\phi_{\s}{}^\a + z_2\, \phi_\m{}^{\a\b} \nabla_{\!\a} \nabla_{\!\b} \phi_{\n\r\s} + g_{\m\n} \big( z_3 \nabla_{\!\r} \phi_\a \nabla_{\!\s} \phi^\a \\[3pt]
& + z_4\, \phi^{\a\b\g} \nabla_{\!\r} \nabla_{\!\s} \phi_{\a\b\g} \big) + z_5\, R_{\m\n} \phi_\r \phi_\s + z_6\, g_{\m\n} R_{\r\s} \phi_\a \phi^\a \, .
\end{split}
\ee
%

\subsection{Gauge transformations}\label{app:gauge3}

Adding the vertex \eqref{ansatz334} to the quadratic Lagrangians \eqref{L3quad} and \eqref{L4quad} induces the following deformations of the gauge transformations, that are necessary to preserve the gauge invariance of the action up to quadratic order in the higher-spin fields. The spin-3 transformation of the rank-3 tensor receives the correction
\be \label{dx-phi}
\begin{split}
\d^{(1)}_3 \phi^{\m\n\r} & = a_1\, \vf_\a{}^{\m\n\r} \nabla\!\cdot \x^\a + a_2\, \vf_{\a\b}{}^{(\m\n} \nabla^{\r)} \x^{\a\b} + a_3\, \vf_{\a}{}^{\b(\m\n} \nabla_{\!\b\,} \x^{\r)\a} + a_4\, \vf^{(\m\n} \nabla\!\cdot \x^{\r)} \\[3pt]
& + a_5\, \vf_\a{}^{(\m} \nabla^\n \x^{\r)\a} + a_6\, \vf^{\a(\m} \nabla_{\!\a\,} \x^{\n\r)} + b_1\, \x^{\a\b} \nabla_{\!\a\,} \vf_\b{}^{\m\n\r} + b_2\, \x^{\a\b} \nabla^{(\m} \vf^{\n\r)}{}_{\a\b} \\[3pt]
& + b_3\, \x^{\a(\m} \nabla\!\cdot \vf^{\n\r)}{}_\a + b_4\, \x^{\a(\m} \nabla_{\!\a\,} \vf^{\n\r)} + b_5\, \x^{\a(\m} \nabla^\n \vf^{\r)}{}_\a + b_6\, \x^{(\m\n} \nabla\!\cdot \vf^{\r)} \\
& + g^{(\m\n|} \Big\{ c_1\, \vf^{|\r)}{}_{\a\b\g} \nabla^\a \x^{\b\g} + c_2\, \vf^{|\r)}{}_\a \nabla\!\cdot \x^\a + c_3\, \vf_{\a\b} \nabla^{|\r)} \x^{\a\b} + c_4\, \vf_{\a\b} \nabla^\a \x^{|\r)\b} \\
& + d_1\, \x^{\a\b} \nabla\!\cdot \vf^{|\r)}{}_{\a\b} + d_2\, \x^{\a\b} \nabla_{\!\a\,} \vf^{|\r)}{}_{\b} + d_3\, \x^{\a\b} \nabla^{|\r)} \vf_{\a\b} + d_4\, \x^{|\r)\a} \nabla^\b \vf_{\a\b} \Big\} \, ,
\end{split}
\ee
and the field also acquires a spin-4 gauge transformation:
\be \label{dk-phi}
\begin{split}
\d^{(1)}_4 \phi^{\m\n\r} & = p_1\, \phi_\a{}^{(\m\n} \nabla\!\cdot \k^{\r)\a} + p_2\, \phi_{\a\b}{}^{(\m} \nabla^\n \k^{\r)\a\b} + p_3\, \phi_{\a}{}^{\b(\m} \nabla_{\!\b\,} \k^{\n\r)\a} + p_4\, \phi^{(\m} \nabla\!\cdot \k^{\n\r)} \\[3pt]
& + p_5\, \phi_\a \nabla^{(\m} \k^{\n\r)\a} + p_6\, \phi^\a \nabla_{\!\a\,} \k^{\m\n\r} + q_1\, \k^{\a\b(\m} \nabla_{\!\a\,} \phi^{\n\r)}{}_\b + q_2\, \k^{\a\b(\m} \nabla^\n \phi^{\r)}{}_{\a\b} \\[3pt]
& + q_3\, \k^{\a(\m\n} \nabla\!\cdot \phi^{\r)}{}_{\a} + q_4\, \k^{\a(\m\n} \nabla_{\!\a\,} \phi^{\r)}{} + q_5\, \k^{\a(\m\n} \nabla^{\r)} \phi_\a + q_6\, \k^{\m\n\r} \nabla\!\cdot \phi \\
& + g^{(\m\n|} \Big\{ v_1\, \phi^{|\r)}{}_{\a\b} \nabla\!\cdot \k^{\a\b} + \phi_{\a\b\g} \big( v_2 \nabla^{|\r)} \k^{\a\b\g} + v_3 \nabla^\a \k^{|\r)\b\g} \big) + v_4\, \phi_\a \nabla\!\cdot \k^{|\r)\a} \\
& + \k^{\a\b\g} \big( w_1 \nabla_{\!\a\,} \phi^{|\r)}{}_{\b\g} + w_2 \nabla^{|\r)} \phi_{\a\b\g} \big) + \k^{|\r)\a\b} \big( w_3 \nabla\!\cdot \phi_{\a\b} + w_4 \nabla_{\!\a\,} \phi_\b \big) \Big\} \, .
\end{split}
\ee
In a similar fashion, the rank-4 tensor acquires a spin-3 gauge transformation:
\be \label{dx-vf}
\begin{split}
\d^{(1)}_3 \vf^{\m\n\r\s} & = \tilde{a}_1\, \phi^{(\m\n\r} \nabla\!\cdot \x^{\s)} + \tilde{a}_2\, \phi_\a{}^{(\m\n} \nabla^\r \x^{\s)\a} + \tilde{a}_3\, \phi^{\a(\m\n} \nabla_{\!\a\,} \x^{\r\s)} + \tilde{a}_4\, \phi^{(\m} \nabla^\n \x^{\r\s)} \\
& + \tilde{b}_1\, \x^{\a(\m} \nabla_{\!\a\,} \phi^{\n\r\s)} + \tilde{b}_2\, \x^{\a(\m} \nabla^\n \phi^{\r\s)}{}_\a + \tilde{b}_3\, \x^{(\m\n} \nabla\!\cdot \phi^{\r\s)} + \tilde{b}_4\, \x^{(\m\n} \nabla^\r \phi^{\s)} \\
& + g^{(\m\n|} \Big\{ \tilde{c}_1\, \phi^{|\r\s)}{}_\a \nabla_{\!\b\,} \x^{\a\b} + \tilde{c}_2\, \phi_{\a\b}{}^{|\r} \nabla^{\s)} \x^{\a\b} + \tilde{c}_3\, \phi_{\a}{}^{\b|\r} \nabla_{\!\b\,} \x^{\s)\a} + \tilde{c}_4\, \phi^{|\r} \nabla\!\cdot \x^{\s)} \\
& + \tilde{c}_5\, \phi_\a \nabla^{|\r} \x^{\s)\a} + \tilde{c}_6\, \phi^\a \nabla_{\!\a\,} \x^{|\r\s)} + \tilde{d}_1\, \x^{\a\b} \nabla_{\!\a\,} \phi_\b{}^{|\r\s)} + \tilde{d}_2\, \x^{\a\b} \nabla^{|\r} \phi^{\s)}{}_{\a\b} \\
& + \tilde{d}_3\, \x^{\a|\r} \nabla\!\cdot \phi^{\s)}{}_\a + \tilde{d}_4\, \x^{\a|\r} \nabla_{\!\a\,} \phi^{\s)} + \tilde{d}_5\, \x^{\a|\r} \nabla^{\s)} \phi_\a + \tilde{d}_6\, \x^{|\r\s)} \nabla\!\cdot \phi \Big\} \\
& + g^{(\m\n} g^{\r\s)} \Big\{ \tilde{e}_1\, \phi_{\a\b\g} \nabla^\a \x^{\b\g} + \tilde{e}_2\, \phi_\a \nabla\!\cdot \x^\a + \tilde{e}_3\, \x^{\a\b} \nabla\!\cdot \phi_{\a\b} + \tilde{e}_4\, \x^{\a\b} \nabla_{\!\a\,} \phi_\b \Big\} \, ,
\end{split}
\ee
where preservation of the double trace constraint imposes
\begin{subequations}
\begin{align}
\tilde{e}_1 & = - \frac{1}{15} \left( 2a_2 + 2a_3 + 5c_2 + 5c_3 \right) , \\
\tilde{e}_2 & = - \frac{1}{15} \left( 3a_1 + a_2 + 2a_4 + 5c_1 + 5c_4 + 5c_5 \right) , \\
\tilde{e}_3 & = - \frac{1}{15} \left( 2b_2 + 2b_3 + 5d_2 + 5d_3 \right) , \\
\tilde{e}_4 & = - \frac{1}{15} \left( 3b_1 + b_2 + 2b_4 + 5d_1 + 5d_4 + 5d_5 \right) .
\end{align}
\end{subequations}
The gauge transformation of the metric is also deformed by terms of the form $\d^{(1)}_3 g = \phi \vf \x$ and $\d^{(1)}_4 g = \phi^2 \k$ that we refrain from displaying explicitly.

\subsection{Coefficients in the action}\label{app:coeff3action}

In the main body of the paper we never need the precise form of the vertex, since we extract all information from the gauge transformations. We display it anyway for completeness and to show that it is unique (up to an overall coupling constant which we denote by $\g$). If one wants a gauge invariant action, the coefficients in $J_1$ must be fixed as follows:
\begingroup
\allowdisplaybreaks
\begin{alignat}{5}
A_1 & = 0 \, , & \quad 
A_2 & = \g - 2r_1 \, , & \quad 
A_3 & = -\, \frac{9\g}{2} - r_6  \, , \nn \\[5pt]
A_4 & = -\,6\g - r_6 \, , & \quad 
A_5 & = 6\g - 2r_1\, , & \quad 
B_1 & = -\,\frac{3\g}{2} \, , \nn \\[5pt]
B_2 & = 11\g -2r_1 \, , & \quad 
B_3 & = \frac{13\g}{4} - \frac{r_2}{2}  \, , & \quad 
B_4 & = -\,r_7 \, , \nn \\[5pt]
B_5 & = -\,\frac{23\g}{4} - \frac{3r_2}{2} \, , & \quad 
B_6 & = -\,\frac{9\g}{2} - r_8 \, , & \quad 
B_7 & = -\,\frac{\g}{4} - \frac{3r_2}{2} \, , \nn \\[5pt]
B_8 & = -\,\frac{\g}{4} - \frac{r_2}{2} \, ,  & \quad
B_9 & = \frac{3\g}{4} + \frac{r_2}{2} - r_3  \, , & \quad
B_{10} & = \frac{3\g}{8} - \frac{3r_2-2r_3}{4} \, , \nn \\[5pt]
B_{11} & = -\,\frac{3\g}{2} + r_1 - r_4 \, , & \quad
B_{12} & =  6\g - 2r_4\, , & \quad
B_{13} & =  -\,\frac{21\g}{4} - \frac{4r_1 + 3r_2 - 4r_4}{2}\, , \nn \\[5pt]
B_{14} & = \frac{3\g}{2} - r_1 + r_4 - \frac{r_7}{2} \, , & \quad 
B_{15} & = -\,3\g - 3r_5 \, , & \quad
B_{16} & = -\frac{3\g}{4} - \frac{3r_2}{2} + 3r_5 - r_8 \, , & \nn \\[5pt]
B_{17} & = \frac{3\g}{4} - \frac{r_2}{2} \, , & \quad 
B_{18} & = \frac{3\g}{2} - \frac{r_7}{2} \, . & \quad &
\end{alignat}
Note that they all depend only on the coupling constant $\g$ and on the coefficients $r_i$ of \eqref{J2_3}, which parameterise the freedom to redefine the higher-spin fields. The same is true for the remaining coefficients of the terms with two derivatives:
\begin{alignat}{5}
C_1 & = -\,\frac{7\g}{2} + 2r_1 \, , & \qquad\qquad 
C_2 & = 10\g + \frac{r_6}{2} \, , \nn \\[5pt]
C_3 & = -\,3\g \, , & \qquad\qquad
C_4 & = -\,12\g + 2r_1 \, , \nn \\[5pt] 
C_5 & = \frac{13\g}{2} + \frac{r_6}{2} \, , & \qquad\qquad 
C_6 & = \frac{\g}{4} + \frac{3r_2}{2} \, , \nn \\[5pt] 
C_7 & = -\,\frac{29\g}{2} + 2r_1 \, , & \qquad\qquad
C_8 & = 21\g + 3r_2 \, , \nn \\[5pt]
C_9 & = -\,\frac{\g}{4} +\frac{3r_2}{2} \, , & \qquad\qquad 
C_{10} & = -\,3\g + \frac{r_7}{2} \, , \nn \\[5pt]
C_{11} & = 3\g + \frac{r_7}{2} \, , & \qquad\qquad 
C_{12} & = -\,7\g + \frac{r_8}{2} \, , \nn \\[5pt] 
C_{13} & = 4\g + \frac{r_8}{2} \, , & \qquad\qquad 
C_{14} & = \frac{9\g}{2} + \frac{3r_2}{2}-\frac{r_7}{4}+\frac{r_{10}}{2} \, , \nn \\[5pt]
C_{15} & = -\,\frac{15\g}{2} + r_1 + \frac{r_7}{2} - r_{10} \, , & \qquad\qquad
C_{16} & = \frac{21\g}{2} + \frac{3r_6}{2} - r_9 \, , \nn \\[5pt] 
C_{17} & = -\,9\g + 4r_1 + 2r_6 -2r_9 \, , & \qquad\qquad
C_{18} & = -\,\frac{21\g}{2} - 2r_6 + \frac{r_7}{2} + 2r_9 \, , \nn \\[5pt] 
C_{19} & = \frac{21\g}{4} + \frac{3r_2}{2} - r_6 + r_9 \, , & \qquad\qquad
C_{20} & = 3\g + \frac{r_6}{2} \, , \nn \\[5pt] 
C_{21} & = -\,3\g + r_1 \, , & \qquad\qquad
C_{22} & = -\,\frac{3\g}{2} + \frac{r_{12}}{2} \, , \nn \\[5pt] 
C_{23} & = \frac{27\g}{4} + \frac{3r_2}{2} + r_8 - r_{12} \, , & \qquad\qquad
C_{24} & = -\,\frac{9\g}{4} + \frac{3r_2}{2} + \frac{r_7}{2} - r_{11} \, , \nn \\[5pt] 
C_{25} & = -\,\frac{9\g}{2} + \frac{3r_7}{2} - 2 r_{11} \, , & \qquad\qquad
C_{26} & = \frac{9\g}{2} - r_7 + \frac{r_8}{2} + 2r_{11} \, , \nn \\[5pt] 
C_{27} & = \frac{3\g}{2} - \frac{r_7}{2} + r_{11} \, . & &
\end{alignat}%
\endgroup
Substituting $\nabla_{\!\m} \to \pr_\m$ and keeping the same coefficients one obtains the 3--3--4 vertex in flat space, while the mass-like term that appear in $AdS$ depends also on the coefficients of the terms with the Ricci tensor in \eqref{J3_3}:
\begin{align}
D_1 &= 33\,\g - r_1 - 6r_4 + 2\,( t_1 + t_3 + 3\,t_8 ) \, , \nn \\[5pt]
D_2 &= -\,\frac{237}{8}\,\g + 2\,r_1 - \frac{5}{4}\,r_2 - \frac{11}{2}\,r_3 - 2\,r_4 - 6\,r_5 + 2\,( t_2 + t_4 + t_5 + 3\,t_9 ) \, , \nn \\[5pt]
D_3 &= -\,15\,\g - 2\,r_1 + \frac{3}{4}\,r_6 -6\,r_9 + 2\,( t_6 + t_{10} + t_{15} + 3 t_{18} ) \, , \nn \\[5pt]
D_4 &= 30\,\g - \frac{1}{2}\,( 3\,r_2 + 2\,r_4 - 4\,r_6 - r_7 + 4\,r_9 + 11\,r_{10} + 12\,r_{11} ) \nn \\
& +\, 2\,( t_7 + t_{11} + t_{12} + t_{16} + 3\,t_{19} ) \, , \nn \\[5pt]
D_5 &= \frac{33}{4}\,\g - 3\,r_5 + r_6 + r_7 + \frac{r_8}{4} - r_9 - 2\,r_{11} - \frac{11}{2}\,r_{12} + 2\,( t_{13} + t_{17} + 3\,t_{20} ) \, .
\end{align}
In order to obtain these results, one has to take into account the dimensional dependent identities that involve the tensors which appear in the gauge transformation of the action. If not, one would discover that, as in $D > 3$, the only solution is a ``fake'' vertex which can be eliminated by a field redefinition. 

\subsection{Coefficients in the gauge transformations}\label{app:coeff3gauge}

The coefficients $a_2$ and $a_5$ in $\d^{(1)}_3\phi^{\m\n\r}$ are not fixed since they parameterise redefinitions of the gauge parameter of the type
\be \label{xi->4xi}
\xi^{\m\n} \to \a_1\, \vf^{\m\n\a\b} \x_{\a\b} + \a_2\, \vf_{\a}{}^{(\m} \x^{\n)\a} \, .
\ee
The remaining coefficients in front of the terms where the derivative acts on the gauge parameter read
\begin{subequations} \label{d3phi3_par}
\begin{alignat}{5}
a_1 & = -\,\frac{3}{4}\,\g + \frac{r_2}{2} - r_5 \, , & \qquad\qquad 
a_3 & = -\,3\,\g + r_1 - r_4 \, , \\[5pt]
a_4 & = -\,\frac{3}{2}\,\g + \frac{r_7}{2} - r_{11} \, , & \qquad\qquad 
a_6 & = 3\,\g + \frac{r_6-r_9}{2} \, , 
\end{alignat}
and 
\begin{align}
c_1 & = \frac{9}{8}\,\g - \frac{1}{4}\,( 4\,r_1 + r_2 - 2\,r_3 - 4\,r_4 ) \, , \\[5pt]
c_2 & = -\,\frac{15}{12}\,\g - \frac{1}{6}\,( 3\,r_2 - 6\,r_5 + 2\,r_6 + 2\,r_7 + 2\,r_8 - 2\,r_9 - 4\,r_{11} - 2\,r_{12} ) \, , \\[5pt]
c_3 & = \frac{15}{12}\,\g - \frac{a_2+a_5}{3} - \frac{r_7 - 2\,r_{10}}{12} \, , \\[5pt]
c_4 & = -\,\frac{21}{6}\,\g - \frac{1}{6}\,( 2\,r_1 - 2\,r_4 + 4\,r_6 + r_7 - 4\,r_9 - 2\,r_{10} ) \, .
\end{align}
\end{subequations}
One can set to zero the previous coefficients that do not depend on $a_2$ and $a_5$ (i.e.\ all of them but $c_3$) by fixing the $r_i$ as
\begin{alignat}{10}
r_1 & = \frac{3\g}{2} \, , \qquad & r_2 & = -\, \frac{3\g}{2} \, , \qquad & r_3 & = 3\g \, , \qquad & r_4 & = -\, \frac{3\g}{2} \, , \nn \\[5pt]
r_5 & = -\, \frac{3\g}{2} \, , \qquad & r_6 & = 0 \, , \qquad & r_7 & = 12\g \, , \qquad & r_8 & = -\, \frac{9\g}{2}\, , \label{rnumbers} \\[5pt]
r_9 & = 6\g \, , \qquad & r_{10} & = \frac{15\g}{2}\, , \qquad & r_{11} & = \frac{9\g}{2}\, , \qquad & r_{12} & = -\, \frac{3\g}{2} \, . \nn
\end{alignat}
This observation generalizes what was already noticed at the quadratic order in \cite{metric-like}. In $D = 3$ one can eliminate almost all terms where the derivative acts on the gauge parameter by tuning appropriately field redefinitions and fixing the free parameters in the gauge transformations. The coefficients of the terms where the derivative acts on the field read instead
\begin{subequations} \label{d3phi3_field}
\begin{alignat}{5}
b_1 & = \frac{3}{2}\,\g \, , & \qquad\quad
b_2 & = -\,\frac{9}{4}\,\g + a_2 - \frac{r_1 - r_4}{2} \, , \\[5pt]
b_3 & = 3\,\g \, , & \qquad\quad 
b_4 & = -\,3\,\g \, , \\[5pt]
b_5 & = -\,\frac{21}{2}\,\g + a_5 - r_6 + r_9 \, , & \qquad\quad 
b_6 & = \frac{3}{4}\,\g \, ,
\end{alignat}
and
\begin{alignat}{5}
d_1 & = -\,\frac{3}{2}\,\g \, , & \qquad\quad
d_2 & = 3\,\g , \\[5pt]
d_3 & = \frac{45}{12}\,\g - \frac{a_2+a_5}{3} + \frac{1}{6}\,( r_1 - r_4 + 2\,r_6 - 2\,r_9 ) \, , & \qquad\quad 
d_4 & = 0 \, . 
\end{alignat}
\end{subequations}

The structure of $\d_4^{(1)}\phi^{\m\n\r}$ is similar: one can leave free the coefficients $p_2$ and $p_5$ that account for the mixing with linearised spin-3 transformations generated by the field dependent parameter
\be \label{xi->3k}
\x^{\m\n} = \a_3\, \phi^{\a\b(\m} \k^{\n)}{}_{\a\b} + \a_4\, \phi_\a{} \k^{\m\n\a} \, .
\ee
The remaining coefficients in front of the terms where the derivative acts on the gauge parameter read
\begin{alignat}{5}
p_1 & = 6\,\g + r_6 - r_9 \, , & \qquad\qquad 
p_3 & = -\,3\,\g + r_1 - r_4 \, , \\[5pt]
p_4 & = -\,\frac{3}{2}\,\g + \frac{r_7}{2} - r_{11} \, , & \qquad\qquad 
p_6 & = -\,\frac{3}{8}\,\g + \frac{r_2}{4} - \frac{r_5}{2} \, , 
\end{alignat}
and 
\begin{align}
v_1 & = -\,\frac{21}{6}\,\g - \frac{1}{6}\,( 2\,r_1 - 2\,r_4 + 4\,r_6 + r_7 - 4\,r_9 - 2\,r_{10} )   \, , \\[5pt]
v_2 & = \frac{9}{24}\,\g - \frac{p_2}{3} - \frac{1}{12}\,( r_2 - 2\,r_3 ) \, , \\[5pt]
v_3 & = \frac{9}{8}\,\g - \frac{1}{4}\,( 4\,r_1 + r_2 - 2\,r_3 - 4\,r_4 ) \, , \\[5pt]
v_4 & = -\,\frac{15}{12}\,\g - \frac{1}{6}\,( 3\,r_2 - 6\,r_5 + 2\,r_6 + 2\,r_7 + 2\,r_8 - 2\,r_9 - 4\,r_{11} - 2\,r_{12} ) \, .
\end{align}
All coefficients but $v_2$ (which depends on $p_2$) also vanish if one fixes the $r_i$ as in \eqref{rnumbers}. The coefficients of the terms where the derivative acts on the field read instead
\begin{alignat}{5}
q_1 & =  3\,\g \, , & \qquad\quad 
q_2 & =  -\,3\,\g + p_2 - r_1 + r_4 \, , \\[5pt]
q_3 & =  6\,\g \, , & \qquad\quad 
q_4 & =  -\,6\,\g \, , \\[5pt]
q_5 & =  -\,\frac{39}{8}\,\g + p_5 - \frac{3}{4}\,(r_2-2\,r_5) \, , & \qquad\quad
q_6 & =  \frac{3}{2}\,\g \, , \\[5pt]
w_1 & = 0 \, , & \qquad\quad
w_2 & = -\,\frac{p_2}{3} + \frac{r_1-r_4}{3} \, , \\[5pt]
w_3 & = -\,3\,\g \, , & \qquad\quad
w_4 & = 6\,\g \, . 
\end{alignat}

In $\d_3^{(1)} \vf^{\m\n\r\s}$ one can leave free the coefficients $\tilde{a}_2$ and $\tilde{a}_4$ that account for the mixing with linearised spin-4 transformations generated by the field dependent parameter
\be \label{k->3xi}
\k^{\m\n\r} = \b_1\, \phi^{\a(\m\n} \x^{\r)}{}_\a + \b_2\, \phi^{(\m} \x^{\n\r)} \, . 
\ee
The remaining coefficients in front of the terms where the derivative acts on the gauge parameter read
\begin{alignat}{5}
\tilde{a}_1 & = -\,\frac{3}{4}\,\g - \frac{r_2}{2} \, , & \qquad 
\tilde{a}_3 & = \frac{3}{4}\,\g - \frac{r_1}{2} \, , \\[5pt]
\tilde{c}_1 & = -\,\frac{21}{20}\,\g + \frac{1}{10}\,( 2\,r_1 + 3\,r_2 + r_7 ) \, , & \qquad 
\tilde{c}_2 & = \frac{9}{10}\,\g - \frac{2}{5}\,\tilde{a}_2 + \frac{r_6}{10} \, , \\[5pt]
\tilde{c}_3 & = -\,\frac{6}{5}\,\g + \frac{4\,r_1 + r_6}{5} \, , & \qquad
\tilde{c}_4 & = \frac{9}{5}\,\g + \frac{3\,r_2 + r_8}{5} \, , \\[5pt]
\tilde{c}_5 & = -\,\frac{3}{10}\,\g - \frac{\tilde{a}_2 + 2\,\tilde{a}_4}{5} + \frac{r_7}{10}\, , & \qquad 
\tilde{c}_6 & = -\,\frac{21}{40}\,\g + \frac{1}{20}\,( 2\,r_1 + 3\,r_2 + r_7 ) \, , \\[5pt]
\tilde{e}_1 & = -\,\frac{2\,r_1 + r_6}{10} \, , & \qquad 
\tilde{e}_2 & = -\,\frac{r_1 + 3\,r_2 + r_7 + r_8}{15} \, .
\end{alignat}
They all vanish apart from $\tilde{c}_2$, $\tilde{c}_5$, $e_1$ and $e_2$ if one fixes the $r_i$ as in \eqref{rnumbers}. The coefficients of the terms where the derivative acts on the field read instead
\begin{alignat}{5}
\tilde{b}_1 & = \frac{3}{2}\,\g \, , & \qquad 
\tilde{b}_2 & = -\,6\,\g + \tilde{a}_2 + r_1 \, , \\[5pt]
\tilde{b}_3 & = 0 \, , & \qquad 
\tilde{b}_4 & = \frac{9}{8}\,\g + \tilde{a}_4 + \frac{3}{4}\,r_2 \, , \\[5pt]
\tilde{d}_1 & = -\,\frac{3}{10}\,\g \, , & \qquad
\tilde{d}_2 & = \frac{21}{10}\,\g - \frac{2}{5}\,\tilde{a}_2 - \frac{2}{5}\,r_1 \, , \\[5pt]
\tilde{d}_3 & = -\,\frac{3}{5}\,\g \, , & \qquad
\tilde{d}_4 & = \frac{3}{5}\,\g \, , \\[5pt]
\tilde{d}_5 & = \frac{27}{20}\,\g - \frac{\tilde{a}_2 + 2\,\tilde{a}_4}{5} - \frac{2\,r_1 + 3\,r_2}{10} \, , & \qquad
\tilde{d}_6 & = -\,\frac{3}{20}\,\g \, , \\[5pt]
\tilde{e}_3 & = \frac{3}{10}\,\g \, , & \qquad
\tilde{e}_4 & = -\,\frac{3}{5}\,\g \, .
\end{alignat}

\section{4--4--4 cubic vertex}\label{app:vertex2}

\subsection{Action}\label{app:action4}

We decompose the general ansatz as
\be \label{ansatz444}
\cL_{4-4-4} = \vf^{\,\m\n\r\s} \left( \cJ_1 + \cJ_2 + \cJ_3 + \cJ_4 \right)_{\m\n\r\s} \, ,
\ee
where $\cJ_4$ contains the terms that one can set to zero thanks to the dimensional dependent identities which involve three rank-4 tensors and two derivatives. $\cJ_2$ collects the terms that one can shift independently with field redefinitions of the form $\vf \to \vf^2$, while $\cJ_3$ collects all terms with the Ricci tensor, that one can eliminate with a field redefinition of the metric of the form $g \to \vf^3$. $\cJ_1$ contains again the non-trivial part of the vertex, which is fixed up to an overall coupling constant if one imposes that the action \eqref{I4} be invariant up to quadratic order in the higher-spin fields:
\begin{align}
& (\cJ_1)_{\m\n\r\s} = \tilde{A}_1 \nabla_{\!\a} \vf_{\b\g\m\n} \nabla^\b \vf^{\a\g}{}_{\r\s} + \tilde{A}_2 \nabla_{\!\m} \vf_{\n\a\b\g} \nabla^\a \vf^{\b\g}{}_{\r\s} + \nabla\cdot \vf_{\m\n\a} \big( \tilde{B}_1 \nabla_{\!\r} \vf_{\s}{}^\a + \tilde{B}_2 \nabla^\a \vf_{\r\s} \big) \nn \\[3pt]
& + \nabla_{\!\m} \vf_{\n\r\a\b} \big( \tilde{B}_3 \nabla\cdot \vf_\s{}^{\a\b} + \tilde{B}_4 \nabla_{\!\s} \vf^{\a\b} + \tilde{B}_5 \nabla^\a \vf^\b{}_\s \big) + \nabla_{\!\a} \vf_{\b\m\n\r} \big( \tilde{B}_6 \nabla\!\cdot  \vf_\s{}^{\a\b} + \tilde{B}_7 \nabla_{\!\s} \vf^{\a\b} \nn \\[3pt]
& + \tilde{B}_8 \nabla^{\b} \vf^\a{}_\s \big) + \tilde{B}_9 \nabla_{\!\m} \vf_{\n\r\s\a} \nabla\!\cdot \vf^\a + \nabla_{\!\m} \vf_{\n\a} \big( \tilde{B}_{10} \nabla_{\!\r} \vf_{\s}{}^\a + \tilde{B}_{11} \nabla^\a \vf_{\r\s} \big) + \tilde{B}_{12} \nabla_{\!\m} \vf_{\n\r} \nabla\!\cdot \vf_\s \nn \\
& + g_{\m\n} \Big\{ \tilde{C}_1 \nabla_{\!\a} \vf_{\b\g\d\r} \nabla^{\b} \vf_\s{}^{\a\g\d} + \tilde{C}_2 \nabla_{\!\r} \vf_{\a\b\g\d} \nabla^{\a} \vf_\s{}^{\b\g\d} + \tilde{C}_3 \nabla_{\!\a} \vf_{\b\g\r\s} \nabla\!\cdot \vf^{\a\b\g} \nn \\
& + \tilde{C}_4 \nabla_{\!\a} \vf_{\b\r} \nabla\!\cdot \vf_\s{}^{\a\b}  + \tilde{C}_5 \nabla_{\!\b} \vf_{\a\g} \nabla^\a \vf_{\r\s}{}^{\b\g} + \tilde{C}_6 \nabla_{\!\r} \vf_{\s\a\b\g} \nabla^{\a} \vf^{\b\g} + \tilde{C}_7 \nabla_{\!\r} \vf_{\a\b} \nabla_{\!\s} \vf^{\a\b} \nn \\[3pt]
& + \tilde{C}_8 \nabla_{\!\r} \vf_{\s\a} \nabla\!\cdot \vf^\a + \tilde{C}_9 \nabla_{\!\a} \vf_{\b\r} \nabla^\b \vf_\s{}^\a + \tilde{C}_{10} \nabla_{\!\a} \vf_{\r\s} \nabla\!\cdot \vf^{\a} \Big\} \nn \\
& + \frac{1}{\ell^2} \Big\{ \tilde{D}_1\, \vf_{\m\n\a\b} \vf_{\r\s}{}^{\a\b} + \tilde{D}_2\, \vf_{\m\n\r\a} \vf_\s{}^\a + g_{\m\n} \big( \tilde{D}_3\, \vf_{\r\s\a\b} \vf^{\a\b} + \tilde{D}_4\, \vf_{\r\a} \vf_\s{}^\a \big) \Big\} \, , \label{J1_4}
\end{align}
where we labelled with $\tilde{A}_i$ the terms that would appear also in the traceless and transverse gauge. A generic quadratic field redefinition of the rank-4 tensor which preserves the double-trace constraint reads
\be \label{redef4->44}
\begin{split}
\vf^{\m\n\r\s} \to & \, \tilde{f}_1\, \vf_{\a\b}{}^{(\m\n} \vf^{\r\s)\a\b} + \tilde{f}_2\, \vf_\a{}^{(\m} \vf^{\n\r\s)\a} + \tilde{f}_3\, \vf^{(\m\n} \vf^{\r\s)} \\
& + g^{(\m\n|}\! \left\{ \tilde{f}_4\, \vf_{\a\b\g}{}^{|\r} \vf^{\s)\a\b\g} + \tilde{f}_5\, \vf^{|\r\s)\a\b} \vf_{\a\b} + \tilde{f}_6\, \vf_\a{}^{|\r} \vf^{\s)\a} \right\} \\
& - \frac{1}{15}\, g^{(\m\n} g^{\r\s)} \Big\{ (2\tilde{f}_1 + 5\tilde{f}_4)\, \vf_{\a\b\g\d} \vf^{\a\b\g\d} \\
& + (\tilde{f_1} + 3\tilde{f}_2 + 2\tilde{f}_3 + 5(\tilde{f}_5+\tilde{f}_6))\, \vf_{\a\b} \vf^{\a\b} \Big\} \, .
\end{split}
\ee
As a result, $\cJ_2$ contains 6 contributions:
\be \label{J2_4}
\begin{split}
(\cJ_2)_{\m\n\r\s} & = \tilde{r}_1 \nabla_{\!\a} \vf_{\b\g\m\n} \nabla^\a \vf^{\b\g}{}_{\r\s} + \tilde{r}_2 \nabla_{\!\a} \vf_{\b\m\n\r} \nabla^{\a} \vf^\b{}_\s + \tilde{r}_3 \nabla_{\!\a} \vf_{\m\n} \nabla^\a \vf_{\r\s} \\
& + g_{\m\n} \big( \tilde{r}_4 \nabla_{\!\a} \vf_{\b\g\d\r} \nabla^{\a} \vf_\s{}^{\b\g\d} + \tilde{r}_5 \nabla_{\!\a} \vf_{\b\g} \nabla^\a \vf_{\r\s}{}^{\b\g} + \tilde{r}_6 \nabla_{\!\a} \vf_{\b\r} \nabla^\a \vf_\s{}^\b \big)\, .
\end{split}
\ee
The independent terms with the Ricci tensors are instead
\begin{align}
& (\cJ_3)_{\m\n\r\s} = R_{\a\b} \big( \tilde{t}_1\, \vf_{\m\n\g}{}^{\a} \vf_{\r\s}{}^{\b\g} + \tilde{t}_2\, \vf_{\m\n\r\g} \vf_\s{}^{\a\b\g} \big) + R_{\m\a} \big( \tilde{t}_3\, \vf_{\n\r\s\b} \vf^{\a\b} + \tilde{t}_4\, \vf^{\a\b}{}_{\n\r} \vf_{\s\b} \big) \nn \\[3pt]
& + \tilde{t}_5\, R_{\m\n} \vf_{\r\s\a\b} \vf^{\a\b} + R\, \big( \tilde{t}_6\, \vf_{\m\n\a\b} \vf_{\r\s}{}^{\a\b} + \tilde{t}_7\, \vf_{\m\n\r\a} \vf_\s{}^\a \big) \nn \\
& + g_{\m\n} \Big\{ \tilde{t}_8\, R_{\a\b} \vf_{\r\s\g}{}^\a \vf^{\b\g} + \tilde{t}_9\, R_{\r\s} \vf_{\a\b\g\d} \vf^{\a\b\g\d} + \tilde{t}_{10}\, R_{\r\a} \vf_{\s\b} \vf^{\a\b} + \tilde{t}_{11}\, R\, \vf_{\r\s\a\b} \vf^{\a\b} \Big\} \, , \label{J3_4}
\end{align}
while there are 7 independent identities that involve two covariant derivatives and two tensors of rank-4. One can thus eliminate from the general ansatz the following terms:
\be
\begin{split}
& (\cJ_4)_{\m\n\r\s} = \tilde{z}_1 \nabla_{\!\m} \vf_{\n\a\b\g} \nabla_{\!\r} \vf_\s{}^{\a\b\g} + \tilde{z}_2 \nabla_{\!\a} \vf_{\m\n\r\s} \nabla\!\cdot  \vf^\a + g_{\m\n} \big( \tilde{z}_3 \nabla_{\!\r} \vf_{\a\b\g\d} \nabla_{\!\s} \vf^{\a\b\g\d} \\ 
& + \tilde{z}_4 \nabla_{\!\r} \vf^{\a\b} \nabla_{\!\a} \vf_{\b\s} \big) + \tilde{z}_5\, R_{\m\n} \vf_{\r\a} \vf_\s{}^\a + g_{\m\n} \big( \tilde{z}_6\, R_{\r\s} \vf_{\a\b} \vf^{\a\b} + \tilde{z}_7\, R\, \vf_{\r\a} \vf_\s{}^\a \big) \, .
\end{split}
\ee
Eq.~\eqref{ansatz444} does not contain all possible contractions of three rank-4 tensors and two derivatives. The reason is the symmetry under exchanges of the three identical rank-4 tensors: to eliminate the freedom to integrate by parts one also have to set to zero some terms and we choose to eliminate from the ansatz
\be \label{J4_4}
\begin{split}
& (\cJ_5)_{\m\n\r\s} = T_1 \nabla\cdot \vf_{\m\n\a} \nabla\cdot \vf_{\r\s}{}^\a + T_2 \nabla\cdot \vf_{\m\n\r} \nabla\cdot \vf_\s + g_{\m\n} \big(\, T_3 \nabla_{\!\r} \vf_{\s\a\b\g} \nabla\!\cdot \vf^{\a\b\g} \\[3pt]
& + T_4 \nabla\!\cdot \vf_{\r\s\a} \nabla\!\cdot \vf^\a + T_5 \nabla\!\cdot \vf_{\a\b\r} \nabla\!\cdot \vf_\s{}^{\a\b} + T_6 \nabla_{\!\r} \vf^{\a\b} \nabla\!\cdot \vf_{\s\a\b} + T_7 \nabla\!\cdot \vf_{\r} \nabla\!\cdot \vf_\s \,\big) \, .
\end{split}
\ee

\subsection{Gauge transformations}\label{app:gauge4}

At cubic level the previous vertex is  insensitive to the presence of a rank-3 tensor in the spectrum. It thus only induces the following deformation of the spin-4 gauge transformation of the rank-4 tensor:
\begin{align}
\d^{(1)}_\k \vf^{\m\n\r\s} & = \tilde{p}_1\, \vf_\a{}^{(\m\n\r} \nabla\!\cdot \k^{\s)\a} + \tilde{p}_2\, \vf_{\a\b}{}^{(\m\n} \nabla^\r \k^{\s)\a\b} + \tilde{p}_3\, \vf_{\a}{}^{\b(\m\n} \nabla_{\!\b\,} \k^{\r\s)\a} + \tilde{p}_4\, \vf^{(\m\n} \nabla\!\cdot \k^{\r\s)} \nn \\[3pt]
& + \tilde{p}_5\, \vf_\a{}^{(\m} \nabla^{\n} \k^{\r\s)\a} + \tilde{p}_6\, \vf^{\a(\m} \nabla_{\!\a\,} \k^{\n\r\s)} + \tilde{q}_1\, \k^{\a\b(\m} \nabla_{\!\a\,} \vf^{\n\r\s)}{}_\b + \tilde{q}_2\, \k^{\a\b(\m} \nabla^\n \vf^{\r\s)}{}_{\a\b} \nn \\[3pt]
& + \tilde{q}_3\, \k^{\a(\m\n} \nabla\!\cdot \vf^{\r\s)}{}_\a + \tilde{q}_4\, \k^{\a(\m\n} \nabla_{\!\a\,} \vf^{\r\s)} + \tilde{q}_5\, \k^{\a(\m\n} \nabla^\r \vf^{\s)}{}_\a + \tilde{q}_6\, \k^{(\m\n\r} \nabla\!\cdot \vf^{\s)} \\
& + g^{(\m\n|} \Big\{ \tilde{v}_1\, \vf_{\a\b\g}{}^{|\r} \nabla^{\s)} \k^{\a\b\g} + \tilde{v}_2\, \vf_{\a\b}{}^{\g|\r} \nabla_{\!\g\,} \k^{\s)\a\b} + \tilde{v}_3\, \vf^{|\r\s)}{}_{\a\b} \nabla\!\cdot \k^{\a\b} + \tilde{v}_4\, \vf_\a{}^{|\r} \nabla\!\cdot \k^{\s)\a} \nn \\
& + \tilde{v}_5\, \vf_{\a\b} \nabla^{|\r} \k^{\s)\a\b} + \tilde{v}_6\, \vf_{\a\b} \nabla^\a \k^{|\r\s)\b} + \tilde{w}_1\, \k^{\a\b\g} \nabla^{|\r} \vf^{\s)}{}_{\a\b\g} + \tilde{w}_2\, \k^{\a\b|\r} \nabla\!\cdot \vf^{\s)}{}_{\a\b} \nn \\[3pt]
& + \tilde{w}_3\, \k^{\a\b\g} \nabla_{\!\a\,} \vf^{|\r\s)}{}_{\b\g} + \tilde{w}_4\, \k^{\a\b|\r} \nabla_{\!\a\,} \vf^{\s)}{}_\b + \tilde{w}_5\, \k^{\a\b|\r} \nabla^{\s)} \vf_{\a\b} + \tilde{w}_6\, \k^{\a|\r\s)} \nabla\!\cdot \vf_{\a} \Big\} \nn \\[3pt]
& + g^{(\m\n} g^{\r\s)} \Big\{ \tilde{y}_1\, \vf_{\a\b\g\d} \nabla^{\a} \k^{\b\g\d} + \tilde{y}_2\, \vf_{\a\b} \nabla\!\cdot \k^{\a\b} + \k^{\a\b\g} \big( \tilde{y}_3\,  \nabla\!\cdot \vf_{\a\b\g} + \tilde{y}_4\, \nabla_{\a} \vf_{\b\g} \big) \Big\} \, , \nn
\end{align}
where preservation of the double trace constraint imposes
\begin{subequations}
\begin{align}
\tilde{y}_1 & = - \frac{1}{15} \left( 2\tilde{p}_2 + 2\tilde{p}_3 + 5\tilde{v}_1 + 5\tilde{v}_2 \right) , \\
\tilde{y}_2 & = - \frac{1}{15} \left( 3\tilde{p}_1 + \tilde{p}_2 +2\tilde{p}_4 + 2\tilde{p}_5 + 5\tilde{v}_3 + 5 \tilde{v}_4 + 5\tilde{v}_5 \right) , \\
\tilde{y}_3 & = - \frac{1}{15} \left( 2\tilde{q}_2 + 2\tilde{q}_3 + 5\tilde{w}_1 + 5\tilde{w}_2 \right) , \\
\tilde{y}_4 & = - \frac{1}{15} \left( 3\tilde{q}_1 + \tilde{q}_2 + 2\tilde{q}_4 + 2\tilde{q}_5 + 5\tilde{w}_3 + 5\tilde{w}_4 + 5\tilde{w}_5  \right) .
\end{align}
\end{subequations}

\subsection{Coefficients in the action}\label{app:coeff4}

If one wants a gauge invariant action, the coefficients in $\cJ_1$ must be fixed as follows:
\begingroup
\allowdisplaybreaks
\begin{alignat}{5}
\tilde{A}_1 & = \r -2\,\tilde{r}_1 \, , & \qquad 
\tilde{A}_2 & = \frac{5}{4}\,\r - 2\,\tilde{r}_1 \, , \nn \\[5pt]
\tilde{B}_1 & = -\,\frac{29}{8}\,\r +2\,\tilde{r}_1 \, , & \qquad 
\tilde{B}_2 & = -\,\r + \tilde{r}_1 - \tilde{r}_3 \, , \nn \\[5pt]
\tilde{B}_3 & = -\,\r \, , & \qquad 
\tilde{B}_4 & = -\,\frac{\r}{8} + \tilde{r}_1 \, , \nn \\[5pt]
\tilde{B}_5 & = -\,\frac{5}{8}\,\r + 2\,\tilde{r}_1 - 3\,\tilde{r}_4 \, , & \qquad 
\tilde{B}_6 & = \frac{\r}{4} - \tilde{r}_2 + \tilde{r}_4 \, , \nn \\[5pt]
\tilde{B}_7 & = -\,\frac{3}{8}\,\r + \frac{\tilde{r}_2 - \tilde{r}_4}{2} \, , & \qquad 
\tilde{B}_8 & = \frac{\r}{8} + \frac{\tilde{r}_2 - 3\,\tilde{r}_4}{2} \, , \nn \\[5pt]
\tilde{B}_9 & = \frac{\r}{4} \, , & \qquad 
\tilde{B}_{10} & = 3\,\r + 3\,\tilde{r}_4 \, , \nn \\[5pt]
\tilde{B}_{11} & = \frac{21}{8}\,\r + \tilde{r}_3 \, , & \qquad 
\tilde{B}_{12} & = -\,\r + 3\,\tilde{r}_4 \, , \nn \\[5pt]
\tilde{C}_1 & = -\,3\,\tilde{r}_4 \, , & \qquad 
\tilde{C}_2 & = \frac{\r}{4} - \tilde{r}_4 \, , \nn \\[5pt]
\tilde{C}_3 & = -\,\frac{\r}{8} - \tilde{r}_3 \, , & \qquad 
\tilde{C}_4 & = \frac{9}{16}\,\r - \tilde{r}_3 + 3\,\tilde{r}_4 + \tilde{r}_5 - \frac{3}{4}\,\tilde{r}_6 \, , \nn \\[5pt]
\tilde{C}_5 & = \frac{5}{8}\,\r + 3\,\tilde{r}_3 - 2\,\tilde{r}_5 \, , & \qquad 
\tilde{C}_6 & = -\,\r + 2\,\tilde{r}_3 + 3\,\tilde{r}_4 - 2\,\tilde{r}_5 \, , \nn \\[5pt]
\tilde{C}_7 & = \frac{17}{32}\,\r - \frac{12\,(\tilde{r}_3-\tilde{r}_5) - \tilde{r}_6}{8} \, , & \qquad 
\tilde{C}_8 & = -\,\frac{33}{32}\,\r - \frac{12\,(\tilde{r}_3-\tilde{r}_5) + \tilde{r}_6}{8} \, , \nn \\[5pt]
\tilde{C}_9 & = -\,\frac{23}{32}\,\r + \frac{12\,(\tilde{r}_3-\tilde{r}_5) + \tilde{r}_6}{8} \, , & \qquad 
\tilde{C}_{10} & = \frac{3}{16}\,\r - \frac{8\,(\tilde{r}_3-\tilde{r}_5) - \tilde{r}_6}{4} \, . 
\end{alignat}
Substituting $\nabla_{\!\m} \to \pr_\m$ and keeping the same coefficients one obtains the 4--4--4 vertex in flat space, while the mass-like term that appear in $AdS$ depends also on the coefficients of the terms with the Ricci tensor in \eqref{J3_4}:
\begin{align}
\tilde{D}_1 & = -\,\frac{15}{2}\,\r + 2\,( 6\,\tilde{r}_1 + \tilde{t}_1 + 3\,\tilde{t}_6 ) \, , \nn \\[5pt]
\tilde{D}_2 & = \frac{11}{2}\,\r + \frac{1}{2}\,( 9\,\tilde{r}_2 + 39\,\tilde{r}_4 + 4\,\tilde{t}_2 + 4\,\tilde{t}_3 + 4\,\tilde{t}_4 + 12\,\tilde{t}_7 ) \, , \nn \\[5pt]
\tilde{D}_3 & = -\,\frac{7}{4}\,\r - 3\,\tilde{r}_3 - 3\,\tilde{r}_4 + 12\,\tilde{r}_5 + 2\,\tilde{t}_5 + 2\,\tilde{t}_8 + 6\,\tilde{t}_{11} \, , \nn \\[5pt]
\tilde{D}_4 & = -\,\frac{101}{16}\,\r + 3\,\tilde{r}_3 + 3\,\tilde{r}_4 - 3\,\tilde{r}_5 + \frac{15}{4}\,\tilde{r}_6 + 2\,\tilde{t}_{10} \, .
\end{align}

\subsection{Coefficients in the gauge transformations}\label{app:coeffgauge4}


The coefficients $\tilde{p}_2$ and $\tilde{p}_5$ in $\d^{(1)}_4\vf^{\m\n\r\s}$ are not fixed since they parameterise redefinitions of the gauge parameter of the type
\be \label{k->4k}
\k^{\m\n\r} \to \b_3\, \vf^{\a\b(\m\n} \k^{\r)}{}_{\a\b} + \b_4\, \vf_{\a}{}^{(\m} \k^{\n\r)\a} \, .
\ee
The remaining coefficients in front of the terms where the derivative acts on the gauge parameter read
\be
\tilde{p}_1 = \frac{\r}{4} + \tilde{r}_4 \, ,  \qquad
\tilde{p}_3 = \frac{\r}{8} + \tilde{r}_1 \, , \qquad
\tilde{p}_4 = -\,\frac{\r}{8} - \tilde{r}_3 + \tilde{r}_5 \, , \qquad
\tilde{p}_6 = \frac{\r}{8} + \frac{\tilde{r}_4}{2} \, , 
\ee
and 
\begin{align}
\tilde{v}_1 & = -\,\frac{3}{40}\,\r - \frac{4}{10}\,\tilde{p}_2 - \frac{\tilde{r}_2 - \tilde{r}_4}{10} \, , \\[5pt]
\tilde{v}_2 & = \frac{9}{40}\,\r - \frac{3}{10}\,( 4\,\tilde{r}_1 + \tilde{r}_2 - \tilde{r}_4 ) \, , \\[5pt]
\tilde{v}_3 & = \frac{\r}{20} - \frac{\tilde{r}_1 + \tilde{r}_3 + 3\,\tilde{r}_4}{5} \, , \\[5pt]
\tilde{v}_4 & = -\,\frac{3}{80}\,\r + \frac{3}{20}\,( 4\,\tilde{r}_3 - 8\,\tilde{r}_4 - 4\,\tilde{r}_5 - \tilde{r}_6 ) \, , \\[5pt]
\tilde{v}_5 & = -\,\frac{3}{40}\,\r - \frac{\tilde{p}_2 + 2\,\tilde{p}_5}{5} - \frac{\tilde{r}_3}{5} \, , \\[5pt]
\tilde{v}_6 & = \frac{\r}{20} - \frac{\tilde{r}_1 + \tilde{r}_3 + 3\,\tilde{r}_4}{5} \, , \\[5pt] 
\tilde{y}_1 & = -\,\frac{\r}{15} + \frac{2}{15}\,( 2\,\tilde{r}_1 + \tilde{r}_2 - \tilde{r}_4 ) \, , \\[5pt]
\tilde{y}_2 & = -\,\frac{\r}{80} + \frac{1}{60}\,( 4\,\tilde{r}_1 + 4\,\tilde{r}_3 + 24\,\tilde{r}_4 + 4\,\tilde{r}_5 + 3\,\tilde{r}_6 ) \, .\end{align}
In analogy with what we have seen in Sect.~\ref{app:coeff3gauge} one can set to zero all previous coefficients but $\tilde{v}_1$, $\tilde{v_5}$ (which depend on $\tilde{p}_2$ and $\tilde{p}_5$), $\tilde{y}_1$ and $\tilde{y}_2$ provided that one fixes the $\tilde{r}_i$ as
\be \label{tildercoeff}
\tilde{r}_1 = -\, \frac{\r}{8} \, , \quad 
\tilde{r}_2 = \r \, , \quad 
\tilde{r}_3 = \frac{9\r}{8} \, , \quad 
\tilde{r}_4 = -\, \frac{\r}{4} \, , \quad
\tilde{r}_5 = \frac{5\r}{4} \, , \quad 
\tilde{r}_6 = \frac{5\r}{4} \, . 
\ee
The coefficients of the terms where the derivative acts on the field read instead
\begin{alignat}{5}
\tilde{q}_1 & = -\,\frac{\r}{2} \, , & \qquad 
\tilde{q}_2 & = \frac{9}{8}\,\r + \tilde{p}_2 - \tilde{r}_1 \, , & \qquad
\tilde{q}_3 & = -\,\r \, , \\[5pt]
\tilde{q}_4 & = \r \, , & \qquad
\tilde{q}_5 & = \frac{9}{8}\,\r + \tilde{p}_5 - \frac{3}{2}\,\tilde{r}_4 \, , & \qquad
\tilde{q}_6 & = -\,\frac{\r}{4} \, ,
\end{alignat}
and
\begin{alignat}{5}
\tilde{w}_1 & = -\,\frac{7\r}{20} - \frac{2}{5}\,\tilde{p}_2 + \frac{2}{5}\,\tilde{r}_1 \, , & \quad
\tilde{w}_2 & = \frac{9}{10}\,\r \, , & \quad
\tilde{w}_3 & = \frac{\r}{10} \, , & \quad
\tilde{w}_4 & = -\,\frac{8}{5}\,\r \, , \\[5pt]
\tilde{w}_5 & = -\,\frac{31\r}{40} - \frac{\tilde{p}_2 + 2\,\tilde{p}_5}{5} + \frac{\tilde{r}_1 + 3\,\tilde{r}_4}{5} \, , & \quad
\tilde{w}_6 & = \frac{\r}{10} \, , & \quad
\tilde{y}_3 & = -\,\frac{\r}{5} \, , & \quad
\tilde{y}_4 & = \frac{\r}{2} \, .
\end{alignat}
\endgroup

\end{appendix}


\end{document}